\documentclass[journal,twocolumn,10pt,twoside]{IEEEtran}
\usepackage{cite}

\usepackage{xcolor}
\usepackage{bbm}
\usepackage[utf8]{inputenc} 
\usepackage[T1]{fontenc}
\usepackage{url}              
\usepackage{cite}             

\usepackage[cmex10]{amsmath}  
\usepackage{mleftright}       
\mleftright                   

\usepackage{graphicx}         
\usepackage{booktabs}         
\newtheorem{theorem}{Theorem}
\newtheorem{definition}{Definition}
\newtheorem{lemma}{Lemma}
\newtheorem{proposition}{Proposition}
\newtheorem{corollary}{Corollary}
\newtheorem{remark}{Remark}

\DeclareMathAlphabet\mathbfcal{OMS}{cmsy}{b}{n}
\begin{document}
\title{Rate-Distortion-Perception Tradeoff Based on the Conditional-Distribution Perception Measure}

\author{Sadaf Salehkalaibar,   Jun Chen, Ashish Khisti, Wei Yu
\thanks{Manuscript submitted to IEEE Transactions on Information Theory on 22 January 2024; revised 19 July 2024; accepted 22 July 2024.
Date of publication 25 September 2024; date of current version 22 November
2024. This work was supported by Huawei Technologies Canada. (\textit{Corresponding author: Sadaf Salehkalaibar.})}
 \thanks{Sadaf Salehkalaibar was with the Department of Electrical and Computer Engineering at the University of Toronto, Toronto, Canada. She is now with the Department of Computer Science at the University of Manitoba, Winnipeg, R3T 5V6, Canada (email: Sadaf.Salehkalaibar@umanitoba.ca),}
 \thanks{Jun Chen is with the Department of Electrical and Computer Engineering at McMaster University, Hamilton, ON L8S 4K1, Canada (email:   chenjun@mcmaster.ca),}
 \thanks{Ashish Khisti and Wei Yu are with the Department of Electrical and Computer Engineering at the University of Toronto, Toronto, M5S 3G4, Canada (email:\{akhisti, weiyu\}@ece.utoronto.ca).}
 \thanks{Communicated by Y. Kochman, Associate Editor for Signal Processing and
Source Coding.}
	}

\maketitle
\begin{abstract}
This paper studies the rate-distortion-perception (RDP) tradeoff for a memoryless source model in the asymptotic limit of large block-lengths. The perception measure is based on a divergence between the  distributions of the source and reconstruction sequences \emph{conditioned} on the encoder output, first proposed by Mentzer et al. We consider the case when there is no shared randomness between the encoder and the decoder and derive a single-letter characterization of the RDP function for the case of discrete memoryless sources. This is in contrast to the marginal-distribution metric case (introduced by Blau and Michaeli), whose RDP characterization remains open when there is no shared randomness.  The achievability scheme is based on  lossy source coding with a posterior reference map. For the case of continuous valued sources under the squared error distortion measure and the squared quadratic Wasserstein perception measure, we also derive a single-letter characterization and show that  the decoder can be restricted to 
a noise-adding mechanism. Interestingly, the RDP function characterized for the case of zero perception loss coincides with that of the marginal metric, and further zero perception loss can be achieved with a 3-dB penalty in minimum distortion. Finally we specialize to the case of Gaussian sources, and derive the RDP function for  Gaussian vector case and propose a reverse water-filling type solution. We also partially characterize the RDP function for a mixture of Gaussian vector sources.
\end{abstract}


\section{Introduction}

Rate-distortion-perception (RDP) tradeoff~\cite{blau2019rethinking}, a generalization of the classical rate-distortion function~\cite{Cover1} to incorporate distribution constraints on the reconstruction, provides a theoretical framework for a variety of deep neural compression systems that exhibit an inherent tradeoff between reconstruction fidelity and realism~\cite{blau2018perception}.  In this framework, the perception loss is measured through a  suitable divergence metric between the source and reconstruction distributions, with perfect realism corresponding to the case when the source and reconstruction distributions are identical.  The work of Blau and Michaeli~\cite{blau2019rethinking} 
establishes that when distortion loss  is measured using mean squared error, perfect realism can be achieved with no more than $3$-dB increase in the minimum distortion.   The work of Theis and Wagner~\cite{Theis-Wagner}  establishes an operational interpretation of the RDP function.  The special case of (scalar) Gaussian sources has been studied in \cite{Jun-Ashish2021} where it is shown that Gaussian distributions attain the RDP function. Furthermore, a natural notion of universality is established, whereby any reconstruction corresponding to a boundary point of the distortion-perception region can be transformed into another reconstruction associated with a different boundary point.  The case when there is limited or no shared randomness between the encoder and decoder has been studied in~\cite{Matsumoto18, Matsumoto19, wagner2022rate,Jun-JSAIT} (see also~\cite{Saldi}).  To the best of the authors' knowledge, unlike the setting with (unlimited) shared   randomness,  a computable characterization of RDP function  remains largely open in these settings\footnote{The perception constraint can be applied at either the symbol level or the sequence level. Under the symbol-wise perception constraint, the RDP function
does not depend on the available amount of shared randomness \cite{Jun-JSAIT}. On the other hand, the RDP function under the sequence-wise perception constraint is still not fully characterized when there is limited or no shared randomness. In this work, we focus on the sequence-wise perception constraint unless specified otherwise.
}. 
The extension of RDP function to the case when correlated side  information is available to either the encoder or the decoder has been studied in~\cite{gunduz1, gunduz2}.
The application of RDP function to neural compression has been studied in e.g.,~\cite{video1, image-comp1, image-comp2, image-comp3, image-comp4, GAN, SSIM,  Huan-Liu} and references therein. 
The RDP tradeoff is practically important as it provides guiding principles for optimizing data compression systems by balancing three critical factors: the compression rate, the reconstruction distortion, and the perceptual quality perceived by human observers or other end-users. Understanding and leveraging this tradeoff enables the creation of efficient, high-quality, and user-friendly compression technologies that cater to the evolving demands of various industries.

While the perception loss metric in prior works \cite{blau2018perception} is based on the divergence between  the source and reconstruction distributions, a different choice is proposed in~\cite{video1, GAN}, which empirically observe that a perception loss metric that measures the divergence between the source and reconstruction distributions  \emph{conditioned} on the output of the encoder results in higher perceptual quality in reconstructions. 
The rationale behind such a perception metric can be intuitively understood as follows. In a lossy source coding system designed optimally in the traditional rate-distortion sense, there is essentially a bijection between the encoder output and the reconstruction. By adopting the encoder from this system and incorporating a  perception metric conditioned on the encoder output, the decoder is compelled to follow the conditional distribution of the source given the traditional rate-distortion reconstruction. Specifically, in image coding, the conditional metric forces the decoder to adjust fine details that reduce blurriness, without deviating significantly from the minimum-distortion reconstruction. Moreover, the conditional metric provides finer control over the perceptual quality than the marginal counterpart, as it enforces consistency between the source and reconstruction distributions for every realization of the encoder output.

To our knowledge, a theoretical study of the RDP function when the perception measure is based on such a conditional metric has not been previously considered\footnote{Recently, a related problem is studied in \cite{xu2023conditional}, where the perception loss is measured through a divergence between the source and reconstruction distributions conditioned on a certain side information; moreover, it is assumed that there is unlimited shared randomness between the encoder and decoder.}.  In this work we make the assumption that there is no shared randomness between the encoder and decoder, and denote this setting as the \emph{conditional-distribution-based perception measure} while denoting the original setting of \cite{blau2018perception} as the \emph{marginal-distribution-based perception measure}. 
The main contributions of this paper are as follows:

\begin{itemize}
\item We characterize the RDP tradeoff for finite alphabet sources  (Theorem~\ref{thm:RDP}) and explicitly derive the tradeoff for the uniform Bernoulli source (Theorem~\ref{thm:binary}). The achievable scheme uses some recent tools developed for lossy source coding with a posterior reference map \cite{ASP23}. It is interesting to note that a complete characterization of the RDP function for the conditional-distribution-based perception measure is possible in the absence of shared randomness, while a similar characterization of the RDP function for the marginal-distribution-based perception measure only exists for the case of zero perception loss (i.e., when the source and reconstruction distributions exactly match)~\cite{wagner2022rate}, if there is no shared  randomness. 
\item The RDP tradeoff is further characterized for continuous alphabet sources (Theorem~\ref{thm:continuous}) under the squared error distortion measure and the squared quadratic Wasserstein perception measure. For the case of zero perception loss, it is shown that the rate-distortion tradeoff coincides with that of the marginal-distribution-based perception measure (Corollary~\ref{cor:CM}). As it has been previously observed for the latter measure \cite{Jun-JSAIT}, \cite{YWYML21}, \cite{TA21},  at a fixed rate, 
the distortion of the optimal reconstruction satisfying zero perception loss is exactly twice that of the traditional rate-distortion reconstruction. 
Furthermore, there is no loss of optimality in replacing any given representation with the minimum mean square error (MMSE) estimate of the source based on that representation and generating the reconstruction from this estimate using a noise-adding mechanism.
\item Shannon's lower bound \cite[Eq. (13.159)]{Cover1} for the RD tradeoff is extended to the RDP setting (Theorem~\ref{thm:lowerbound}). Using this lower bound, we are able to partially characterize the RDP tradeoff for a Gaussian mixture source (Corollary~\ref{cor:Gaussianmixture}) and completely characterize the tradeoff for a  Gaussian vector source (Corollary~\ref{cor:Gaussian}). For  the latter case, a reverse water-filling type solution is also derived. 
\end{itemize}

The rest of this paper is organized as follows. 
Section \ref{sec:formulation} introduces the system model as well as the operational and the informational definitions of RDP function with conditional-distribution-based perception measure. In Section \ref{sec:discrete}, we establish a single-letter characterization of the operational RDP function by showing that it coincides with its informational version; for the uniform Bernoulli source under Hamming distortion measure and divergence induced by Hamming cost function, a more explicit characterization is obtained. In Section \ref{sec:continuous}, we prove that the aforementioned single-letter characterization continues to hold for square-integrable sources under the squared error distortion measure and the squared quadratic Wasserstein perception measure, and matches its marginal-distribution-based counterpart when the perceptual quality is required to be perfect; we also extend the Shannon lower bound to  our setting, which yields a partial characterization of the RDP function for Gaussian mixture sources and a complete solution for Gaussian sources. 
Section \ref{sec:conclusion} includes some concluding remarks.

Notation: Entropy, differential entropy, and mutual information are denoted by $H(\cdot)$, $h(\cdot)$, and $I(\cdot;\cdot)$, repsectively. We use $[a:b]$ to represent the set of integers from $a$ to $b$ for any two integers $a\leq b$, and use $x\wedge y$ to represent the minimum of two real numbers $x$ and $y$ . The cardinality of set $\mathcal{S}$ is written as $|\mathcal{S}|$.
Let $\mbox{Ber}(p)$ and $\mathcal{N}(\mu,\Sigma)$ denote respectively the Bernoulli distribution with parameter $p$ and the Gaussian distribution with mean $\mu$ and covariance matrix $\Sigma$. For two matrices $A$ and $B$, we write $A\succ  B$ ($A\succeq B$) if $A-B$ is positive definite (semidefinite). Throughout this paper, the base of the logarithm function is $e$.

\section{Problem Formulation}\label{sec:formulation}
\begin{figure*}[t]
\centering
\includegraphics[scale=0.5]{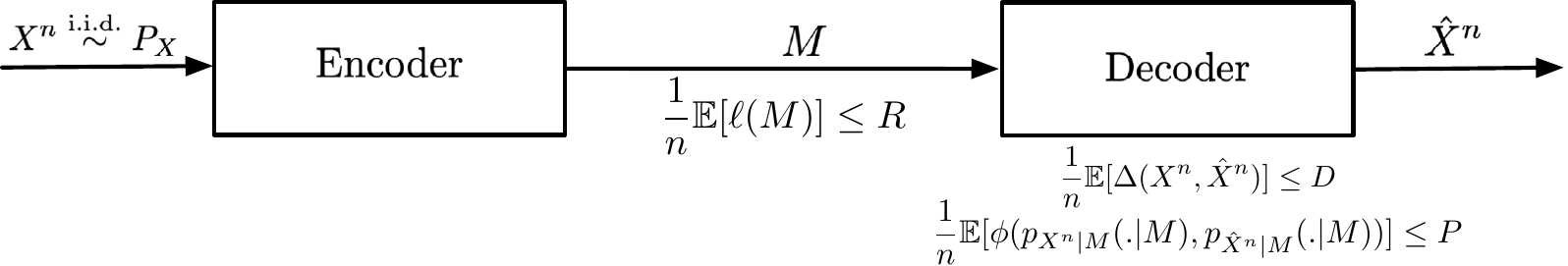}
\caption{System model with the perception measure based on the conditional distribution.}
\label{fig:sys_model}
\end{figure*}

\subsection{System Model}\label{sys-model-section}

Let the source $\{X(t)\}_{t=1}^{\infty}$ be a stationary and memoryless process with marginal distribution $p_X$ over alphabet $\mathcal{X}$ (see Fig.~\ref{fig:sys_model}). A stochastic encoder $f^{(n)}:\mathcal{X}^n\rightarrow\mathcal{M}$ maps a length-$n$ source sequence $X^n$ to a codeword $M$ in a binary prefix code $\mathcal{M}$ according to some conditional distribution $p_{M|X^n}$. A stochastic decoder $g^{(n)}:\mathcal{M}\rightarrow\mathcal{X}^n$ then generates a length-$n$ reconstruction sequence $\hat{X}^n$ based on $M$ according to some conditional distribution $p_{\hat{X}^n|M}$. 
Note that this coding system induces the following joint distribution:
\begin{align}
p_{X^nM\hat{X}^n}=p_{X^n}p_{M|X^n}p_{\hat{X}^n|M}.
\end{align}

Let $\Delta: \mathcal{X}\times \mathcal{X}\to [0,\infty)$ be a distortion measure with $\Delta(x,\hat{x})=0$ if and only if $x=\hat{x}$. Define $\Delta(x^n,\hat{x}^n)=\sum_{t=1}^n\Delta(x(t),\hat{x}(t))$ for $x^n,\hat{x}^n\in\mathcal{X}^n$. Let $\phi:\mathcal{P}\times\mathcal{P}\rightarrow[0,\infty]$ be a divergence with $\phi(p,p')=0$ if and only if $p=p'$ a.s., where $\mathcal{P}$ denotes the set of probability distributions. In this work, $\phi$ will serve the role of perception measure (applied at the sequence level). 
It will become clear that the single-letterization of the RDP function is facilitated by $\phi$ having a certain ``additive" structure. 
For this reason, we focus on a special class of divergences that arise from the theory of optimal transport. Specifically, for any two probability distributions $p_{\tilde{X}^n}$ and $p_{\bar{X}^n}$ over $\mathcal{X}^n$, let 
\begin{align}
	\phi(p_{\tilde{X}^n},p_{\bar{X}^n})=\inf\limits_{p_{\tilde{X}^n\bar{X}^n}\in\Pi(p_{\tilde{X}^n},p_{\bar{X}^n})}\sum\limits_{t=1}^n\mathbbm{E}[c(\tilde{X}(t),
	\bar{X}(t))],\label{eq:divergence}
\end{align}
where $\Pi(p_{\tilde{X}^n},p_{\bar{X}^n})$ denotes the set of all joint distributions whose marginals are $p_{\tilde{X}^n}$ and $p_{\bar{X}^n}$, respectively, and $c:\mathcal{X}\times\mathcal{X}\rightarrow[0,\infty)$ is  a cost function with $c(\tilde{x},\bar{x})=0$ if and only if $\tilde{x}=\bar{x}$. Note that this class is quite rich as it contains all $p$-Wasserstein distances (raised to the $p$-th power).

\begin{proposition}\label{prop:property}
	$\phi$ defined in (\ref{eq:divergence}) has the following properties.
	\begin{enumerate}
		\item Tensorizability:
		\begin{align}
			\phi(p_{\tilde{X}^n},p_{\bar{X}^n})\geq\sum\limits_{t=1}^n\phi(p_{\tilde{X}(t)},p_{\bar{X}(t)})
		\end{align}
		and the equality holds if $p_{\tilde{X}^n}=\prod_{i=1}^np_{\tilde{X}(t)}$ and $p_{\bar{X}^n}=\prod_{i=1}^np_{\bar{X}(t)}$.

		\item Convexity:
		\begin{align}
			&\phi((1-\lambda)p_{\tilde{X}^n}+\lambda p'_{\tilde{X}^n},(1-\lambda)p_{\bar{X}^n}+\lambda p'_{\bar{X}^n})\nonumber\\
   &\leq(1-\lambda)\phi(p_{\tilde{X}^n},p_{\bar{X}^n})+\lambda \phi(p'_{\tilde{X}^n},p'_{\bar{X}^n})
		\end{align}
	for $\lambda\in[0,1]$.
		
		\item Continuity: 
		\begin{align}
			&|\phi(p_{\tilde{X}^n},p_{\bar{X}^n})-\phi(p_{\tilde{Y}^n},p_{\bar{Y}^n})|\nonumber\\
   &\leq nc_{\max}(d_{\text{TV}}(p_{\tilde{X}^n},p_{\tilde{Y}^n})+d_{\text{TV}}(p_{\bar{X}^n},p_{\bar{Y}^n})),
		\end{align}
	where $d_{\text{TV}}$ is the total variation distance, and $c_{\max}=\sup_{x,x'\in\mathcal{X}}c(x,x')$.
	\end{enumerate}
\end{proposition}
\begin{IEEEproof}
	See Appendix \ref{proof:property}.
\end{IEEEproof}	
	



\subsection{Rate-Distortion-Perception Function}

\begin{definition}\label{def:opRDP}
	We say rate $R$ is achievable subject to distortion and percetion constraints $D$ and  $P$ if for some $n$, there exist encoder $f^{(n)}$ and decoder $g^{(n)}$ such that (see Fig.~\ref{fig:sys_model})
	\begin{align}
		&\frac{1}{n}\mathbbm{E}[\ell(M)]\leq R,\\
		&\frac{1}{n}\mathbbm{E}[\Delta(X^n,\hat{X}^n)]\leq D,\\
		&\frac{1}{n}\mathbbm{E}[\phi(p_{X^n|M}(\cdot|M),p_{\hat{X}^n|M}(\cdot|M))]\leq P,\label{eq:cdP}
	\end{align}
	where $\ell(M)$ denotes the length of $M$. The infimum of all such $R$ is denoted by $R_{\text{C}}(D,P)$, which is referred to as the operational RDP function for the conditional-distribution-based perception measure. 
\end{definition}

To the end of characterizing $R_{\text{C}}(D,P)$, we introduce the following informational RDP function:
\begin{align}
	R(D,P)=&\inf\limits_{p_{U\hat{X}|X}}I(X;U)\label{eq:informationRDP}\\
	&\hspace{0.2cm}\mbox{s.t. }X\leftrightarrow U\leftrightarrow\hat{X}\mbox{ form a Markov chain},\label{eq:RDP1}\\
	&\hspace{0.8cm}\mathbbm{E}[\Delta(X,\hat{X})]\leq D,\label{eq:distortion1}\\
	&\hspace{0.8cm}\mathbbm{E}[\phi(p_{X|U}(\cdot|U),p_{\hat{X}|U}(\cdot|U))]\leq P.\label{eq:perception1}
\end{align}
The auxiliary random variable $U$ can be interpreted as a representation of $X$ and corresponds to the encoder output $M$ in the operational definition.

\begin{proposition}\label{prop:convexity}
	$R(D,P)$ is convex in $(D,P)$.
\end{proposition}
\begin{IEEEproof}
	See Appendix \ref{proof:convexity}.
\end{IEEEproof}

\begin{proposition}
	If $|\mathcal{X}|<\infty$, then there is no loss of generality in assuming that the alphabet of $U$, denoted by $\mathcal{U}$, satisfies $|\mathcal{U}|\leq|\mathcal{X}|+2$; moreover, the infimum in the definition of $R(D,P)$ can be attained, thus is a minimum.
\end{proposition}
\begin{IEEEproof}
	Note that $|\mathcal{X}|<\infty$ implies $c_{\max}<\infty$, which in light of part 3) of Proposition \ref{prop:property} further implies the continuity of $\phi(p,p')$ in $(p,p')$.
	Therefore, we can invoke the support lemma \cite[p. 631]{ElGamal} to establish the desired cardinality bound. Moreover, the continuity of $\phi(p,p')$, together with the cardinality bound, implies that the feasible domain for $p_{U\hat{X}|X}$  is compact. As a consequence, the objective function $I(X;U)$, which is continuous in $p_{U\hat{X}|X}$, has a minimum value over this domain.
\end{IEEEproof}

\section{Finite Alphabet Sources}\label{sec:discrete}

We focus on finite alphabet sources in this section. The first main result shows that the operational RDP function coincides with its informational counterpart for such sources. 

\begin{theorem}\label{thm:RDP}
Assume $|\mathcal{X}|<\infty$. For $D\geq 0$ and $P\geq 0$,
	\begin{align}
		R_{\text{C}}(D,P)=R(D,P).
	\end{align}
\end{theorem}
\begin{IEEEproof}
	See Appendix~\ref{app:RDP}. The achievability part of the proof relies on a recent development in information theory known as lossy source coding with a 
	posterior reference map \cite{ASP23}. Roughly speaking, given a backward test channel $p_{X|U}$ (referred to as a posterior reference map), it is possible to construct an encoder of rate approximately $I(X;U)$ such that $p_{X^n|M}$ is close to $p^n_{X|U}$ in total variation distance; the decoder is simply a generator with
	$p_{\hat{X}^n|M}$ simulating  $p^n_{\hat{X}|U}$ for some $p_{\hat{X}|U}$. The problem then boils down to choosing suitable $p_{X|U}$ and $p_{\hat{X}|U}$ to meet the distortion and perception constraints.	
	\end{IEEEproof}
\begin{remark}
	The proof actually indicates that if (\ref{eq:cdP}) is replaced with the following stronger constraint
	\begin{align}
		\frac{1}{n}\phi(p_{X^n|M}(\cdot|m),p_{\hat{X}^n|M}(\cdot|m))\leq P,\quad m\in\mathcal{M},
	\end{align}
	Theorem \ref{thm:RDP} continues to hold.
\end{remark}


The next result provides an explicit characterization of $R_{\text{C}}(D,P)$ for the uniform Bernoulli source (i.e., $X\sim\mbox{Ber}(\frac{1}{2})$) under Hamming distortion measure (i.e., $\Delta(x,\hat{x})=d_H(x,\hat{x})$) and divergence induced by Hamming cost function (i.e., $c(x,\hat{x})=d_H(x,\hat{x})$). 
For $D\geq 0$ and $P\geq 0$, let 
\begin{align}
	\hbar(D,P)=\begin{cases}
		H_b\left(\frac{1+(D\wedge P)-\sqrt{1+(D\wedge P)^2-2D}}{2}\right), & D\in[0,\frac{1}{2}), \\
		\log2, & D\in[\frac{1}{2},\infty),
	\end{cases}
\end{align}
where $H_b$ denotes the binary entropy function. Moreover, let $\overline{\hbar}$ be the upper concave envelope of $\hbar$ over $[0,\infty)^2$. 

\begin{theorem}\label{thm:binary}
	Assume $X\sim\text{Ber}(\frac{1}{2})$, $\Delta(x,\hat{x})=d_H(x,\hat{x})$, and $c(x,\hat{x})=d_H(x,\hat{x})$.  For $D\geq 0$ and $P\geq 0$,
	\begin{align}
		R_{\text{C}}(D,P)=\log 2-\overline{\hbar}(D,P).
	\end{align}
\end{theorem}
\begin{IEEEproof}
	See Appendix \ref{app:binary}. 
\end{IEEEproof}
\begin{remark}
	The upper concave envelope operation is necessary as $\hbar$ itself is not concave in $(D,P)$. See Appendix \ref{app:non-concavity} for some relevant analysis. 
\end{remark}

\begin{remark}
	Theorem \ref{thm:binary} implies that for $D\geq 0$ and $P\geq D$,
	\begin{align}
		R_{\text{C}}(D,P)&\leq\log 2-\hbar(D,P)\nonumber\\
		&=\begin{cases}
			\log2-H_b(D),& D\in[0,\frac{1}{2}),\\
			0,& D\in[\frac{1}{2},\infty).
		\end{cases}
	\end{align}
This upper bound is tight because it coincides with the rate-distortion function of the uniform Bernoulli source under Hamming distortion measure \cite[Theorem 13.3.1]{Cover1}, which is the infimum of  achievable rates when the perception constraint is ignored.
\end{remark}

We now contrast $R_{\text{C}}(D,P)$ with the operational RDP function for the marginal-distribution-based perception measure, denoted as $R_{\text{M}}(D,P)$, which can be defined similarly by replacing (\ref{eq:cdP}) with
\begin{align}
	\frac{1}{n}\phi(p_{X^n},p_{\hat{X}^n})\leq P.\label{eq:mdP}
\end{align}
It follows by part 2) of Proposition \ref{prop:property} that (\ref{eq:cdP}) implies (\ref{eq:mdP}). 
As a consequence, we must have
\begin{align}
	R_{\text{M}}(D,P)\leq R_{\text{C}}(D,P).\label{eq:M<C}
\end{align}
Different from $R_{\text{C}}(D,P)$, a single-letter characterization of $R_{\text{M}}(D,P)$ is unavailable except for the special case $P=0$, where it is known \cite[Section III.B]{Saldi}, \cite[Corollary 1]{wagner2022rate}, \cite[Eq. (16)]{Jun-JSAIT} that 
\begin{align}
	R_{\text{M}}(D,0)=&\inf\limits_{p_{U\hat{X}|X}}\max\{I(X;U), I(\hat{X};U)\}\label{eq:RDP_M}\\
	&\hspace{0.3cm}\mbox{s.t. }X\leftrightarrow U\leftrightarrow\hat{X}\mbox{ form a Markov chain},\label{eq:M1}\\
	&\hspace{0.8cm}\mathbbm{E}[\Delta(X,\hat{X})]\leq D,\label{eq:M2}\\
	&\hspace{0.8cm}p_{\hat{X}}=p_X.\label{eq:M3}
\end{align}
The difficulty in characterizing $R_{\text{M}}(D,P)$ arises from the fact that the i.i.d. form of the reconstruction sequence $\hat{X}^n$ favored by the perception constraint (\ref{eq:mdP}) (see part 1 and part 2 of Proposition \ref{prop:property}) is not necessarily desirable from the rate perspective. This tension disappears when $P=0$ as $\hat{X}^n$ is forced to be an i.i.d. sequence. In contrast, under constraint (\ref{eq:cdP}), 
the conditional i.i.d. form of the source sequence $X^n$ and the reconstruction sequence $\hat{X}^n$ given the codeword $M$ is desirable from both the perception and the rate perspectives. This explains why $R_{\text{C}}(D,P)$ is more amenable to single-letterization as compared to $R_{\text{M}}(D,P)$.

Note that the single-letter characterization of $R_{\text{M}}(D,0)$ in (\ref{eq:RDP_M})--(\ref{eq:M3}) takes a different form than $R(D,P)$ in (\ref{eq:informationRDP})--(\ref{eq:perception1}) with $P=0$.
Setting $P=0$ in (\ref{eq:perception1}) forces $p_{\hat{X}|U}=p_{X|U}$, which, in turn, implies $p_{\hat{X}}=p_X$ and $I(\hat{X};U)=I(X;U)$.
However, it is unclear whether $p_{\hat{X}|U}$ and $p_{X|U}$ induced by the optimal $p_{U\hat{X}|X}$ associated with the single-letter characterization of $R_{\text{M}}(D,0)$   must be identical. This leaves open the question of the strictness of  the inequality in (\ref{eq:M<C})  when $P=0$.

For the special case of $X\sim\mbox{Ber}(\frac{1}{2})$ and $\Delta(x,\hat{x})=d_H(x,\hat{x})$, in light of Theorem \ref{thm:binary}, we have
\begin{align}
	R_{\text{C}}(D,0)&\leq\log 2-\hbar(D,0)\nonumber\\
	&=\begin{cases}
		\log 2-H_b(\frac{1-\sqrt{1-2D}}{2}), &D\in[0,\frac{1}{2}),\\
		0,&D\in[\frac{1}{2},\infty).
	\end{cases}
\end{align}
Interestingly, the corresponding $R_{\text{M}}(D,0)$ is given by \cite[Theorem 7]{Jun-JSAIT}
\begin{align}
	R_{\text{M}}(D,0)=\begin{cases}
		\log 2-H_b(\frac{1-\sqrt{1-2D}}{2}), &D\in[0,\frac{1}{2}),\\
		0,&D\in[\frac{1}{2},\infty).
	\end{cases}
\end{align}
In view of (\ref{eq:M<C}), we must have
\begin{align}
	R_{\text{C}}(D,0)=R_{\text{M}}(D,0)\label{eq:C=M}
\end{align}
for this special case.
As shown in the next section (see Corollary \ref{cor:CM}), this is a general phenomenon under the squared error distortion measure, rather than a coincidence.

\section{Continous Alphabet Sources}\label{sec:continuous}

In this section, we characterize $R_{\text{C}}(D,P)$  for continuous alphabet sources, more specifically, the case where $X=(X_1,X_2,\ldots,X_L)^T$ is a random vector  with $\mathcal{X}=\mathbbm{R}^L$. The following result indicates that Theorem \ref{thm:RDP} continues to hold for square-integrable sources (i.e., $\mathbbm{E}[\|X\|^2]<\infty$) under the squared error
distortion measure (i.e, $\Delta(x,\hat{x})=\|x-\hat{x}\|^2$) and the squared quadratic Wasserstein
perception measure (i.e., $\phi(p,p')=W^2_2(p,p')$, resulting from choosing $c(x,\hat{x})=\|x-\hat{x}\|^2$). As  $D=0$ corresponds to lossless source coding, which is generally impossible for continuous alphabet sources (unless $p_X$ has a discrete support), we focus on the case $D>0$ throughout this section.

\begin{theorem}\label{thm:continuous}
	Assume $\mathbbm{E}[\|X\|^2]<\infty$, $\Delta(x,\hat{x})=\|x-\hat{x}\|^2$, and $\phi(p,p')=W^2_2(p,p')$. For $D>0$ and $P\geq 0$,
	\begin{align}
		R_{\text{C}}(D,P)=R(D,P).\label{eq:continuousRDP}
	\end{align}
Moreover, in this case, 
\begin{align}
	R(D,P)=R'(D,P),\label{eq:alternative}
\end{align}
where
\begin{align}
	R'(D,P)=&\inf\limits_{p_{U'\hat{X}'|X}}I(X;U')\\
	&\hspace{0.5cm}\mbox{s.t. }X\leftrightarrow U'\leftrightarrow\hat{X}'\mbox{ form a Markov chain},\label{eq:RDP'1}\\
	&\hspace{1cm}U'=\mathbbm{E}[X|U']=\mathbbm{E}[\hat{X}'|U']\mbox{ almost surely},\label{eq:RDP'2}\\
	&\hspace{1cm}\mathbbm{E}[\|V\|^2]+\mathbbm{E}[\|\hat{V}\|^2]\leq D,\label{eq:RDP'3}\\
	&\hspace{1cm}\mathbbm{E}[W^2_2(p_{V|U'}(\cdot|U'),p_{\hat{V}|U'}(\cdot|U'))]\leq P,\label{eq:RDP'4}
\end{align}
with $V=X-U'$ and $\hat{V}=\hat{X}'-U'$.
\end{theorem}
\begin{IEEEproof}
	See Appendix \ref{app:continuous}. We establish (\ref{eq:continuousRDP}) by reducing the continuous alphabet case to the finite alphabet case through a delicate quantization argument. 
 The proof of 
(\ref{eq:alternative})
is based on the observation that under the squared error distortion measure and the squared quadratic Wasserstein perception measure, no optimality is lost by replacing $U$ and $\hat{X}$ in the definition of $R(D,P)$ respectively with $U'$ and $\hat{X}'$,  where $U'=\mathbbm{E}[X|U]$ and $\hat{X}'=U'+\hat{X}-\mathbbm{E}[\hat{X}|U]$, then leveraging the induced $p_{X|U'}$ and $p_{\hat{X}'|U'}$ to restore the Markov structure $X\leftrightarrow U'\leftrightarrow \hat{X}'$.
\end{IEEEproof}

\begin{remark}\label{re:strict}
In some scenarios, the source only takes on values from a strict subset $\mathcal{X}$ of $\mathbbm{R}^L$ and the reconstruction is also confined to $\mathcal{X}$. The proof of (\ref{eq:continuousRDP}) is not directly applicable to such scenarios as the output of quantizer $\xi$ may live outside $\mathcal{X}$ (except for the special case $P=0$ where the reconstruction is forced to have the same distribution as the source). Nevertheless, it can be shown (see Appendix \ref{app:strict}) using a more delicate argument that (\ref{eq:continuousRDP}) continues to hold in the aforementioned scenarios (correspondingly, $R(D,P)$ is defined with $\hat{X}$ restricted to $\mathcal{X}$). On the other hand, except for the special case $P=0$, the proof of (\ref{eq:alternative}) relies critically on the fact that $\hat{X}$ and $\hat{X}'$ in the definition of $R(D,P)$ and $R'(D,P)$ have the freedom to take on values from $\mathbbm{R}^L$.
\end{remark}

\begin{remark}
The proof of ({\ref{eq:alternative}}) indicates that  there is no loss of optimality in replacing any given representation $U$ with the MMSE estimate  of the source based on $U$, and then adding noise to this estimate to generate the desired reconstruction.
See \cite[Theorem 3]{freirich2021theory} and \cite[Theorem 1]{Yan22} for the marginal-distribution-based counterpart of this result.     
    It is also closely related to the universality of the traditional rate-distortion reconstruction observed in the setting with (unlimited) shared randomness  \cite{Jun-Ashish2021}.
\end{remark}


The following result indicates that $R_{\text{C}}(D,0)$ is always equal to $R_{\text{M}}(D,0)$ under the  squared error distortion measure, and connects them to the classical rate-distortion function.

\begin{corollary}\label{cor:CM}
Assume $\mathbbm{E}[\|X\|^2]<\infty$ and  $\Delta(x,\hat{x})=\|x-\hat{x}\|^2$. For $D>0$,
\begin{align}\label{eq:twoequalities}
	R_{\text{C}}(D,0)=R_{\text{M}}(D,0)=R\left(\frac{D}{2}\right),
\end{align}
where
\begin{align}
	R\left(\frac{D}{2}\right)=&\inf\limits_{p_{\bar{U}|X}}I(X;\bar{U})\\
	&\hspace{0.3cm}\mbox{s.t.}\quad\mathbbm{E}[\|X-\bar{U}\|^2]\leq\frac{D}{2}.\label{eq:conventionalRD}
\end{align}
\end{corollary}
\begin{IEEEproof}
	See Appendix \ref{app:CM}.
\end{IEEEproof}
\begin{remark}
	The second equality in (\ref{eq:twoequalities}) is known \cite[Eq. (20)]{Jun-JSAIT} (see also \cite[Theorem 2]{YWYML21}, \cite[Theorem 2]{TA21}).
\end{remark}
\begin{remark}
	Note that (\ref{eq:C=M}) can be viewed as a special case of Corollary \ref{cor:CM}, because $d_H(x,\hat{x})=\|x-\hat{x}\|^2$ for $x,\hat{x}\in\{0,1\}$.  
\end{remark}
\begin{remark}
	In the definition of $R(\frac{D}{2})$, we allow $\bar{U}$ to have the freedom to take on values from $\mathbbm{R}^L$ even if $X$ is only defined over a strict subset of $\mathbbm{R}^L$. Otherwise, the second equality in (\ref{eq:twoequalities}) might not hold. For example, when $X\sim\mbox{Ber}(\frac{1}{2})$ and $\Delta(x,\hat{x})=\|x-\hat{x}\|^2$, we have $R_{\text{C}}(\frac{1}{2},0)=R_{\text{M}}(\frac{1}{2},0)=0$; on the other hand, $R(\frac{1}{4})=0$ only if $\bar{U}$ is allowed to be equal to $\frac{1}{2}$, which does not belong to $\{0,1\}$. 	
	See Remark \ref{re:strict} for a related discussion.
\end{remark}

The next result extends the Shannon lower bound \cite[Eq. (13.159)]{Cover1} to the RDP setting.
\begin{theorem}\label{thm:lowerbound}
	Assume $\sigma^2_{\ell}=\mathbbm{E}[(X_{\ell}-\mathbbm{E}[X_{\ell}])^2]\in(0,\infty)$, $\ell\in[1:L]$, $\Delta(x,\hat{x})=\|x-\hat{x}\|^2$, and $\phi(p,p')=W^2_2(p,p')$. For $D>0$ and $P\geq 0$,
	\begin{align}
		R_{\text{C}}(D,P)\geq h(X)-\sum\limits_{\ell=1}^L\frac{1}{2}\log(2\pi e\omega_{\ell}),
	\end{align}
where 
\begin{align}
	\omega_{\ell}=\begin{cases}
		\omega\wedge\sigma^2_{\ell},&D^*<\sum_{\ell'=1}^L\sigma^2_{\ell'},\\
		\sigma^2_{\ell},&D^*\geq \sum_{\ell'=1}^L\sigma^2_{\ell'},
	\end{cases}\quad\ell\in[1:L],\label{eq:omega}
\end{align}
with $D^*$ defined as
\begin{align}D^*=\frac{D+\sqrt{(2D-(D\wedge P))(D\wedge P)}}{2}\label{eq:D*}
  \end{align}
  and
$\omega$ being the unique solution to
\begin{align}
	\sum\limits_{\ell=1}^L(\omega\wedge\sigma^2_{\ell})=D^*.
\end{align}
\end{theorem}
\begin{IEEEproof}
	See Appendix \ref{app:lowerbound}.
\end{IEEEproof}

The following result provides a partial characterization of $R_{\text{C}}(D,P)$ for Gaussian mixture sources. Let $\sum_{k=1}^K\beta_k\mathcal{N}(\mu_k,\Sigma_k)$ be a mixture of $K$ Gaussian distributions, $\mathcal{N}(\mu_k,\Sigma_k)$, $k\in[1:K]$, with $\beta_k>0$, $k\in[1:K]$, and $\sum_{k=1}^K\beta_k=1$. We assume $\Sigma_k\succ 0$  and consequently $\lambda_{\min}(\Sigma_k)>0$, $k\in[1:K]$, where 
$\lambda_{\min}(A)$ denotes the minimum eigenvalue of symmetric matrix $A$.

\begin{corollary}\label{cor:Gaussianmixture}
	Assume $X\sim\sum_{k=1}^K\beta_k\mathcal{N}(\mu_k,\Sigma_k)$, $\Delta(x,\hat{x})=\|x-\hat{x}\|^2$, and $\phi(p,p')=W^2_2(p,p')$. For $D>0$ and $P\geq 0$ satisfying
	\begin{align}
		\frac{D^*}{L}\leq\min\{\lambda_{\min}(\Sigma_k)\}_{k=1}^K,
	\end{align}
we have
	\begin{align}
		&R_{\text{C}}(D,P)=h(X)-\frac{L}{2}\log\left(\frac{2\pi e D^*}{L}\right).
	\end{align}
\end{corollary}
\begin{IEEEproof}
	See Appendix \ref{app:Gaussianmixture}.	
\end{IEEEproof}

The next result shows that a variant of the Shannon lower bound in Theorem \ref{thm:lowerbound} is tight for Gaussian sources  $X\sim\mathcal{N}(\mu,\Sigma)$, yielding a complete characterization of $R_{\text{C}}(D,P)$.
It can also be viewed as a strengthened version of Corollary \ref{cor:Gaussianmixture} for degenerate Gaussian mixture sources that  have only one mode.
Let $\Sigma=\Theta^T\Lambda\Theta$ be the eigenvalue decomposition of $\Sigma$, where $\Theta$ is a unitary matrix and $\Lambda$ is a diagonal matrix with the $\ell$-th diagonal entry denoted by $\lambda_{\ell}$, $\ell\in[1:L]$. We assume  $\Sigma\succ 0$ and consequently $\lambda_{\ell}>0$, $\ell\in[1:L]$.

\begin{figure}[t]
\centering
\includegraphics[scale=0.4]{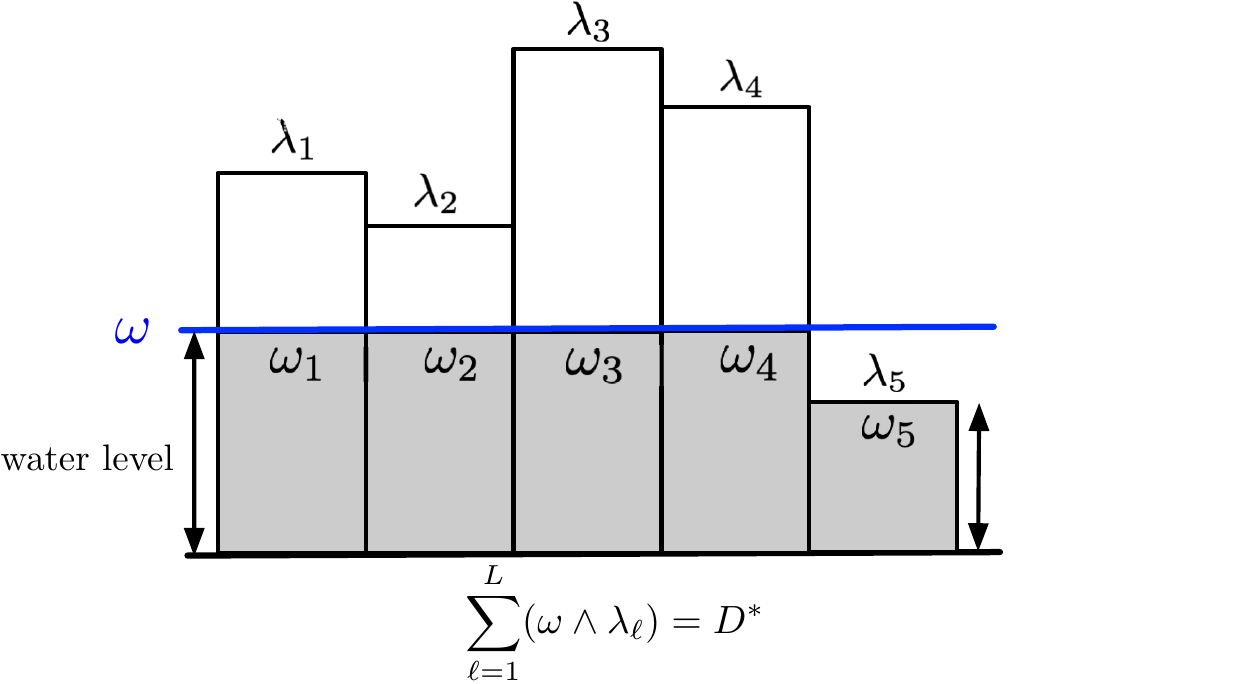}
\caption{Reverse water-filling solution for the Gaussian vector source.}
\label{fig:waterfilling}
\end{figure}

\begin{corollary}\label{cor:Gaussian}
	Assume $X\sim\mathcal{N}(\mu,\Sigma)$, $\Delta(x,\hat{x})=\|x-\hat{x}\|^2$, and $\phi(p,p')=W^2_2(p,p')$. For $D>0$ and $P\geq 0$,
	\begin{align}
		R_{\text{C}}(D,P)=\sum\limits_{\ell=1}^L\frac{1}{2}\log\left(\frac{\lambda_{\ell}}{\omega_{\ell}}\right),
	\end{align}
where $\omega_{\ell}$ is defined in (\ref{eq:omega}) with $\sigma^2_{\ell}$ replaced by $\lambda_{\ell}$, $\ell\in[1:L]$.
\end{corollary}
\begin{IEEEproof}
	See Appendix \ref{app:Gaussian}.	
\end{IEEEproof}
\begin{remark}
	Note that $R_{\text{C}}(D,P)$ degenerates to the rate-distortion function $R(D)$ of quadratic vector Gaussian source coding when $P\geq D$, where $R(D)$ is given by the conventional reverse water-filling formula \cite[Theorem 13.3.3]{Cover1}
	\begin{align}
		R(D)=
		\begin{cases}
		\sum\limits_{\ell=1}^L\frac{1}{2}\log\left(\frac{\lambda_{\ell}}{\omega\wedge\lambda_{\ell}}\right),&D<\sum_{\ell=1}^L\lambda_{\ell},\\
		0,&D\geq\sum_{\ell=1}^L\lambda_{\ell},
		\end{cases}		
	\end{align}
	with $\omega$ being the unique solution to 
	\begin{align}
		\sum\limits_{\ell=1}^L(\omega\wedge\lambda_{\ell})=D.
	\end{align}
	It is also easy to verify that $R_{\text{C}}(D,0)=R(\frac{D}{2})$, which is 
  consistent with Corollary \ref{cor:CM}.  More generally, we have
  \begin{align}
  R_{\text{C}}(D,P)=R(D^*).
  \end{align}
This can be viewed as a manifestation of the universality of the traditional rate-distortion  encoder in this conditional RDP context for the Gaussian source. Indeed, it can be verified that at rate $R(D^*)$, the optimal distortion-perception tradeoff in (\ref{eq:D*}) is achieved by adding appropriate amount of noise to each component of the  traditional rate-distortion reconstruction associated with distortion $D^*$ for $D^*\leq\sum_{\ell=1}^L\lambda_{\ell}$.
\end{remark}
\begin{remark}
  Note that $\omega_{\ell}$ can be interpreted as the water level in the subspace associated with eigenvalue $\lambda_{\ell}$, $\ell\in[1:L]$ (see Fig.~\ref{fig:waterfilling}). Differing from the conventional reverse water-filling formula where the water level coincides with the distortion loss in each subspace when the distortion constraint is active, the situation is more complex here due to the presence of the perception constraint. Specifically, for $\ell\in[1:L]$,   
  the distortion loss $D_{\ell}$ and the perception loss $P_{\ell}$ in the subspace associated with eigenvalue $\lambda_{\ell}$ are given respectively by\footnote{Actually $D_{\ell}$ and $P_{\ell}$ are not uniquely defined  when $D^*>\sum_{\ell=1}^L\lambda_{\ell}$ (which implies $R_{\text{C}}(D,P)=0$).}
  \begin{align}
  	&D_{\ell}=\frac{D}{D^*}\omega_{\ell},\\
  	&P_{\ell}=\frac{D\wedge P}{D^*}\omega_{\ell}.
  \end{align}
  Recently, reverse water-filling type formulas  have been derived in \cite{vectorGaussianRDP} for the Gaussian RDP function under the squared error distortion measure and the marginal perception metric based on either the
  Kullback-Leibler divergence or the squared quadratic Wasserstein distance in the setting with unlimited shared randomness. The formulas in \cite{vectorGaussianRDP} differ significantly from that of $R_{\text{C}}(R,P)$ in the sense that the water level in each subspace remains below the corresponding eigenvalue as long as the perception constraint is active. This difference can be attributed to the loss of universality of the traditional rate-distortion encoder in the setting of \cite{vectorGaussianRDP}. 
   Moreover, from the perspective of the decoder, 
  the solutions in \cite{vectorGaussianRDP} can be interpreted as some appropriate \emph{scalings} of the MMSE estimate of the source components given the corresponding encoder's output (together with the shared random seed) to satisfy the (marginal) perception constraint, while the solution in this paper corresponds to \emph{adding noises} to the MMSE estimate of the source components given the respective encoder's output to satisfy the (conditional) perception constraint.
  Therefore, the optimal compression, reconstruction, and rate-allocation strategies all  depend very much on the specific problem formulation. 
\end{remark}

\section{Conclusion}\label{sec:conclusion}

This paper characterizes the RDP tradeoff for both finite and continuous alphabet sources when the perception measure is based on the divergence between the distributions of the source and reconstruction sequences conditioned on the encoder output. For the Gaussian vector source, a novel reverse water-filling type solution is obtained under the squared error distortion measure and the squared quadratic Wasserstein perception measure.
In contrast to the conventional reverse water-filling solution, here the water level depends on both distortion and perception losses. Throughout this work, we have focused on the setting when no shared randomness is assumed between the encoder and the decoder.  When shared randomness is available, the analysis of the proposed conditional-distribution-based perception measure appears significantly harder and is left for future research. It is also worthwhile to verify our theoretical findings through empirical experiments in the future and to explore the application of conditional-distribution-based perception measures in more general network source coding problems.


\appendices
\section{Proof of Proposition~\ref{prop:property}}\label{proof:property}

\underline{\textit{Tensorizability}}: Note that
\begin{align}
	\phi(p_{\tilde{X}^n},p_{\bar{X}^n})&=\inf\limits_{p_{\tilde{X}^n\bar{X}^n}\in\Pi(p_{\tilde{X}^n},p_{\bar{X}^n})}\sum\limits_{t=1}^n\mathbbm{E}[c(\tilde{X}(t),
	\bar{X}(t))]\nonumber\\
	&\stackrel{(a)}{\geq}\sum\limits_{t=1}^n\inf\limits_{p_{\tilde{X}_i\bar{X}(t)}\in\Pi(p_{\tilde{X}(t)},p_{\bar{X}(t)})}\mathbbm{E}[c(\tilde{X}(t),	\bar{X}(t))]\nonumber\\
	&=\sum\limits_{t=1}^n\phi(p_{\tilde{X}(t)},p_{\bar{X}(t)}),\nonumber
\end{align}
where ($a$) becomes an equality if $p_{\tilde{X}^n}=\prod_{i=1}^np_{\tilde{X}(t)}$ and $p_{\bar{X}^n}=\prod_{i=1}^np_{\bar{X}(t)}$ because $\prod_{t=1}^np_{\tilde{X}(t)\hat{X}(t)}\in\Pi(p_{\tilde{X}^n},p_{\bar{X}^n})$ for any $p_{\tilde{X}(t)\bar{X}(t)}\in\Pi(p_{\tilde{X}(t)},p_{\bar{X}(t)})$, $t\in[1:n]$.

\underline{\textit{Convexity}}: Note that
\begin{align}
	&(1-\lambda)\phi(p_{\tilde{X}^n},p_{\bar{X}^n})+\lambda \phi(p'_{\tilde{X}^n},p'_{\bar{X}^n})\nonumber\\
	&=(1-\lambda)\inf\limits_{p_{\tilde{X}^n\bar{X}^n}\in\Pi(p_{\tilde{X}^n},p_{\bar{X}^n})}\sum\limits_{t=1}^n\mathbbm{E}[c(\tilde{X}(t),\bar{X}(t))]\nonumber\\
 &\quad+\lambda\inf\limits_{p'_{\tilde{X}^n\bar{X}^n}\in\Pi(p'_{\tilde{X}^n},p'_{\bar{X}^n})}\sum\limits_{t=1}^n\mathbbm{E}[c(\tilde{X}(t),\bar{X}(t))]\nonumber\\
	&=\inf\limits_{\stackrel{(1-\lambda)p_{\tilde{X}^n\bar{X}^n}+\lambda p'_{\tilde{X}^n\bar{X}^n}:}{p_{\tilde{X}^n\bar{X}^n}\in\Pi(p_{\tilde{X}^n},p_{\bar{X}^n}),p'_{\tilde{X}^n\bar{X}^n}\in\Pi(p'_{\tilde{X}^n},p'_{\bar{X}^n})}}\sum\limits_{t=1}^n\mathbbm{E}[c(\tilde{X}(t),\bar{X}(t))]\nonumber\\
	&\stackrel{(b)}{\geq}\phi((1-\lambda)p_{\tilde{X}^n}+\lambda p'_{\tilde{X}^n},(1-\lambda)p_{\bar{X}^n}+\lambda p'_{\bar{X}^n}),\label{eq:convexity}
\end{align}
where ($b$) is due to the fact that $(1-\lambda)p_{\tilde{X}^n\bar{X}^n}+\lambda p'_{\tilde{X}^n\bar{X}^n}\in\Pi((1-\lambda)p_{\tilde{X}^n}+\lambda p'_{\tilde{X}^n},(1-\lambda)p_{\bar{X}^n}+\lambda p'_{\bar{X}^n})$
for any $p_{\tilde{X}^n\bar{X}^n}\in\Pi(p_{\tilde{X}^n},p_{\bar{X}^n})$ and $p'_{\tilde{X}^n\bar{X}^n}\in\Pi(p'_{\tilde{X}^n},p'_{\bar{X}^n})$.

\underline{\textit{Continuity}}: Construct a Markov chain $\tilde{Y}^n\leftrightarrow\tilde{X}^n\leftrightarrow\bar{X}^n\leftrightarrow\bar{Y}^n$ with $p_{\tilde{X}^n\tilde{Y}^n}\in\Pi(p_{\tilde{X}^n},p_{\tilde{Y}^n})$, $p_{\tilde{X}^n\bar{X}^n}\in\Pi(p_{\tilde{X}^n},p_{\bar{X}^n})$, and $p_{\bar{X}^n\bar{Y}^n}\in\Pi(p_{\bar{X}^n},p_{\bar{Y}^n})$. We have
\begin{align}
	&\sum\limits_{t=1}^n\mathbbm{E}[c(\tilde{X}(t),\bar{X}(t))]\nonumber\\
	&\geq\mathbbm{P}\{\tilde{X}^n=\tilde{Y}^n,\bar{X}^n=\bar{Y}^n\}\nonumber\\
 &\quad\times\sum\limits_{t=1}^n\mathbbm{E}[c(\tilde{X}(t),\bar{X}(t))|\tilde{X}^n=\tilde{Y}^n,\bar{X}^n=\bar{Y}^n]\nonumber\\
	&=\mathbbm{P}\{\tilde{X}^n=\tilde{Y}^n,\bar{X}^n=\bar{Y}^n\}\nonumber\\
 &\quad\times\sum\limits_{t=1}^n\mathbbm{E}[c(\tilde{Y}(t),\bar{Y}(t))|\tilde{X}^n=\tilde{Y}^n,\bar{X}^n=\bar{Y}^n]\nonumber\\
	&=\sum\limits_{t=1}^n\mathbbm{E}[c(\tilde{Y}(t),\bar{Y}(t))]-\mathbbm{P}\{\tilde{X}^n\neq\tilde{Y}^n\mbox{ or }\bar{X}^n\neq\bar{Y}^n\}\nonumber\\
 &\quad\times\sum\limits_{t=1}^n\mathbbm{E}[c(\tilde{Y}(t),\bar{Y}(t))|\tilde{X}^n\neq\tilde{Y}^n\mbox{ or }\bar{X}^n\neq\bar{Y}^n]\nonumber\\
	&\geq\sum\limits_{t=1}^n\mathbbm{E}[c(\tilde{Y}(t),\bar{Y}(t))]-nc_{\max}\mathbbm{P}\{\tilde{X}^n\neq\tilde{Y}^n\mbox{ or }\bar{X}^n\neq\bar{Y}^n\}\nonumber\\
	&\geq\sum\limits_{t=1}^n\mathbbm{E}[c(\tilde{Y}(t),\bar{Y}(t))]\nonumber\\
 &\quad-nc_{\max}(\mathbbm{P}\{\tilde{X}^n\neq\tilde{Y}^n\}+\mathbbm{P}\{\bar{X}^n\neq\bar{Y}^n\})\nonumber\\
	&\stackrel{(c)}{\geq}\phi(p_{\tilde{Y}^n},p_{\bar{Y}^n})-nc_{\max}(\mathbbm{P}\{\tilde{X}^n\neq\tilde{Y}^n\}+\mathbbm{P}\{\bar{X}^n\neq\bar{Y}^n\}),\label{eq:Y}
\end{align}
where ($c$) is due to $p_{\tilde{Y}^n\bar{Y}^n}\in\Pi(p_{\tilde{Y}^n},p_{\bar{Y}^n})$.
Therefore,
\begin{align}
	&\inf\limits_{p_{\tilde{X}^n\bar{X}^n}\in\Pi(p_{\tilde{X}^n},p_{\bar{X}^n})}\sup\limits_{\stackrel{p_{\tilde{X}^n\tilde{Y}^n}\in\Pi(p_{\tilde{X}^n},p_{\tilde{Y}^n})}{p_{\bar{X}^n\bar{Y}^n}\in\Pi(p_{\bar{X}^n},p_{\bar{Y}^n})}}\sum\limits_{t=1}^n\mathbbm{E}[c(\tilde{X}(t),\bar{X}(t))]\nonumber\\
	&\geq\inf\limits_{p_{\tilde{X}^n\bar{X}^n}\in\Pi(p_{\tilde{X}^n},p_{\bar{X}^n})}\sup\limits_{\stackrel{p_{\tilde{X}^n\tilde{Y}^n}\in\Pi(p_{\tilde{X}^n},p_{\tilde{Y}^n})}{p_{\bar{X}^n\bar{Y}^n}\in\Pi(p_{\bar{X}^n},p_{\bar{Y}^n})}}\phi(p_{\tilde{Y}^n},p_{\bar{Y}^n})\nonumber\\
 &\quad-nc_{\max}(\mathbbm{P}\{\tilde{X}^n\neq\tilde{Y}^n\}+\mathbbm{P}\{\bar{X}^n\neq\bar{Y}^n\}).\label{eq:subed}
\end{align}
Note that
\begin{align}
	&\inf\limits_{p_{\tilde{X}^n\bar{X}^n}\in\Pi(p_{\tilde{X}^n},p_{\bar{X}^n})}\sup\limits_{\stackrel{p_{\tilde{X}^n\tilde{Y}^n}\in\Pi(p_{\tilde{X}^n},p_{\tilde{Y}^n})}{p_{\bar{X}^n\bar{Y}^n}\in\Pi(p_{\bar{X}^n},p_{\bar{Y}^n})}}\sum\limits_{t=1}^n\mathbbm{E}[c(\tilde{X}(t),\bar{X}(t))]\nonumber\\&=\inf\limits_{p_{\tilde{X}^n\bar{X}^n}\in\Pi(p_{\tilde{X}^n},p_{\bar{X}^n})}\sum\limits_{t=1}^n\mathbbm{E}[c(\tilde{X}(t),\bar{X}(t))]\nonumber\\
	&=\phi(p_{\tilde{X}^n},p_{\bar{X}^n})\label{eq:sub1}
\end{align}
and
\begin{align}
	&\inf\limits_{p_{\tilde{X}^n\bar{X}^n}\in\Pi(p_{\tilde{X}^n},p_{\bar{X}^n})}\sup\limits_{\stackrel{p_{\tilde{X}^n\tilde{Y}^n}\in\Pi(p_{\tilde{X}^n},p_{\tilde{Y}^n})}{p_{\bar{X}^n\bar{Y}^n}\in\Pi(p_{\bar{X}^n},p_{\bar{Y}^n})}}\phi(p_{\tilde{Y}^n},p_{\bar{Y}^n})\nonumber\\
 &-nc_{\max}(\mathbbm{P}\{\tilde{X}^n\neq\tilde{Y}^n\}+\mathbbm{P}\{\bar{X}^n\neq\bar{Y}^n\})\nonumber\\
	&=\inf\limits_{p_{\tilde{X}^n\bar{X}^n}\in\Pi(p_{\tilde{X}^n},p_{\bar{X}^n})}\phi(p_{\tilde{Y}^n},p_{\bar{Y}^n})\nonumber\\
 &\quad-nc_{\max}(d_{\text{TV}}(p_{\tilde{X}^n},p_{\tilde{Y}^n})+d_{\text{TV}}(p_{\bar{X}^n},p_{\bar{Y}^n}))\nonumber\\
	&=\phi(p_{\tilde{Y}^n},p_{\bar{Y}^n})-nc_{\max}(d_{\text{TV}}(p_{\tilde{X}^n},p_{\tilde{Y}^n})+d_{\text{TV}}(p_{\bar{X}^n},p_{\bar{Y}^n})).\label{eq:sub2}
\end{align}
Substituting (\ref{eq:sub1}) and (\ref{eq:sub2})
into (\ref{eq:subed}) proves
\begin{align}
	&\phi(p_{\tilde{X}^n},p_{\bar{X}^n})\nonumber\\&\geq\phi(p_{\tilde{Y}^n},p_{\bar{Y}^n})-nc_{\max}(d_{\text{TV}}(p_{\tilde{X}^n},p_{\tilde{Y}^n})+d_{\text{TV}}(p_{\bar{X}^n},p_{\bar{Y}^n})).\label{eq:sym1}
\end{align}
By symmetry, we also have
\begin{align}
	&\phi(p_{\tilde{Y}^n},p_{\bar{Y}^n})\nonumber\\&\geq\phi(p_{\tilde{X}^n},p_{\bar{X}^n})-nc_{\max}(d_{\text{TV}}(p_{\tilde{X}^n},p_{\tilde{Y}^n})+d_{\text{TV}}(p_{\bar{X}^n},p_{\bar{Y}^n})).\label{eq:sym2}
\end{align}
Combining (\ref{eq:sym1}) and (\ref{eq:sym2}) yields the desired result.

\section{Proof of Proposition~\ref{prop:convexity}}\label{proof:convexity}

For any $(D^{(j)},P^{(j)})$ and $\epsilon>0$, there exists $p_{U^{(j)}\hat{X}^{(j)}|X}$ such that 
\begin{align}
	&I(X;U^{(j)})\leq R(D^{(j)},P^{(j)})+\epsilon,\\
	&X\leftrightarrow U^{(j)}\leftrightarrow\hat{X}^{(j)}\mbox{ form a Markov chain},\\
	&\mathbbm{E}[\Delta(X,\hat{X}^{(j)})]\leq D^{(j)},\\
	&\mathbbm{E}[\phi(p_{X|U^{(j)}}(\cdot|U^{(j)}),p_{\hat{X}|U^{(j)}}(\cdot|U^{(j)}))]\leq P^{(j)},\quad j=0,1.
\end{align}
Let $Q\sim\mbox{Ber}(\lambda)$ with $\lambda\in[0,1]$. Construct $V$ and $\hat{X}$ such that $p_{V|Q}(\cdot|j)=p_{U^{(j)}}(\cdot)$ and $p_{\hat{X}|VQ}(\cdot|\cdot,j)=p_{\hat{X}^{(j)}|U^{(j)}}(\cdot|\cdot)$, $j=0,1$. Further let $X$ be jointly distributed with $(V,Q,\hat{X})$ such that $p_{X|VQ\hat{X}}=p_{X|VQ}$ with $p_{X|VQ}(\cdot|\cdot,j)=p_{X|U^{(j)}}(\cdot|\cdot)$, $j=0,1$. Setting $U=(V,Q)$, it can be verified that
\begin{align}
	&I(X;U)=(1-\lambda)I(X;U^{(0)})+\lambda I(X;U^{(1)})\nonumber\\
 &\hspace{0.472in}\leq (1-\lambda)R(D^{(0)},P^{(0)})+\lambda R(D^{(1)},P^{(1)})+\epsilon,\\
	&X\leftrightarrow U\leftrightarrow \hat{X}\mbox{ form a Markov chain},\\
	&\mathbbm{E}[\Delta(X,\hat{X})]=(1-\lambda)\mathbbm{E}[\Delta(X,\hat{X}^{(0)})]+\lambda\mathbbm{E}[\Delta(X,\hat{X}^{(1)})]\nonumber\\
 &\hspace{0.71in}\leq(1-\lambda)D^{(0)}+\lambda D^{(1)},\\
	&\mathbbm{E}[\phi(p_{X|U}(\cdot|U),p_{\hat{X}|U}(\cdot|U))]\nonumber\\
 &=(1-\lambda)\mathbbm{E}[\phi(p_{X|U^{(0)}}(\cdot|U^{(0)}),p_{\hat{X}|U^{(0)}}(\cdot|U^{(0)}))]\nonumber\\
 &\quad+\lambda\mathbbm{E}[\phi(p_{X|U^{(1)}}(\cdot|U^{(1)}),p_{\hat{X}|U^{(1)}}(\cdot|U^{(1)}))]\nonumber\\
	&\leq(1-\lambda)P^{(0)}+\lambda P^{(1)},
\end{align}
which further implies 
\begin{align}
	&R((1-\lambda)D^{(0)}+\lambda D^{(1)},(1-\lambda)P^{(0)}+\lambda P^{(1)})\nonumber\\
 &\leq(1-\lambda)R(D^{(0)},P^{(0)})+\lambda R(D^{(1)},P^{(1)})+\epsilon.
\end{align}
Sending $\epsilon\rightarrow 0$ proves the desired result.

\section{Proof of Theorem~\ref{thm:RDP}}\label{app:RDP}

\underline{\textit{Achievability}}: We first establish the following two technical lemmas.

\begin{lemma}\label{lem:encoder}
	Let $p_U$ and $p_{X|U}$ denote respectively the output distribution and the backward test channnel induced by $p_X$ and $p_{U|X}$. 
	For any $\delta>0$, there exists a stochastic function $\psi^{(n)}:\mathcal{X}^n\rightarrow\mathcal{G}^{(n)}\cup\mathcal{B}^{(n)}\subseteq\mathcal{U}^n$, with   $\mathcal{G}^{(n)}\subseteq\mathcal{T}^{(n)}_{\delta}(p_U)$\footnote{$\mathcal{T}^{(n)}_{\delta}(p_U)$ denotes the set of $\delta$-typical $n$-sequences with respect to $p_U$ \cite[p. 25]{ElGamal}.} and $\mathcal{G}^{(n)}\cap\mathcal{B}^{(n)}=\emptyset$, mapping  $X^n$ to $U^n$ according to some conditional distribution $p_{U^n|X^n}$ such that 
	\begin{align}
		&\frac{1}{n}\lceil\log|\mathcal{G}^{(n)}\cup\mathcal{B}^{(n)}|\rceil\leq I(X;U)+\delta,\\
		&\mathbbm{P}\{U^n\in\mathcal{B}^{(n)}\}\leq \delta,\\
		&d_{\text{TV}}(p_{X^n|U^n}(\cdot|u^n),p^n_{X|U}(\cdot|u^n))\leq\delta,\quad u^n\in\mathcal{G}^{(n)}.
	\end{align}
\end{lemma}
\begin{IEEEproof}
	Let $p_{X|U}$ play the role of posterior reference map in the sense of \cite{ASP23}.
	By \cite[Theorem 3]{ASP23}, for every sufficiently small $\eta>0$,  there exists a stochastic function $\psi^{(n)}:\mathcal{X}^n\rightarrow\mathcal{C}^{(n)}\cup\{\tilde{u}^n\}$, with $\mathcal{C}^{(n)}\subseteq\mathcal{T}^{(n)}_{\eta}(p_U)$ and $\tilde{u}^n\in\mathcal{U}^n\backslash\mathcal{T}^{(n)}_{\eta}(p_U)$, mapping  $X^n$ to $U^n$ according to some conditional distribution $p_{U^n|X^n}$ such that 
	\begin{align}
		&\frac{1}{n}\lceil\log|\mathcal{C}^{(n)}\cup\{\tilde{u}^n\}|\rceil\leq I(X;U)+\eta,\\
		&\mathbbm{P}\{U^n=\tilde{u}^n\}\leq\eta,\\
		&\mathbbm{E}[d_{\text{TV}}(p_{X^n|U^n}(\cdot|U^n),p^n_{X|U}(\cdot|U^n))]\leq\eta.
	\end{align}
	Let $\mathcal{G}^{(n)}=\{u^n\in\mathcal{C}^{(n)}:d_{\text{TV}}(p_{X^n|U^n}(\cdot|u^n),p^n_{X|U}(\cdot|u^n))\leq\delta\}$ and $\mathcal{B}^n=(\mathcal{C}^{(n)}\backslash\mathcal{G}^{(n)})\cup\{\tilde{u}^n\}$. We have
	\begin{align}
		&\mathbbm{P}\{U^n\in\mathcal{B}^{(n)}\}\nonumber\\
  &=\mathbbm{P}\{d_{\text{TV}}(p_{X^n|U^n}(\cdot|U^n),p^n_{X|U}(\cdot|U^n))>\delta\mbox{ or }U^n=\tilde{u}^n\}\nonumber\\
		&\leq\mathbbm{P}\{d_{\text{TV}}(p_{X^n|U^n}(\cdot|U^n),p^n_{X|U}(\cdot|U^n))>\delta\}+\eta\nonumber\\
		&\stackrel{(a)}{\leq}\frac{\eta}{\delta}+\eta,\label{eq:Markov}
	\end{align}
	where ($a$) is due to Markov's inequality. Choosing $\eta\leq\frac{\delta^2}{1+\delta}$ completes the proof of Proposition \ref{lem:encoder}.
\end{IEEEproof}

	\begin{lemma}\label{lem:distortion}
	For $p_{\tilde{X}^n\bar{X}^n}=p_{\tilde{X}^n}p_{\bar{X}^n}$ and $p_{\tilde{Y}^n\bar{Y}^n}=p_{\tilde{Y}^n}p_{\bar{Y}^n}$,
	\begin{align}
		&|\mathbbm{E}[\Delta(\tilde{X}^n,\bar{X}^n)]-\mathbbm{E}[\Delta(\tilde{Y}^n,\bar{Y}^n)]|\nonumber\\
  &\leq2n\Delta_{\max}(d_{\text{TV}}(p_{\tilde{X}^n},p_{\tilde{Y}^n})+d_{\text{TV}}(p_{\bar{X}^n},p_{\bar{Y}^n})),
	\end{align}
	where $\Delta_{\max}=\max_{x,\hat{x}\in\mathcal{X}}\Delta(x,\hat{x})$.
\end{lemma}
\begin{IEEEproof}
	Note that
	\begin{align}
		&\mathbbm{E}[\Delta(\tilde{X}^n,\bar{X}^n)]\nonumber\\
		&=\sum\limits_{\tilde{x}^n,\bar{x}^n\in\mathcal{X}^n}p_{\tilde{X}^n}(\tilde{x}^n)p_{\bar{X}^n}(\bar{x}^n)\Delta(\tilde{x}^n,\bar{x}^n)\nonumber\\
		&\leq \sum\limits_{\tilde{x}^n,\bar{x}^n\in\mathcal{X}^n}(p_{\tilde{Y}^n}(\tilde{x}^n)p_{\bar{Y}^n}(\bar{x}^n)\Delta(\tilde{x}^n,\bar{x}^n)\nonumber\\
  &\hspace{0.43in}+|p_{\tilde{X}^n}(\tilde{x}^n)p_{\bar{X}^n}(\bar{x}^n)-p_{\tilde{Y}^n}(\tilde{x}^n)p_{\bar{Y}^n}(\bar{x}^n)|\Delta(\tilde{x}^n,\bar{x}^n))\nonumber\\
		&=\mathbbm{E}[\Delta(\tilde{Y}^n,\bar{Y}^n)]+\sum\limits_{\tilde{x}^n,\bar{x}^n\in\mathcal{X}^n}|p_{\tilde{X}^n}(\tilde{x}^n)p_{\bar{X}^n}(\bar{x}^n)\nonumber\\&\hspace{1.7in}-p_{\tilde{Y}^n}(\tilde{x}^n)p_{\bar{Y}^n}(\bar{x}^n)|\Delta(\tilde{x}^n,\bar{x}^n)\nonumber\\
		&\leq\mathbbm{E}[\Delta(\tilde{Y}^n,\bar{Y}^n)]\nonumber\\&\quad+n\Delta_{\max}\sum\limits_{\tilde{x}^n,\bar{x}^n\in\mathcal{X}^n}|p_{\tilde{X}^n}(\tilde{x}^n)p_{\bar{X}^n}(\bar{x}^n)-p_{\tilde{Y}^n}(\tilde{x}^n)p_{\bar{Y}^n}(\bar{x}^n)|\nonumber\\
		&=\mathbbm{E}[\Delta(\tilde{Y}^n,\bar{Y}^n)]\nonumber\\
  &\quad+n\Delta_{\max}\sum\limits_{\tilde{x}^n,\bar{x}^n\in\mathcal{X}^n}|p_{\tilde{X}^n}(\tilde{x}^n)p_{\bar{X}^n}(\bar{x}^n)-p_{\tilde{Y}^n}(\tilde{x}^n)p_{\bar{X}^n}(\bar{x}^n)|\nonumber\\
		&\quad+n\Delta_{\max}\sum\limits_{\tilde{x}^n,\bar{x}^n\in\mathcal{X}^n}|p_{\tilde{Y}^n}(\tilde{x}^n)p_{\bar{X}^n}(\bar{x}^n)-p_{\tilde{Y}^n}(\tilde{x}^n)p_{\bar{Y}^n}(\bar{x}^n)|\nonumber\\
		&=\mathbbm{E}[\Delta(\tilde{Y}^n,\bar{Y}^n)]\nonumber\\
  &\quad+n\Delta_{\max}\sum\limits_{\bar{x}^n\in\mathcal{X}^n}p_{\bar{X}^n}(\bar{x}^n)\sum\limits_{\tilde{x}^n\in\mathcal{X}^n}|p_{\tilde{X}^n}(\tilde{x}^n)-p_{\tilde{Y}^n}(\tilde{x}^n)|\nonumber\\
		&\quad+n\Delta_{\max}\sum\limits_{\tilde{x}^n}p_{\tilde{Y}^n}(\tilde{x}^n)\sum\limits_{\bar{x}^n\in\mathcal{X}^n}|p_{\bar{X}^n}(\bar{x}^n)-p_{\bar{Y}^n}(\bar{x}^n)|\nonumber\\
		&=\mathbbm{E}[\Delta(\tilde{Y}^n,\bar{Y}^n)]+2n\Delta_{\max}(d_{\text{TV}}(p_{\tilde{X}^n},p_{\tilde{Y}^n})\nonumber\\
  &\quad+d_{\text{TV}}(p_{\bar{X}^n},p_{\bar{Y}^n})).\label{eq:comb1}
	\end{align}
	By symmetry, we also have
	\begin{align}
		\mathbbm{E}[\Delta(\tilde{Y}^n,\bar{Y}^n)]&\leq\mathbbm{E}[\Delta(\tilde{X}^n,\bar{X}^n)]+2n\Delta_{\max}(d_{\text{TV}}(p_{\tilde{X}^n},p_{\tilde{Y}^n})\nonumber\\&\quad+d_{\text{TV}}(p_{\bar{X}^n},p_{\bar{Y}^n})).\label{eq:comb2}
	\end{align}
	Combining (\ref{eq:comb1}) and (\ref{eq:comb2}) yields the desired result.
\end{IEEEproof}

We shall prove the achievability part of Theorem \ref{thm:RDP} by treating the following three cases separately.
\begin{enumerate}
	\item $D=0$:
	
	It is easy to verify that $R(0,P)=H(X)$. Since $\mathbbm{E}[\Delta(X^n,\hat{X}^n)]=0$ implies $p_{X^n|M}=p_{\hat{X}^n|M}$ and consequently $\mathbbm{E}[\phi(p_{X^n|M}(\cdot|M),p_{\hat{X}^n|M}(\cdot|M))]=0$, the problem boils down to lossless source coding, for which $H(X)$ is known to be the infimum of achievable rates. This proves the achievability part for case 1).
	
	\item $D>0$ and $P=0$:
	
	In light of Proposition \ref{prop:convexity}, $R(D,0)$ is continuous in $D$ for $D>0$. As a consequence, for any $\epsilon>0$, there exists $\rho>0$ satisfying $R(D-\rho,0)\leq R(D,0)+\epsilon$. By the definition of $R(D-\rho,0)$, we can find $p_{U\hat{X}|X}$ such that $I(X;U)=R(D-\rho,0)$, $X\leftrightarrow U\leftrightarrow\hat{X}$ form a Markov chain, $\mathbbm{E}[\Delta(X,\hat{X})]\leq D-\rho$, and $p_{X|U}=p_{\hat{X}|U}$. 
	Let $p_{U^n|X^n}$ be the conditional distribution specified in Lemma \ref{lem:encoder}, and set $p_{U^n\hat{X}^n|X^n}=p_{U^n|X^n}p_{\hat{X}^n|U^n}$ with
	\begin{align} p_{\hat{X}^n|U^n}(\cdot|u^n)=p_{X^n|U^n}(\cdot|u^n),\quad u^n\in\mathcal{G}^{(n)}\cup\mathcal{B}^{(n)}.\label{eq:symmetry}
	\end{align}	
	Construct a bijection $\kappa:\mathcal{G}^{(n)}\cup\mathcal{B}^{(n)}\rightarrow\mathcal{M}$, where $\mathcal{M}$ is a set of binary codewords of length $\lceil\log|\mathcal{G}^{(n)}\cup\mathcal{B}^{(n)}|\rceil$. Define $M=\kappa(U^n)$. For the rate, we have
	\begin{align}
		\frac{1}{n}\mathbbm{E}[\ell(M)]&=\frac{1}{n}\lceil\log|\mathcal{G}^{(n)}\cup\mathcal{B}^{(n)}|\rceil\nonumber\\
		&\leq I(X;U)+\delta\nonumber\\
		&=R(D-\rho,0)+\delta\nonumber\\
		&\leq R(D,0)+\epsilon+\delta.
	\end{align}
	As to the perception loss, it follows by 
	(\ref{eq:symmetry}) that
	$p_{\hat{X}^n|M}=p_{X^n|M}$, which further implies
	\begin{align}
		\frac{1}{n}\mathbbm{E}[\phi(p_{X^n|M}(\cdot|M),p_{\hat{X}^n|M}(\cdot|M))]=0.\label{eq:perception}
	\end{align}
	It remains to analyze the distortion loss. Note that
	\begin{align}
		&\frac{1}{n}\mathbbm{E}[\Delta(X^n,\hat{X}^n)]\nonumber\\&=\frac{1}{n}\mathbbm{E}[\mathbbm{E}[\Delta(X^n,\hat{X}^n)|U^n]]\nonumber\\
		&=\frac{1}{n}\sum\limits_{u^n\in\mathcal{G}^{(n)}}p_{U^n}(u^n)\mathbbm{E}[\Delta(X^n,\hat{X}^n)|U^n=u^n]\nonumber\\
  &\quad+\frac{1}{n}\sum\limits_{u^n\in\mathcal{B}^{(n)}}p_{U^n}(u^n)\mathbbm{E}[\Delta(X^n,\hat{X}^n)|U^n=u^n].\label{eq:tbs}
	\end{align}
	For $u^n\in\mathcal{G}^{(n)}$,
	\begin{align}
		&\mathbbm{E}[\Delta(X^n,\hat{X}^n)|U^n=u^n]\nonumber\\
		&\stackrel{(b)}{\leq}\sum\limits_{t=1}^n\mathbbm{E}[\Delta(X,\hat{X})|U=u(t)]\nonumber\\
  &\quad+4n\Delta_{\max}d_{\text{TV}}(p_{X^n|U^n}(\cdot|u^n),p^n_{X|U}(\cdot|u^n))\nonumber\\
		&\leq\sum\limits_{t=1}^n\mathbbm{E}[\Delta(X,\hat{X})|U=u(t)]+4n\Delta_{\max}\delta\nonumber\\
		&\stackrel{(c)}{\leq} n(1+\delta)\mathbbm{E}[\mathbbm{E}[\Delta(X,\hat{X})|U]]+4n\Delta_{\max}\nonumber\\
		&=n(1+\delta)\mathbbm{E}[\Delta(X,\hat{X})]+4n\Delta_{\max}\delta\nonumber\\
		&\leq n(1+\delta)(D-\rho)+4n\Delta_{\max}\delta,\label{eq:subst1}
	\end{align}
	where ($b$) is due to Lemma \ref{lem:distortion}, and ($c$) is due to the typical average lemma \cite[p. 26]{ElGamal}.
	For $u^n\in\mathcal{B}^{(n)}$,
	\begin{align}
		\mathbbm{E}[\Delta(X^n,\hat{X}^n)|U^n=u^n]\leq n\Delta_{\max}.\label{eq:subst2}
	\end{align}
	Substituting (\ref{eq:subst1}) and (\ref{eq:subst2}) into (\ref{eq:tbs}) gives
	\begin{align}
		&\frac{1}{n}\mathbbm{E}[\Delta(X^n,\hat{X}^n)]\nonumber\\
  &\leq((1+\delta)(D-\rho)+4\Delta_{\max}\delta)\mathbbm{P}\{U^n\in\mathcal{G}^{(n)}\}\nonumber\\
  &\quad+\Delta_{\max}\mathbbm{P}\{U^n\in\mathbbm{B}^{(n)}\}\nonumber\\
		&\leq(1+\delta)(D-\rho)+5\Delta_{\max}\delta\nonumber\\
		&=D-\rho+(D-\rho+5\Delta_{\max})\delta.
	\end{align}
	Choosing $\delta\leq\max\{\epsilon,\frac{\rho}{D-\rho+5\Delta_{\max}}\}$ ensures
	\begin{align}
		&\frac{1}{n}\mathbbm{E}[\ell(M)]\leq R(D,0)+2\epsilon,\label{eq:rate}\\
		&\frac{1}{n}\mathbbm{E}[\Delta(X^n,\hat{X}^n)]\leq D.\label{eq:distortion}
	\end{align}
	In view of (\ref{eq:perception}), (\ref{eq:rate}),  (\ref{eq:distortion}), and the fact that $\epsilon>0$ is arbitrary, the achievability part for case 2) is proved.

	\item $D>0$ and $P>0$:
	
	In light of Proposition \ref{prop:convexity}, $R(D,P)$ is continuous in $(D,P)$ for $D>0$ and $P>0$. As a consequence, for any $\epsilon>0$, there exists $\rho>0$ satisfying $R(D-\rho,P-\rho)\leq R(D,P)+\epsilon$. By the definition of $R(D-\rho,P-\rho)$, we can find $p_{U\hat{X}|X}$ such that $I(X;U)=R(D-\rho,P-\rho)$, $X\leftrightarrow U\leftrightarrow\hat{X}$ form a Markov chain, $\mathbbm{E}[\Delta(X,\hat{X})]\leq D-\rho$, and $\mathbbm{E}[\phi_{X|U}(\cdot|U),p_{\hat{X}|U}(\cdot|U)]\leq P-\rho$.  
	Let $p_{U^n|X^n}$ be the conditional distribution specified in Lemma \ref{lem:encoder}, and set $p_{U^n\hat{X}^n|X^n}=p_{U^n|X^n}p_{\hat{X}^n|U^n}$ with
	\begin{align}
		&p_{\hat{X}^n|U^n}(\cdot|u^n)=p^n_{\hat{X}|U}(\cdot|u^n),\quad u^n\in\mathcal{G}^{(n)},\\
		&p_{\hat{X}^n|U^n}(\cdot|u^n)=p_{X^n|U^n}(\cdot|u^n),\quad u^n\in\mathcal{B}^{(n)}.
	\end{align}
	Construct a bijection $\kappa:\mathcal{G}^{(n)}\cup\mathcal{B}^{(n)}\rightarrow\mathcal{M}$, where $\mathcal{M}$ is a set of binary codewords of length $\lceil\log|\mathcal{G}^{(n)}\cup\mathcal{B}^{(n)}|\rceil$. Define $M=\kappa(U^n)$. The rate can be bounded as follows:
	\begin{align}
		\frac{1}{n}\mathbbm{E}[\ell(M)]&=\frac{1}{n}\lceil\log|\mathcal{G}^{(n)}\cup\mathcal{B}^{(n)}|\rceil\nonumber\\
		&\leq I(X;U)+\delta\nonumber\\
		&=R(D-\rho,P-\rho)+\delta\nonumber\\
		&\leq R(D,P)+\epsilon+\delta.
	\end{align}
	As to the distoriton loss, we have
	\begin{align}
		&\frac{1}{n}\mathbbm{E}[\Delta(X^n,\hat{X}^n)]\nonumber\\
  &=\frac{1}{n}\mathbbm{E}[\mathbbm{E}[\Delta(X^n,\hat{X}^n)|U^n]]\nonumber\\
		&=\frac{1}{n}\sum\limits_{u^n\in\mathcal{G}^{(n)}}p_{U^n}(u^n)\mathbbm{E}[\Delta(X^n,\hat{X}^n)|U^n=u^n]\nonumber\\
  &\quad+\frac{1}{n}\sum\limits_{u^n\in\mathcal{B}^{(n)}}p_{U^n}(u^n)\mathbbm{E}[\Delta(X^n,\hat{X}^n)|U^n=u^n].\label{eq:tbs2}
	\end{align}
	For $u^n\in\mathcal{G}^{(n)}$,
	\begin{align}
		&\mathbbm{E}[\Delta(X^n,\hat{X}^n)|U^n=u^n]\nonumber\\
		&\stackrel{(d)}{\leq}\sum\limits_{t=1}^n\mathbbm{E}[\Delta(X,\hat{X})|U=u(t)]\nonumber\\
  &\quad+2n\Delta_{\max}d_{\text{TV}}(p_{X^n|U^n}(\cdot|u^n),p^n_{X|U}(\cdot|u^n))\nonumber\\
		&\leq\sum\limits_{t=1}^n\mathbbm{E}[\Delta(X,\hat{X})|U=u(t)]+2n\Delta_{\max}\delta\nonumber\\
		&\stackrel{(e)}{\leq} n(1+\delta)\mathbbm{E}[\mathbbm{E}[\Delta(X,\hat{X})|U]]+2n\Delta_{\max}\delta\nonumber\\
		&=n(1+\delta)\mathbbm{E}[\Delta(X,\hat{X})]+2n\Delta_{\max}\delta\nonumber\\
		&\leq n(1+\delta)(D-\rho)+2n\Delta_{\max}\delta,\label{eq:subst12}
	\end{align}
	where ($d$) is due to Lemma \ref{lem:distortion}, and ($e$) is due to the typical average lemma \cite[p. 26]{ElGamal}.
	For $u^n\in\mathcal{B}^{(n)}$,
	\begin{align}
		\mathbbm{E}[\Delta(X^n,\hat{X}^n)|U^n=u^n]\leq n\Delta_{\max}.\label{eq:subst22}
	\end{align}
	Substituting (\ref{eq:subst12}) and (\ref{eq:subst22}) into (\ref{eq:tbs2}) gives
	\begin{align}
		&\frac{1}{n}\mathbbm{E}[\Delta(X^n,\hat{X}^n)]\nonumber\\&\leq((1+\delta)(D-\rho)+2\Delta_{\max}\delta)\mathbbm{P}\{U^n\in\mathcal{G}^{(n)}\}\nonumber\\&\quad+\Delta_{\max}\mathbbm{P}\{U^n\in\mathbbm{B}^{(n)}\}\nonumber\\
		&\leq(1+\delta)(D-\rho)+3\Delta_{\max}\delta\nonumber\\
		&=D-\rho+(D-\rho+3\Delta_{\max})\delta.
	\end{align}
	It remains to analyze the perception loss. Note that
	\begin{align}
		&\frac{1}{n}\mathbbm{E}[\phi(p_{X^n|M}(\cdot|M),p_{\hat{X}^n|M}(\cdot|M))]\nonumber\\
		&=\frac{1}{n}\sum\limits_{u^n\in\mathcal{G}^{(n)}}p_{U^n}(u^n)\phi(p_{X^n|U^n}(\cdot|u^n),p_{\hat{X}^n|U^n}(\cdot|u^n))\nonumber\\
  &\quad+\frac{1}{n}\sum\limits_{u^n\in\mathcal{B}^{(n)}}p_{U^n}(u^n)\phi(p_{X^n|U^n}(\cdot|u^n),p_{\hat{X}^n|U^n}(\cdot|u^n))\nonumber\\
		&=\frac{1}{n}\sum\limits_{u^n\in\mathcal{G}^{(n)}}p_{U^n}(u^n)\phi(p_{X^n|U^n}(\cdot|u^n),p_{\hat{X}^n|U^n}(\cdot|u^n))\nonumber\\
		&\stackrel{(f)}{\leq}\frac{1}{n}\sum\limits_{u^n\in\mathcal{G}^{(n)}}p_{U^n}(u^n)(\phi(p^n_{X|U}(\cdot|u^n),p^n_{\hat{X}|U}(\cdot|u^n))\nonumber\\
  &\hspace{0.8in}+nc_{\max}d_{\text{TV}}(p_{X^n|U^n}(\cdot|u^n),p^n_{X|U}(\cdot|u^n)))\nonumber\\
		&\leq\frac{1}{n}\sum\limits_{u^n\in\mathcal{G}^{(n)}}p_{U^n}(u^n)\phi(p^n_{X|U}(\cdot|u^n),p^n_{\hat{X}|U}(\cdot|u^n))\nonumber\\&\quad+c_{\max}\delta\nonumber\\
		&\stackrel{(g)}{=}\frac{1}{n}\sum\limits_{u^n\in\mathcal{G}^{(n)}}p_{U^n}(u^n)\sum\limits_{t=1}^n\phi(p_{X|U}(\cdot|u(t)),p_{\hat{X}|U}(\cdot|u(t)))\nonumber\\
  &\quad+c_{\max}\delta\nonumber\\
		&\stackrel{(h)}{\leq}\sum\limits_{u^n\in\mathcal{G}^{(n)}}p_{U^n}(u^n)(1+\delta)\mathbbm{E}[p_{X|U}(\cdot|U),p_{\hat{X}|U}(\cdot|U)]\nonumber\\
  &\quad+c_{\max}\delta\nonumber\\
		&\leq(1+\delta)\mathbbm{E}[p_{X|U}(\cdot|U),p_{\hat{X}|U}(\cdot|U)]+c_{\max}\delta\nonumber\\
		&\leq(1+\delta)(P-\rho)+c_{\max}\delta\nonumber\\
		&=P-\rho+(P-\rho+c_{\max})\delta,
	\end{align}
where ($f$) is due to part 3) of Proposition \ref{prop:property}, ($g$) is due to part 1) of Proposition \ref{prop:property}, and ($h$) is due to the typical average lemma \cite[p. 26]{ElGamal}. Choosing $\delta\leq\max\{\epsilon,\frac{\rho}{D-\rho+3\Delta_{\max}},\frac{\rho}{P-\rho+c_{\max}}\}$ ensures
\begin{align}
	&\frac{1}{n}\mathbbm{E}[\ell(M)]\leq R(D,P)+2\epsilon,\\
	&\frac{1}{n}\mathbbm{E}[\Delta(X^n,\hat{X}^n)]\leq D,\\
	&\frac{1}{n}\mathbbm{E}[\phi(p_{X^n|M}(\cdot|M),p_{\hat{X}^n|M}(\cdot|M))]\leq P.
\end{align}
Since $\epsilon>0$ is arbitrary, the achievability part for case 3) is proved.
\end{enumerate}	

\underline{\textit{Converse}}: Let $R$ be an achievable rate subject to distortion constraint $D$ and perception constraint $P$. By Definition \ref{def:opRDP},  we can find encoder $f^{(n)}$ and decoder $g^{(n)}$ satisfying
\begin{align}
	&\frac{1}{n}\mathbbm{E}[\ell(M)]\leq R,\label{eq:R}\\
	&\frac{1}{n}\mathbbm{E}[\Delta(X^n,\hat{X}^n)]\leq D,\label{eq:D}\\
	&\frac{1}{n}\mathbbm{E}[\phi(p_{X^n|M}(\cdot|M),p_{\hat{X}^n|M}(\cdot|M))]\leq P.\label{eq:P}
\end{align}
Let $U(t)=M$, $t\in[1:n]$. Moreover, let $T$ be uniformly distributed over $[1:n]$ and independent of $(X^n, M, \hat{X}^n)$. Note that
\begin{align}
	\frac{1}{n}E[\ell(M)] &\geq  \frac{1}{n}H(M)\nonumber\\
	&\geq  \frac{1}{n}I(X^n;M)\nonumber\\
	&= \frac{1}{n}\sum_{t=1}^n I(X(t);M|X^{t-1})\nonumber\\
	&= \frac{1}{n}\sum_{t=1}^n I(X(t);M,X^{t-1})\nonumber\\
	&\geq  \frac{1}{n}\sum_{t=1}^n I(X(t);M)\nonumber\\
	&= \frac{1}{n}\sum_{t=1}^n I(X(t);U(t))\nonumber\\
	&=I(X(T);U(T)|T)\nonumber\\
	&\stackrel{(i)}{=}I(X(T);U(T),T),\label{eq:boundR}
\end{align}
where ($i$) is due to the fact that $X(T)$ is independent of $T$ as the conditional distribution of $X(T)$ given $T$ is the same as the marginal distribution of $X(T)$ (which is $p_X$).
In addition, we have
\begin{align}
	\frac{1}{n}\mathbbm{E}[\Delta(X^n,\hat{X}^n)]&=\frac{1}{n}\sum\limits_{t=1}^n\mathbbm{E}[\Delta(X(t),\hat{X}(t))]\nonumber\\
	&=\mathbbm{E}[\mathbbm{E}[\Delta(X(T),\hat{X}(T))|T]]\nonumber\\
	&=\mathbbm{E}[\Delta(X(T),\hat{X}(T))]\label{eq:boundD}
\end{align}
and
\begin{align}
	&\frac{1}{n}\mathbbm{E}[\phi(p_{X^n|M}(\cdot|M),p_{\hat{X}^n|M}(\cdot|M))]\nonumber\\&\stackrel{(j)}{\geq}\frac{1}{n}\sum\limits_{t=1}^n\mathbbm{E}[\phi(p_{X(t)|M}(\cdot|M),p_{\hat{X}(t)|M}(\cdot|M))]\nonumber\\
	&=\frac{1}{n}\sum\limits_{t=1}^n\mathbbm{E}[\phi(p_{X(t)|U(t)}(\cdot|U(t)),p_{\hat{X}(t)|U(t)}(\cdot|U(t)))]\nonumber\\
	&=\mathbbm{E}[\phi(p_{X(T)|U(T),T}(\cdot|U(T),T),p_{\hat{X}(T)|U(T),T}(\cdot|U(T),T))],\label{eq:boundP}
\end{align}
where ($j$) follows by part 1) of Proposition \ref{prop:property}. It is clear that $X(T)\leftrightarrow (U(T),T)\leftrightarrow \hat{X}(T)$ form a Markov chain. Setting $U=(U(T),T)$ and combining (\ref{eq:boundR}), (\ref{eq:boundD}), (\ref{eq:boundP}) with (\ref{eq:R}), (\ref{eq:D}), (\ref{eq:P}) completes the proof.

\section{Proof of Theorem \ref{thm:binary}}\label{app:binary}

We first establish the following technical lemma.
\begin{lemma}\label{lem:envelope}
	For $D\geq 0$ and $P\geq 0$,
	\begin{align}
		\hbar(D,P)=&\max\limits_{a,\hat{a}}H_b(a)\label{eq:maximization}\\
		&\hspace{0.2cm}\text{s.t.}\quad (1-a)\hat{a}+a(1-\hat{a})\leq D,\\
		&\hspace{0.7cm}\quad|a-\hat{a}|\leq P,\\
		&\hspace{0.7cm}\quad 0\leq a\leq 1,\\
		&\hspace{0.7cm}\quad 0\leq \hat{a}\leq 1.
	\end{align}
\end{lemma}
\begin{IEEEproof}
	By symmetry, it suffices to consider $a\in[0,\frac{1}{2}]$. As a consequence, there is no loss of optimality in restricting $\hat{a}\in[0,a]$. We shall treat the following cases separately.
	
	\begin{enumerate}
		\item $D\in[\frac{1}{2},\infty)$:
		
		It is clear that $H_b(a)\leq\log 2$ and this upper bound is attained at $(a,\hat{a})=(\frac{1}{2},\frac{1}{2})$.
		
		\item $D\in[0,\frac{1}{2})$ and $P\in[D,\infty)$:
		
		Since $a\leq (1-a)\hat{a}+a(1-\hat{a})\leq D<\frac{1}{2}$, it follows that $H_b(a)\leq H_b(D)$. This upper bound is attained at $(a,\hat{a})=(D,0)$. 
		
		\item $D\in[0,\frac{1}{2})$ and $P\in[0,D)$:
		
		Setting $(1-a)\hat{a}+a(1-\hat{a})=D$ and $a-\hat{a}=P$ gives
		\begin{align}
			&a=\frac{1+P-\sqrt{1+P^2-2D}}{2},\label{eq:a}\\
			&\hat{a}=\frac{1-P-\sqrt{1+P^2-2D}}{2}.\label{eq:b}
		\end{align}
		Since $H_b(\frac{1+P-\sqrt{1+P^2-2D}}{2})$ is a monotonically increasing function of $(D,P)$ for $D\in[0,\frac{1}{2})$ and $P\in[0,D)$, 
		the maximum value of the optimization problem in (\ref{eq:maximization}) is attained at $(a,\hat{a})$ given by (\ref{eq:a}) and (\ref{eq:b}).
	\end{enumerate} 
	This completes the proof of Lemma \ref{lem:envelope}.
\end{IEEEproof}

Now we proceed to prove Theorem \ref{thm:binary}. Let $p_{U\hat{X}|X}$ a conditional distribution that	satisfies (\ref{eq:RDP1}), (\ref{eq:distortion1}), and (\ref{eq:perception1}). For any $u\in\mathcal{U}$,
\begin{align}
	&H(X|U=u)=H_b(P_{X|U}(1|u)),\\
	&\mathbbm{E}[d_{H}(X,\hat{X})|U=u]\nonumber\\&=(1-p_{X|U}(1|u))p_{\hat{X}|U}(1|u)+p_{X|U}(1|u)(1-p_{\hat{X}|U}(1|u)),\\
	&\mathbbm{E}[d_{\text{TV}}(p_{X|U}(\cdot|u),p_{\hat{X}|U}(\cdot|u))]=|p_{X|U}(1|u)-p_{\hat{X}|U}(1|u)|.
\end{align}
In light of Lemma \ref{lem:envelope},
\begin{align}
	&H(X|U=u)\nonumber\\&\leq\hbar(\mathbbm{E}[d_H(X,\hat{X})|U=u],d_{\text{TV}}(p_{X|U}(\cdot|u),p_{\hat{X}|U}(\cdot|u))).
\end{align}
Therefore,
\begin{align}
	&I(X;U)\nonumber\\&=\log2-H(X|U)\nonumber\\
	&\geq\log2-\mathbbm{E}[\hbar(\mathbbm{E}[d_H(X,\hat{X})|U],d_{\text{TV}}(p_{X|U}(\cdot|U),p_{\hat{X}|U}(\cdot|U)))]\nonumber\\
	&\geq\log2-\overline{\hbar}((\mathbbm{E}[d_H(X,\hat{X})],\mathbbm{E}[d_{\text{TV}}(p_{X|U}(\cdot|U),p_{\hat{X}|U}(\cdot|U))]))\nonumber\\
	&\stackrel{(a)}{\geq}\log2-\overline{h}(D,P),\label{eq:monotonicity}
\end{align}
where ($a$) is due to the fact that $\overline{\hbar}(D,P)$ is non-decreasing in $(D,P)$ (as $\hbar(D,P)$ is non-decreasing in $(D,P)$). This proves $R(D,P)\geq\log 2-\overline{h}(D,P)$.

According to Carath\'{e}odory's theorem, for any $D\geq 0$ and $P\geq 0$, we can write
\begin{align}
	\overline{\hbar}(D,P)=\sum\limits_{i=1}^3\alpha^{(i)}\hbar(D^{(i)},P^{(i)}),
\end{align}
where $\alpha^{(i)}\geq 0$, $D^{(i)}\geq 0$, $P^{(i)}\geq 0$,  $i\in[1:3]$, with $\sum_{i=1}^3\alpha^{(i)}=1$, $\sum_{i=1}^3\alpha^{(i)}D^{(i)}=D$, and $\sum_{i=1}^3\alpha^{(i)}P^{(i)}=P$. By Lemma \ref{lem:envelope}, there exist $(a^{(i)},\hat{a}^{(i)})\in[0,1]^2$ such that $H_b(a^{(i)})=\hbar(D^{(i)},P^{(i)})$, $(1-a^{(i)})\hat{a}^{(i)}+a^{(i)}(1-\hat{a}^{(i)})\leq D^{(i)}$,  and $|a^{(i)}-\hat{a}^{(i)}|\leq P^{(i)}$, $i\in[1:3]$. 
We shall use $(\alpha^{(1)},\alpha^{(2)},\alpha^{(3)})$, $(a^{(1)},a^{(2)},a^{(3)})$, and $(\hat{a}^{(1)},\hat{a}^{(2)},\hat{a}^{(3)})$ to define $p_U$, $p_{X|U}$, and $p_{\hat{X}|U}$, respectively. However, directly setting $p_U(i)=\alpha^{(i)}$, $p_{X|U}(1|i)=a^{(i)}$, and $p_{\hat{X}|U}(1|i)=\hat{a}^{(i)}$, $i\in[1:3]$, might not preserve $p_X$. For this reason, we adopt the following symmetrical construction. 
Specifically, set $\mathcal{U}=[1:6]$ and define a Markov chain $X\leftrightarrow U\leftrightarrow\hat{X}$ with
\begin{align}
	&p_{U}(u)=\frac{\alpha^{(u)}}{2},\quad u\in[1:3],\\
	&p_U(u)=\frac{\alpha^{(u-3)}}{2},\quad u\in[4:6],\\
	&p_{X|U}(1|u)=a^{(u)},\quad u\in[1:3],\\
	&p_{X|U}(1|u)=1-a^{(u-3)},\quad u\in[4:6],\\
	&p_{\hat{X}|U}(1|u)=\hat{a}^{(u)},\quad u\in[1:3],\\
	&p_{\hat{X}|U}(1|u)=1-\hat{a}^{(u-3)},\quad u\in[4:6].
\end{align}
Note that
\begin{align}
	p_X(1)&=\sum\limits_{u=1}^6p_{U}(u)p_{X|U}(1|u)\nonumber\\
	&=\frac{1}{2}\sum\limits_{u=1}^3\alpha^{(u)}a^{(u)}+\frac{1}{2}\sum\limits_{u=4}^6\alpha^{(u-3)}(1-a^{(u-3)})\nonumber\\
	&=\frac{1}{2}\sum\limits_{u=1}^3\alpha^{(u)}a^{(u)}+\frac{1}{2}\sum\limits_{u=1}^3\alpha^{(u)}(1-a^{(u)})\nonumber\\
	&=\frac{1}{2}\sum\limits_{i=1}^3\alpha^{(u)}\nonumber\\
	&=\frac{1}{2}.
\end{align}
So our construction preserves the source distribution. It can be verified that  
\begin{align}
	&H(X|U)\nonumber\\&=\sum\limits_{u=1}^6p_U(u)H(X|U=u)\nonumber\\
	&=\frac{1}{2}\sum\limits_{u=1}^3\alpha^{(u)}H_b(a^{(u)})+\frac{1}{2}\sum\limits_{u=4}^6\alpha^{(u-3)}H_b(1-a^{(u-3)})\nonumber\\
	&=\sum\limits_{u=1}^3\alpha^{(u)}H_b(a^{(u)})\nonumber\\
	&=\sum\limits_{u=1}^3\alpha^{(u)}\hbar(D^{(i)},P^{(i)})\nonumber\\
	&=\overline{\hbar}(D,P).
\end{align}
Moreover,
\begin{align}
	&\mathbbm{E}[d_H(X,\hat{X})]\nonumber\\&=\sum\limits_{u=1}^6p_U(u)\mathbbm{E}[d_H(X,\hat{X})|U=u]\nonumber\\
	&=\frac{1}{2}\sum\limits_{u=1}^3\alpha^{(u)}((1-a^{(u)})\hat{a}^{(u)}+a^{(u)}(1-\hat{a}^{(u)}))\nonumber\\&\quad+\frac{1}{2}\sum\limits_{u=4}^6\alpha^{(u-3)}(a^{(u-3)}(1-\hat{a}^{(u-3)})+(1-a^{(u-3)})\hat{a}^{(u-3)})\nonumber\\
	&=\sum\limits_{u=1}^3\alpha^{(u)}((1-a^{(u)})\hat{a}^{(u)}+a^{(u)}(1-\hat{a}^{(u)}))\nonumber\\
	&\leq\sum\limits_{u=1}^3\alpha^{(u)}D^{(u)}\nonumber\\
	&= D
\end{align}
and
\begin{align}
	&\mathbbm{E}[d_{\text{TV}}(p_{X|U}(\cdot|U),p_{\hat{X}|U}(\cdot|U))]\nonumber\\&=\sum\limits_{u=1}^6p_U(u)d_{\text{TV}}(p_{X|U}(\cdot|u),p_{\hat{X}|U}(\cdot|u))\nonumber\\
	&=\frac{1}{2}\sum\limits_{u=1}^3\alpha^{(u)}|a^{(u)}-\hat{a}^{(u)}|+\frac{1}{2}\sum\limits_{u=4}^6\alpha^{(u-3)}|\hat{a}^{(u-3)}-a^{(u-3)}|\nonumber\\
	&=\sum\limits_{u=1}^3\alpha^{(u)}|a^{(u)}-\hat{a}^{(u)}|\nonumber\\
	&\leq\sum\limits_{u=1}^3\alpha^{(u)}P^{(u)}\nonumber\\
	&=P.
\end{align}
Therefore, we must have $R(D,P)=\log2-\overline{\hbar}(D,P)$. Invoking Theorem \ref{thm:RDP} completes the proof.

\section{On the Concavity of $\hbar$}\label{app:non-concavity}

For $D\in(0,\frac{1}{2})$ and $P\in(0,D)$,
\begin{align}
	&\frac{\partial\hbar}{\partial D}(D,P)=\frac{1}{2\upsilon}\log\left(\frac{1-P+\upsilon}{1+P-\upsilon}\right),\\
	&\frac{\partial\hbar}{\partial P}(D,P)=\frac{\upsilon-P}{2\upsilon}\log\left(\frac{1-P+\upsilon}{1+P-\upsilon}\right).
\end{align}
where $\upsilon=\sqrt{1+P^2-2D}$; moreover,
\begin{align}
	&\frac{\partial^2\hbar}{\partial D^2}(D,P)\nonumber\\&=\frac{1}{2\upsilon^3}\log\left(\frac{1-P+\upsilon}{1+P-\upsilon}\right)-\frac{1}{\upsilon^2(1-(\upsilon-P)^2)},\\
	&\frac{\partial^2\hbar}{\partial P^2}(D,P)\nonumber\\&=-\frac{1-2D}{2\upsilon^3}\log\left(\frac{1-P+\upsilon}{1+P-\upsilon}\right)-\frac{(\upsilon-P)^2}{\upsilon^2(1-(\upsilon-P)^2)},\\
	&\frac{\partial^2\hbar}{\partial D\partial P}(D,P)\nonumber\\&=-\frac{P}{2\upsilon^3}\log\left(\frac{1-P+\upsilon}{1+P-\upsilon}\right)-\frac{\upsilon-P}{\upsilon^2(1-(\upsilon-P)^2)}.
\end{align}
Evaluating the Hessian matrix of $\hbar$ at $(D,P)=(\frac{1}{2}-\epsilon, \epsilon)$ gives
\begin{align}
	&\left(\begin{matrix}
		\frac{\partial^2\hbar}{\partial D^2}\left(\frac{1}{2}-\epsilon,\epsilon\right)&\frac{\partial^2\hbar}{\partial D\partial P}\left(\frac{1}{2}-\epsilon,\epsilon\right)\\
		\frac{\partial^2\hbar}{\partial D\partial P}\left(\frac{1}{2}-\epsilon,\epsilon\right)&\frac{\partial^2\hbar}{\partial P^2}\left(\frac{1}{2}-\epsilon,\epsilon\right)
	\end{matrix}\right)\nonumber\\&=\left(\begin{matrix}
	-\frac{1}{2\sqrt{2\epsilon}}+o\left(\frac{1}{\sqrt{\epsilon}}\right)& -\frac{1}{\sqrt{2\epsilon}}+o\left(\frac{1}{\sqrt{\epsilon}}\right)\\
	-\frac{1}{\sqrt{2\epsilon}}+o\left(\frac{1}{\sqrt{\epsilon}}\right) & -2+o(1)
\end{matrix}\right).\label{eq:Hessian}
\end{align}
Note that
\begin{align}
&\lim\limits_{\epsilon\rightarrow 0}(a,b)	\left(\begin{matrix}
		\frac{\partial^2\hbar}{\partial D^2}\left(\frac{1}{2}-\epsilon,\epsilon\right)&\frac{\partial^2\hbar}{\partial D\partial P}\left(\frac{1}{2}-\epsilon,\epsilon\right)\\
		\frac{\partial^2\hbar}{\partial D\partial P}\left(\frac{1}{2}-\epsilon,\epsilon\right)&\frac{\partial^2\hbar}{\partial P^2}\left(\frac{1}{2}-\epsilon,\epsilon\right)
	\end{matrix}\right)\left(\begin{matrix}
	a\\
	b
\end{matrix}\right)\nonumber\\
&=\begin{cases}
	\infty,& a^2+4ab<0,\\
	-\infty,& a^2+4ab>0.
\end{cases}
\end{align}
Therefore, $\hbar$ is neither concave nor convex. One can also reach the same conclusion by observing that the eigvalues of the Hessian matrix in (\ref{eq:Hessian}) are 
\begin{align}
	\frac{-1\pm\sqrt{17}}{4\sqrt{2\epsilon}}+o\left(\frac{1}{\sqrt{\epsilon}}\right).
\end{align}

\section{Proof of Theorem \ref{thm:continuous}}\label{app:continuous}

\subsection{Proof of (\ref{eq:continuousRDP})}
It suffices to show
\begin{align}
	R_{\text{C}}(D,P)\leq R(D,P)\label{eq:achievability}
\end{align}
since the converse part in the proof of Theorem \ref{thm:RDP} continues to hold for continuous alphabet sources.
We shall treat the following two cases separately.

\subsubsection{ $D>0$ and $P=0$}
	
	In light of Proposition \ref{prop:convexity}, $R(D,0)$ is continuous in $D$ for $D>0$. So for any $\epsilon>0$, there exists $\rho>0$ satisfying $R(D-\rho,0)\leq R(D,0)+\epsilon$. By the definition of $R(D-\rho,0)$, we can find $p_{U\hat{X}|X}$ such that $I(X;U)\leq R(D-\rho,0)+\epsilon$, $X\leftrightarrow U\leftrightarrow\hat{X}$ form a Markov chain, $\mathbbm{E}[\|X-\hat{X}\|^2]\leq D-\rho$, and $p_{X|U}=p_{\hat{X}|U}$. 
	Let $\xi$ be a vector quantizer that maps each point in $\mathbbm{R}^L$  to its nearest neighbor in $\mathcal{Y}=\frac{1}{\sqrt{N}}[-N:N]^L$, where
	$N$ is a positive integer. Moreover, let $Y=\xi(X)$ and $\hat{Y}=\xi(\hat{X})$. Note that $Y\leftrightarrow X\leftrightarrow U\leftrightarrow\hat{X}\leftrightarrow\hat{Y}$ form a Markov chain and $p_{Y|U}=p_{\hat{X}|U}$. As a consequence, we have	
	\begin{align}
		I(Y;U)&\leq I(X;U)\nonumber\\
		&\leq R(D-\rho,0)+\epsilon\nonumber\\
		&\leq R(D,0)+2\epsilon\label{eq:DPI}
		\end{align}
	and
	\begin{align}
		\mathbbm{E}[W^2_2(p_{Y|U}(\cdot|U),p_{\hat{Y}|U}(\cdot|U))]=0.\label{eq:zeroP}
\end{align}
	Moreover, 
	\begin{align}
		&\mathbbm{E}[\|Y-\hat{Y}\|^2]\nonumber\\&=\mathbbm{E}\left[\sum\limits_{\ell=1}^L|Y_{\ell}-\hat{Y}_{\ell}|^2\right]\nonumber\\
		&\leq\mathbbm{E}\left[\sum\limits_{\ell=1}^L\left(|X_{\ell}-\hat{X}_{\ell}|+\frac{1}{\sqrt{N}}\right)^2\right]\nonumber\\
		&=\mathbbm{E}\left[\sum\limits_{\ell=1}^L\left(|X_{\ell}-\hat{X}_{\ell}|^2+\frac{2}{\sqrt{N}}|X_{\ell}-\hat{X}_{\ell}|+\frac{1}{N}\right)\right]\nonumber\\
		&\leq\mathbbm{E}[\|X-\hat{X}\|^2]+2\sqrt{\frac{L}{N}\mathbbm{E}[\|X-\hat{X}\|^2]}+\frac{L}{N}\nonumber\\
		&\leq D-\rho+2\sqrt{\frac{L}{N}(D-\rho)}+\frac{L}{N}\nonumber\\
		&=:D'.\label{eq:D1}
	\end{align}
Combining (\ref{eq:DPI}), (\ref{eq:zeroP}), and (\ref{eq:D1}) yields
\begin{align}
	R(D',0|p_Y)\leq R(D,0)+2\epsilon,
\end{align}
	where $R(D,P|p_Y)$ denotes the counterpart of $R(D,P)$ for source $Y$.
	
According to Theorem \ref{thm:RDP}, there exists 
$p_{Y^nM\hat{Y}^n}=p^n_Yp_{M|Y^n}p_{\hat{Y}^n|M}$ such that
$\frac{1}{n}\mathbbm{E}[\ell(M)]\leq R(D',0|p_Y)+\epsilon$, $\frac{1}{n}\mathbbm{E}[\|Y^n-\hat{Y}^n\|^2]\leq D'$, and $p_{Y^n|M}=p_{\hat{Y}^n|M}$. Construct $p_{X^nY^nM\hat{Y}^n\hat{X}^n}=p_{X^n|Y^n}p_{Y^nM\hat{Y}^n}p_{\hat{X}^n|\hat{Y}^n}$ with $p_{X^n|Y^n}=p_{\hat{X}^n|\hat{Y}^n}=p^n_{X|Y}$. Clearly, we have $p_{X^n|M}=p_{\hat{X}^n|M}$ and consequently
\begin{align}
	\frac{1}{n}\mathbbm{E}[W^2_2(p_{X^n|M}(\cdot|M),p_{\hat{X}^n|M}(\cdot|M))]=0.\label{eq:Ppart}
\end{align}
Moreover, 
\begin{align}
	&\frac{1}{n}\mathbbm{E}[\|X^n-\hat{X}^n\|^2]\nonumber\\&=\frac{1}{n}\sum\limits_{t=1}^n\sum\limits_{\ell=1}^L\mathbbm{E}[|X_{\ell}(t)-\hat{X}_{\ell}(t)|^2]\nonumber\\
	&=\frac{1}{n}\sum\limits_{t=1}^n\sum\limits_{\ell=1}^L\sum\limits_{i=1}^4\mathbbm{P}\{(Y(t),\hat{Y}(t))\in\mathcal{S}_{\ell,i}\}\nonumber\\&\hspace{0.2in}\times\mathbbm{E}[|X_{\ell}(t)-\hat{X}_{\ell}(t)|^2|(Y(t),\hat{Y}(t))\in\mathcal{S}_{\ell,i}],\label{eq:ssum}
\end{align}
where 
\begin{align}
	&\mathcal{S}_{\ell,1}=\{(y,\hat{y})\in\mathcal{Y}^2: |y_{\ell}|\neq\sqrt{N}, |\hat{y}_{\ell}|\neq\sqrt{N}\},\\
	&\mathcal{S}_{\ell,2}=\{(y,\hat{y})\in\mathcal{Y}^2: |y_{\ell}|=\sqrt{N}, |\hat{y}_{\ell}|=\sqrt{N}\},\\
	&\mathcal{S}_{\ell,3}=\{(y,\hat{y})\in\mathcal{Y}^2: |y_{\ell}|=\sqrt{N}, |\hat{y}_{\ell}|\neq\sqrt{N}\},\\
	&\mathcal{S}_{\ell,4}=\{(y,\hat{y})\in\mathcal{Y}^2: |y_{\ell}|\neq\sqrt{N}, |\hat{y}_{\ell}|=\sqrt{N}\}.
\end{align}
For $(y,\hat{y})\in\mathcal{S}_{\ell,1}$,
\begin{align}
	&\mathbbm{E}[|X_{\ell}(t)-\hat{X}_{\ell}(t)|^2|Y(t)=y,\hat{Y}(t)=\hat{y}]\nonumber\\&\leq\left(|y_{\ell}-\hat{y}_{\ell}|+\frac{1}{\sqrt{N}}\right)^2.
\end{align}
Therefore, 
\begin{align}
	&\mathbbm{P}\{(Y(t),\hat{Y}(t))\in\mathcal{S}_{\ell,1}\}\nonumber\\&\times\mathbbm{E}[|X_{\ell}(t)-\hat{X}_{\ell}(t)|^2|(Y(t),\hat{Y}(t))\in\mathcal{S}_{\ell,1}]\nonumber\\
	&=\sum\limits_{(y,\hat{y})\in\mathcal{S}_{\ell,1}}p_{Y(t)\hat{Y}(t)}(y,\hat{y})\nonumber\\&\hspace{0.65in}\times\mathbbm{E}[|X_{\ell}(t)-\hat{X}_{\ell}(t)|^2|Y(t)=y,\hat{Y}(t)=\hat{y}]\nonumber\\
	&\leq\sum\limits_{(y,\hat{y})\in\mathcal{S}_{\ell,1}}p_{Y(t)\hat{Y}(t)}(y,\hat{y})\left(|y_{\ell}-\hat{y}_{\ell}|+\frac{1}{\sqrt{N}}\right)^2\nonumber\\
	&\leq\sum\limits_{(y,\hat{y})\in\mathcal{Y}^2}p_{Y(t)\hat{Y}(t)}(y,\hat{y})\left(|y_{\ell}-\hat{y}_{\ell}|+\frac{1}{\sqrt{N}}\right)^2\nonumber\\
	&=\mathbbm{E}[|Y_{\ell}(t)-\hat{Y}_{\ell}(t)|^2]+\frac{2}{\sqrt{N}}\mathbbm{E}[|Y_{\ell}(t)-\hat{Y}_{\ell}(t)|]+\frac{1}{N}\nonumber\\
	&\leq\mathbbm{E}[|Y_{\ell}(t)-\hat{Y}_{\ell}(t)|^2]+\frac{2}{\sqrt{N}}\sqrt{\mathbbm{E}[|Y_{\ell}(t)-\hat{Y}_{\ell}(t)|^2]}\nonumber\\
 &\quad+\frac{1}{N}.\label{eq:s1}
\end{align}
For $(y,\hat{y})\in\mathcal{S}_{\ell,2}$,
\begin{align}
	&\mathbbm{E}[|X_{\ell}(t)-\hat{X}_{\ell}(t)|^2|Y(t)=y,\hat{Y}=\hat{y}]\nonumber\\&\leq2\mathbbm{E}[X^2_{\ell}(t)+\hat{X}^2_{\ell}(t)|Y(t)=y,\hat{Y}(t)=\hat{y}]\nonumber\\
	&\stackrel{(a)}{=}2\mathbbm{E}[X^2_{\ell}(t)|Y(t)=y]+2\mathbbm{E}[\hat{X}^2_{\ell}(t)|\hat{Y}(t)=\hat{y}],\label{eq:XYMarkov}
\end{align}
where ($a$) follows by the fact that $X_{\ell}(t)\rightarrow Y(t)\leftrightarrow\hat{Y}(t)\leftrightarrow\hat{X}_{\ell}(t)$ form a Markov chain.
Therefore, 
\begin{align}
	&\mathbbm{P}\{(Y(t),\hat{Y}(t))\in\mathcal{S}_{\ell,2}\}\nonumber\\&\times\mathbbm{E}[|X_{\ell}(t)-\hat{X}_{\ell}(t)|^2|(Y(t),\hat{Y}(t))\in\mathcal{S}_{\ell,2}]\nonumber\\
	&=\sum\limits_{(y,\hat{y})\in\mathcal{S}_{\ell,2}}p_{Y(t)\hat{Y}(t)}(y,\hat{y})\nonumber\\&\hspace{0.7in}\times\mathbbm{E}[|X_{\ell}(t)-\hat{X}_{\ell}(t)|^2|Y(t)=y,\hat{Y}(t)=\hat{y}]\nonumber\\
	&\leq2\sum\limits_{(y,\hat{y})\in\mathcal{S}_{\ell,2}}p_{Y(t)\hat{Y}(t)}(y,\hat{y})(\mathbbm{E}[X^2_{\ell}(t)|Y(t)=y]\nonumber\\&\hspace{1.65in}+\mathbbm{E}[\hat{X}^2_{\ell}(t)|\hat{Y}(t)=\hat{y}])\nonumber\\
	&\leq2\sum\limits_{y\in\mathcal{Y}:y_{\ell}\in\{-\sqrt{N},\sqrt{N}\}}p_{Y_{\ell}}(y)\mathbbm{E}[X^2_{\ell}(t)|Y(t)=y]\nonumber\\&\quad+2\sum\limits_{\hat{y}\in\mathcal{Y}:\hat{y}_{\ell}\in\{-\sqrt{N},\sqrt{N}\}}p_{\hat{Y}(t)}(\hat{y})\mathbbm{E}[\hat{X}^2_{\ell}(t)|\hat{Y}(t)=\hat{y}]\nonumber\\
	&\stackrel{(b)}{=}4\sum\limits_{y\in\mathcal{Y}:y_{\ell}\in\{-\sqrt{N},\sqrt{N}\}}p_{Y}(y)\mathbbm{E}[X^2_{\ell}|Y=y]\nonumber\\
	&\leq4\delta_{\ell,N},\label{eq:s2}
\end{align}
where 
\begin{align}
	\delta_{\ell,N}=\mathbbm{P}\left\{|X_{\ell}|\geq\frac{N-1}{\sqrt{N}}\right\}\mathbbm{E}\left[X^2_{\ell}\left||X_{\ell}|\geq\frac{N-1}{\sqrt{N}}\right.\right],
\end{align}
and ($b$) is due to $p_{X_{\ell}(t)Y(t)}=p_{\hat{X}_{\ell}(t)\hat{Y}(t)}=p_{X_{\ell}Y}$.
For $(y,\hat{y})\in\mathcal{S}_{\ell,3}$,
\begin{align}
	&\mathbbm{E}[|X_{\ell}(t)-\hat{X}_{\ell}(t)|^2|Y(t)=y,\hat{Y}(t)=\hat{y}]\nonumber\\&\leq\mathbbm{E}[(2|X_{\ell}(t)|)^2|Y(t)=y,\hat{Y}(t)=\hat{y}]\nonumber\\
	&\stackrel{(c)}{=}4\mathbbm{E}[X^2_{\ell}(t)|Y(t)=y],\label{eq:XYYMarkov}
\end{align}
where ($c$) follows by the fact that $X_{\ell}(t)\leftrightarrow Y(t)\leftrightarrow \hat{Y}(t)$ form a Markov chain. Therefore,
\begin{align}
&\mathbbm{P}\{(Y(t),\hat{Y}(t))\in\mathcal{S}_{\ell,3}\}\nonumber\\&\times\mathbbm{E}[|X_{\ell}(t)-\hat{X}_{\ell}(t)|^2|(Y(t),\hat{Y}(t))\in\mathcal{S}_{\ell,3}]\nonumber\\	
&=\sum\limits_{(y,\hat{y})\in\mathcal{S}_{\ell,3}}p_{Y(t)\hat{Y}(t)}(y,\hat{y})\nonumber\\&\hspace{0.7in}\times\mathbbm{E}[|X_{\ell}(t)-\hat{X}_{\ell}(t)|^2|Y(t)=y,\hat{Y}(t)=\hat{y}]\nonumber\\
&\leq4\sum\limits_{(y,\hat{y})\in\mathcal{S}_{\ell,3}}p_{Y(t)\hat{Y}(t)}(y,\hat{y})\mathbbm{E}[X^2_{\ell}(t)|Y(t)=y]\nonumber\\
&\leq4\sum\limits_{y\in\mathcal{Y}: y_{\ell}\in\{-\sqrt{N},\sqrt{N}\}}p_{Y(t)}(y)\mathbbm{E}[X^2_{\ell}(t)|Y(t)=y]\nonumber\\
&\stackrel{(d)}{=}4\sum\limits_{y\in\mathcal{Y}: y_{\ell}\in\{-\sqrt{N},\sqrt{N}\}}p_{Y}(y)\mathbbm{E}[X^2_{\ell}|Y=y]\nonumber\\
&\leq4\delta_{\ell,N},\label{eq:s3}
\end{align}
where ($d$) is due to $p_{X_{\ell}(t)Y(t)}=p_{X_{\ell}Y}$.
By symmetry
\begin{align}
	&\mathbbm{P}\{(Y(t),\hat{Y}(t))\in\mathcal{S}_{\ell,4}\}\nonumber\\&\times\mathbbm{E}[|X_{\ell}(t)-\hat{X}_{\ell}(t)|^2|(Y(t),\hat{Y}(t))\in\mathcal{S}_{\ell,4}]\nonumber\\&\leq4\delta_{\ell,N}.\label{eq:s4}
\end{align}
Substituting (\ref{eq:s1}), (\ref{eq:s2}), (\ref{eq:s3}), and (\ref{eq:s4}) into (\ref{eq:ssum}) gives
\begin{align}
	&\frac{1}{n}\mathbbm{E}[\|X^n-\hat{X}^n\|^2]\nonumber\\&\leq\frac{1}{n}\sum\limits_{t=1}^n\sum\limits_{\ell=1}^L\Bigg(\mathbbm{E}[|Y_{\ell}(t)-\hat{Y}_{\ell}(t)|^2]\nonumber\\&\hspace{0.5in}+\frac{2}{\sqrt{N}}\sqrt{\mathbbm{E}[|Y_{\ell}(t)-\hat{Y}_{\ell}(t)|^2]}+\frac{1}{N}+12\delta_{\ell,N}\Bigg)\nonumber\\
	&\leq\frac{1}{n}\mathbbm{E}[\|Y^n-\hat{Y}^n\|^2]+2\sqrt{\frac{L}{nN}\mathbbm{E}[\|Y^n-\hat{Y}^n\|^2]}+\frac{L}{N}\nonumber\\&\quad+12\sum\limits_{\ell=1}^L\delta_{\ell,N}\nonumber\\
	&\leq D'+2\sqrt{\frac{L}{N}D'}+\frac{L}{N}+12\sum\limits_{\ell=1}^L\delta_{\ell,N}.
\end{align}
Since $\mathbbm{E}[\|X\|^2]<\infty$, it follows by the dominated convergence theorem that $\sum_{\ell=1}^L\delta_{\ell,N}$ tends to zero as $N\rightarrow\infty$. Moreover, $D'+2\sqrt{\frac{L}{N}D'}+\frac{L}{N}$ converges to $D-\rho$ as $N\rightarrow\infty$. So when $N$ is sufficiently large, we have
\begin{align}
	\frac{1}{n}\mathbbm{E}[\|X^n-\hat{X}^n\|^2]\leq D.\label{eq:Dpart}
\end{align}
Combining (\ref{eq:Ppart}) and (\ref{eq:Dpart}) as well as the fact that 
\begin{align}
	\frac{1}{n}\mathbbm{E}[\ell(M)]\leq R(D',0|p_Y)+\epsilon\leq R(D,0)+3\epsilon
\end{align}
yields $R_{\text{C}}(D,0)\leq R(D,0)+3\epsilon$. Since $\epsilon>0$ is arbitrary, this proves (\ref{eq:achievability}) for $P=0$.

\subsubsection{ $D>0$ and $P>0$}
	
		In light of Proposition \ref{prop:convexity}, $R(D,P)$ is continuous in $(D,P)$ for $D>0$ and $P>0$. So for any $\epsilon>0$, there exists $\rho>0$ satisfying $R(D-\rho,P-\rho)\leq R(D,P)+\epsilon$. By the definition of $R(D-\rho,P-\rho)$, we can find $p_{U\hat{X}|X}$ such that $I(X;U)\leq R(D-\rho,P-\rho)+\epsilon$, $X\leftrightarrow U\leftrightarrow\hat{X}$ form a Markov chain, $\mathbbm{E}[\|X-\hat{X}\|^2]\leq D-\rho$, and $\mathbbm{E}[W^2_2(p_{X|U}(\cdot|U),p_{\hat{X}|U}(\cdot|U))]\leq P-\rho$. Let 
		$Y=\xi(X)$ and $\hat{Y}=\xi(\hat{X})$.
		 Note that $Y\leftrightarrow X\leftrightarrow U\leftrightarrow\hat{X}\leftrightarrow\hat{Y}$ form a Markov chain. As a consequence,
		\begin{align}
			I(Y;U)&\leq I(X;U)\nonumber\\
			&\leq R(D-\rho,P-\rho)+\epsilon\nonumber\\
			&\leq R(D,P)+2\epsilon.\label{eq:combR}
		\end{align}
The derivation of (\ref{eq:D1}) continues to hold here, i.e.,
	\begin{align}
		&\mathbbm{E}[\|Y-\hat{Y}\|^2]\nonumber\\&\leq\mathbbm{E}[\|X-\hat{X}\|^2]+2\sqrt{\frac{L}{N}\mathbbm{E}[\|X-\hat{X}\|^2]}+\frac{L}{N}\label{eq:hold}\\
		&\leq D-\rho+2\sqrt{\frac{L}{N}(D-\rho)}+\frac{L}{N}\nonumber\\
		&= D'.\label{eq:combD}
	\end{align}
Since (\ref{eq:hold}) is valid for any jointly distributed $(X,\hat{X})$ and the induced $(Y,\hat{Y})$ (via quantizer $\xi$), it follows that
\begin{align}
	&W^2_2(p_{Y|U}(\cdot|u),p_{\hat{Y}|U}(\cdot|u))\nonumber\\&\leq W^2_2(p_{X|U}(\cdot|u),p_{\hat{X}|U}(\cdot|u))\nonumber\\&\quad+2\sqrt{\frac{L}{N}W^2_2(p_{X|U}(\cdot|u),p_{\hat{X}|U}(\cdot|u))}+\frac{L}{N},\quad u\in\mathcal{U}.
\end{align}
Therefore, we have
	\begin{align}
		&\mathbbm{E}[W^2_2(p_{Y|U}(\cdot|U),p_{\hat{Y}|U}(\cdot|U))]\nonumber\\
		&\leq \mathbbm{E}[W^2_2(p_{X|U}(\cdot|U),p_{\hat{X}|U}(\cdot|U))]\nonumber\\&\quad+2\sqrt{\frac{L}{N}\mathbbm{E}[W^2_2(p_{X|U}(\cdot|U),p_{\hat{X}|U}(\cdot|U))]}+\frac{L}{N}\nonumber\\
		&\leq P-\rho+2\sqrt{\frac{L}{N}(P-\rho)}+\frac{L}{N}=:P'.\label{eq:combP}
	\end{align}
Combining (\ref{eq:combR}), (\ref{eq:combD}), and (\ref{eq:combP}) yields
\begin{align}
	R(D',P'|p_Y)\leq R(D,P)+2\epsilon.
\end{align}

According to Theorem \ref{thm:RDP}, there exists 
$p_{Y^nM\hat{Y}^n}=p^n_Yp_{M|Y^n}p_{\hat{Y}^n|M}$ such that
$\frac{1}{n}\mathbbm{E}[\ell(M)]\leq R(D',P'|p_Y)+\epsilon$, $\frac{1}{n}\mathbbm{E}[\|Y^n-\hat{Y}^n\|^2]\leq D'$, and $\mathbbm{E}[W^2_2(p_{Y|U}(\cdot|U),p_{\hat{Y}|U}(\cdot|U))]\leq P'$. Construct $p_{X^nY^nM\hat{Y}^n}=p_{X^n|Y^n}p_{Y^nM\hat{Y}^n}$ with $p_{X^n|Y^n}=p^n_{X|Y}$ and let $\hat{X}^n=\hat{Y}^n$. Note that
\begin{align}
	&\frac{1}{n}\mathbbm{E}[\|X^n-\hat{X}^n\|^2]\nonumber\\&=\frac{1}{n}\mathbbm{E}[\|X^n-\hat{Y}^n\|^2]\nonumber\\
	&=\frac{1}{n}\sum\limits_{t=1}^n\sum\limits_{\ell=1}^L\mathbbm{P}\{|Y_{\ell}(t)|\neq\sqrt{N}\}\nonumber\\&\hspace{0.73in}\times\mathbbm{E}[|X_{\ell}(t)-\hat{Y}_{\ell}(t)|^2||Y_{\ell}(t)|\neq\sqrt{N}]\nonumber\\
	&\quad+\frac{1}{n}\sum\limits_{t=1}^n\sum\limits_{\ell=1}^L\mathbbm{P}\{|Y_{\ell}(t)|=\sqrt{N}\}\nonumber\\&\hspace{0.55in}\times\mathbbm{E}[|X_{\ell}(t)-\hat{Y}_{\ell}(t)|^2||Y_{\ell}(t)|=\sqrt{N}].\label{eq:case2sub}
\end{align}
For $y_{\ell}\in\frac{1}{\sqrt{N}}[1-N:N-1]$,
\begin{align}
	&\mathbbm{E}[|X_{\ell}(t)-\hat{Y}_{\ell}(t)|^2|Y_{\ell}(t)=y_{\ell}]\nonumber\\&\leq\mathbbm{E}\left[\left.\left(|y_{\ell}-\hat{Y}_{\ell}(t)|+\frac{1}{\sqrt{N}}\right)^2\right|Y_{\ell}(t)=y_{\ell}\right].
\end{align}
Therefore,
\begin{align}
	&\mathbbm{P}\{|Y_{\ell}(t)|\neq\sqrt{N}\}\mathbbm{E}[|X_{\ell}(t)-\hat{Y}_{\ell}(t)|^2||Y_{\ell}(t)|\neq\sqrt{N}]\nonumber\\
	&=\sum\limits_{y_{\ell}\in\frac{1}{\sqrt{N}}[1-N:N-1]}p_{Y_{\ell}(t)}(y_{\ell})\nonumber\\&\hspace{1.1in}\times\mathbbm{E}[|X_{\ell}(t)-\hat{Y}_{\ell}(t)|^2|Y_{\ell}(t)=y_{\ell}]\nonumber\\
	&=\sum\limits_{y_{\ell}\in\frac{1}{\sqrt{N}}[1-N:N-1]}p_{Y(t)}(y)\nonumber\\&\hspace{0.7in}\times\mathbbm{E}\left[\left.\left(|y_{\ell}-\hat{Y}_{\ell}(t)|+\frac{1}{\sqrt{N}}\right)^2\right|Y_{\ell}(t)=y_{\ell}\right]\nonumber\\
	&\leq\sum\limits_{y_{\ell}\in\frac{1}{\sqrt{N}}[-N:N]}p_{Y(t)}(y)\nonumber\\&\hspace{0.7in}\times\mathbbm{E}\left[\left.\left(|y_{\ell}-\hat{Y}_{\ell}(t)|+\frac{1}{\sqrt{N}}\right)^2\right|Y_{\ell}(t)=y_{\ell}\right]\nonumber\\
	&=\mathbbm{E}[|Y_{\ell}(t)-\hat{Y}_{\ell}(t)|^2]+\frac{2}{\sqrt{N}}\mathbbm{E}[|Y_{\ell}(t)-\hat{Y}_{\ell}(t)|]+\frac{1}{N}\nonumber\\
	&\leq\mathbbm{E}[|Y_{\ell}(t)-\hat{Y}_{\ell}(t)|^2+\frac{2}{\sqrt{N}}\sqrt{\mathbbm{E}[|Y_{\ell}(t)-\hat{Y}_{\ell}(t)|^2]}+\frac{1}{N}.\nonumber\\\label{eq:case2sub1}
\end{align}
For $y_{\ell}\in\{-\sqrt{N},\sqrt{N}\}$,
\begin{align}
	&\mathbbm{E}[|X_{\ell}(t)-\hat{Y}_{\ell}(t)|^2|Y_{\ell}(t)=y_{\ell}]\nonumber\\&\leq\mathbbm{E}[(3|X_{\ell}(t)|)^2|Y_{\ell}(t)=y_{\ell}]\nonumber\\
	&=9\mathbbm{E}[X^2_{\ell}(t)|Y_{\ell}(t)=y_{\ell}].
\end{align}
Therefore,
\begin{align}
	&\mathbbm{P}\{|Y_{\ell}(t)|=\sqrt{N}\}\mathbbm{E}[|X_{\ell}(t)-\hat{Y}_{\ell}(t)|^2||Y_{\ell}(t)|=\sqrt{N}]\nonumber\\
	&=\sum\limits_{y_{\ell}\in\{-\sqrt{N},\sqrt{N}\}}p_{Y_{\ell}(t)}(y_{\ell})\nonumber\\&\hspace{0.97in}\times\mathbbm{E}[|X_{\ell}(t)-\hat{Y}_{\ell}(t)|^2|Y_{\ell}(t)=y_{\ell}]\nonumber\\
	&\leq9\sum\limits_{y\in\{-\sqrt{N},\sqrt{N}\}}p_{Y_{\ell}(t)}(y_{\ell})\mathbbm{E}[X^2_{\ell}(t)|Y_{\ell}(t)=y_{\ell}]\nonumber\\
	&\leq 9\delta_{\ell,N}.\label{eq:case2sub2}
\end{align}
Substituting (\ref{eq:case2sub1}) and (\ref{eq:case2sub2}) into (\ref{eq:case2sub}) gives
\begin{align}
	&\frac{1}{n}\mathbbm{E}[\|X^n-\hat{X}^n\|]\nonumber\\&\leq\frac{1}{n}\sum\limits_{t=1}^n\sum\limits_{\ell=1}^L\Bigg(\mathbbm{E}[|Y_{\ell}(t)-\hat{Y}_{\ell}(t)|^2\nonumber\\&\hspace{0.55in}+\frac{2}{\sqrt{N}}\sqrt{\mathbbm{E}[|Y_{\ell}(t)-\hat{Y}_{\ell}(t)|^2]}+\frac{1}{N}+9\delta_{\ell,N}\Bigg)\nonumber\\
	&\leq\frac{1}{n}\mathbbm{E}[\|Y^n-\hat{Y}^n\|^2]+2\sqrt{\frac{L}{nN}\mathbbm{E}[\|Y^n-\hat{Y}^n\|^2]}+\frac{L}{N}\nonumber\\
 &\quad+9\sum\limits_{\ell=1}^L\delta_{\ell,N}\label{eq:valid}\\
	&\leq D'+2\sqrt{\frac{L}{N}D'}+\frac{L}{N}+9\sum\limits_{\ell=1}^L\delta_{\ell,N}.
\end{align}
Since (\ref{eq:valid}) is valid for any jointly distributed $(Y^n,\hat{Y}^n)$ and the induced $(X^n,\hat{X}^n)$ (via Markov coupling $X^n\leftrightarrow Y^n\leftrightarrow\hat{Y}^n$ with $p_{X^n|Y^n}=p^n_{X|Y}$ and setting $\hat{X}^n=\hat{Y}^n$), it follows that
\begin{align}
	&\frac{1}{n}W^2_2(p_{X^n|M}(\cdot|m),p_{\hat{X}^n|M}(\cdot|m))\nonumber\\
	&\leq \frac{1}{n}W^2_2(p_{Y^n|M}(\cdot|m),p_{\hat{Y}^n|M}(\cdot|m))\nonumber\\&\quad+2\sqrt{\frac{L}{nN}W^2_2(p_{Y^n|M}(\cdot|m),p_{\hat{Y}^n|M}(\cdot|m))}+\frac{L}{N}\nonumber\\&\quad+9\sum\limits_{\ell=1}^L\delta_{\ell,N},\quad m\in\mathcal{M}.
\end{align}
Therefore, we have
\begin{align}
	&\frac{1}{n}\mathbbm{E}[W^2_2(p_{X^n|M}(\cdot|M),p_{\hat{X}^n|M}(\cdot|M))]\nonumber\\
	&\leq \frac{1}{n}\mathbbm{E}[W^2_2(p_{Y^n|M}(\cdot|M),p_{\hat{Y}^n|M}(\cdot|M))]\nonumber\\&\quad+2\sqrt{\frac{L}{nN}\mathbbm{E}[W^2_2(p_{Y^n|M}(\cdot|M),p_{\hat{Y}^n|M}(\cdot|M))]}+\frac{L}{N}\nonumber\\&\quad+9\sum\limits_{\ell=1}^L\delta_{\ell,N}\nonumber\\
	&\leq P'+2\sqrt{\frac{L}{N}P'}+\frac{L}{N}+9\sum\limits_{\ell=1}^L\delta_{\ell,N}.
\end{align}
Since $\mathbbm{E}[\|X\|^2]<\infty$, it follows by the dominated convergence theorem that $\sum_{\ell=1}^L\delta_{\ell,N}$ tends to zero as $N\rightarrow\infty$. Moreover, $D'+2\sqrt{\frac{L}{N}D'}+\frac{L}{N}$ and $P'+2\sqrt{\frac{L}{N}P'}+\frac{L}{N}$ converge respectively to $D-\rho$ and $P-\rho$ as $N\rightarrow\infty$. So when $N$ is sufficiently large, we have
\begin{align}
	&\frac{1}{n}\mathbbm{E}[\|X^n-\hat{X}^n\|^2]\leq D,\label{eq:Dpart2}\\
	&\frac{1}{n}\mathbbm{E}[W^2_2(p_{X^n|M}(\cdot|M),p_{\hat{X}^n|M}(\cdot|M))]\leq P.\label{eq:Ppart2}
\end{align}
Combining (\ref{eq:Dpart2}) and (\ref{eq:Ppart2}) as well as the fact that 
\begin{align}
	\frac{1}{n}\mathbbm{E}[\ell(M)]\leq R(D',P'|p_Y)+\epsilon\leq R(D,P)+3\epsilon
\end{align}
yields $R_{\text{C}}(D,0)\leq R(D,0)+3\epsilon$. Since $\epsilon>0$ is arbitrary, this proves (\ref{eq:achievability}) for $P>0$.

\subsection{Proof of (\ref{eq:alternative})} 
For any $(U,\hat{X})$ satisfying (\ref{eq:RDP1}), let $U'=\mathbbm{E}[X|U]$ and construct $\hat{X}'$ such that $p_{U'\hat{V}|X}=p_{U'|X}p_{\hat{V}|U'}$ and $p_{\hat{V}|U'}=p_{\hat{X}-\mathbbm{E}[\hat{X}|U]|U'}$. Clearly, $X\leftrightarrow U'\leftrightarrow \hat{X}'$ form a Markov chain, and
\begin{align}
	\mathbbm{E}[\hat{X}'|U']&=U'+\mathbbm{E}[\hat{V}|U']\nonumber\\
	&=U'+\mathbbm{E}[\hat{X}-\mathbbm{E}[\hat{X}|U]|U']\nonumber\\
	&\stackrel{(e)}{=}U'+\mathbbm{E}[\hat{X}-\mathbbm{E}[\hat{X}|U']|U']\nonumber\\
	&=U'\mbox{ almost surely},
\end{align}
where ($e$) is because $\hat{X}\leftrightarrow U\leftrightarrow U'$ form a Markov chain.  It follows by the data processing inequality \cite[Theorem 2.8.1]{Cover1} that	
\begin{align}
	I(X;U)\geq I(X;U').
\end{align}
Moreover, we have
\begin{align}
	D&\geq\mathbbm{E}[\|X-\hat{X}\|^2]\nonumber\\
	&\stackrel{(f)}{=}\mathbbm{E}[\|X-\mathbbm{E}[X|U]\|^2]+\mathbbm{E}[\|\mathbbm{E}[X|U]-\mathbbm{E}[\hat{X}|U]\|^2]\nonumber\\&\quad+\mathbbm{E}[\|\hat{X}-\mathbbm{E}[\hat{X}|U]\|^2]\nonumber\\
	&\geq \mathbbm{E}[\|X-\mathbbm{E}[X|U]\|^2]+\mathbbm{E}[\|\hat{X}-\mathbbm{E}[\hat{X}|U]\|^2]\nonumber\\
	&=\mathbbm{E}[\|V\|^2]+\mathbbm{E}[\|\hat{V}\|]
\end{align}
and
\begin{align}
	P&\geq\mathbbm{E}[W^2_2(p_{X|U}(\cdot|U),p_{\hat{X}|U}(\cdot|U))]\nonumber\\
	&=\mathbbm{E}[W^2_2(p_{X-\mathbbm{E}[X|U]|U}(\cdot|U),p_{\hat{X}-\mathbbm{E}[\hat{X}|U]|U}(\cdot|U))]\nonumber\\&\quad+\mathbbm{E}[\|\mathbbm{E}[X|U]-\mathbbm{E}[\hat{X}|U]\|^2]\nonumber\\
	&\geq\mathbbm{E}[W^2_2(p_{X-\mathbbm{E}[X|U]|U}(\cdot|U),p_{\hat{X}-\mathbbm{E}[\hat{X}|U]|U}(\cdot|U))]\nonumber\\
	&=\mathbbm{E}[\mathbbm{E}[W^2_2(p_{X-\mathbbm{E}[X|U]|U}(\cdot|U),p_{\hat{X}-\mathbbm{E}[\hat{X}|U]|U}(\cdot|U))|U']]\nonumber\\
	&\stackrel{(g)}{\geq}\mathbbm{E}[W^2_2(p_{X-\mathbbm{E}[X|U]|U'}(\cdot|U'),p_{\hat{X}-\mathbbm{E}[\hat{X}|U]|U'}(\cdot|U'))]\nonumber\\
	&=\mathbbm{E}[W^2_2(p_{V|U'}(\cdot|U'),p_{\hat{V}|U'}(\cdot|U'))],
\end{align}
where ($f$) follows by the fact that $X-\mathbbm{E}[X|U]$, $\mathbbm{E}[X|U]-\mathbbm{E}[\hat{X}|U]$, and $\hat{X}-\mathbbm{E}[\hat{X}|U]$ are uncorrelated (which is a consequence of  the Markov structure $X\leftrightarrow U\leftrightarrow\hat{X}$) while ($g$) is due to the Markov structures $X\leftrightarrow U\leftrightarrow U'$ and $\hat{X}\leftrightarrow U\leftrightarrow U'$ as well as part 2) of Proposition \ref{prop:property}. This proves $R(D,P)\geq R'(D,P)$.

On the other hand, for any $(U',\hat{X}')$ satisfying (\ref{eq:RDP'1}) and (\ref{eq:RDP'2}),
\begin{align}
	\mathbbm{E}[\|X-\hat{X}'\|^2]&=\mathbbm{E}[\|V\|^2]+\mathbbm{E}[\|\hat{V}^2\|]
\end{align}
and
\begin{align}
	&\mathbbm{E}[W^2_2(p_{X|U'}(\cdot|U'),p_{\hat{X}'|U'}(\cdot|U'))]\nonumber\\&=\mathbbm{E}[W^2_2(p_{V|U'}(\cdot|U'),p_{\hat{V}|U'}(\cdot|U'))].
\end{align}
Therefore, we must have $R(D,P)\leq R'(D,P)$. This completes the proof of Theorem \ref{thm:continuous}.

\section{Proof of  (\ref{eq:continuousRDP}) for Restricted $\mathcal{X}$}\label{app:strict}

The proof of (\ref{eq:continuousRDP}) in Appendix \ref{app:continuous} assumes that the reconstruction has the freedom to take on values from $\mathbbm{R}^L$ (so does $\hat{X}$ in the definition of $R(D,P)$) even if the support of $p_X$ might not fully cover $\mathbbm{R}^L$.

Here we shall show that (\ref{eq:continuousRDP}) continues to hold when the source and reconstruction are confined to a strict subset $\mathcal{X}$ of $\mathbbm{R}^L$  (correspondingly, $R(D,P)$ is defined with $\hat{X}$ restricted to $\mathcal{X}$). It suffices to show
\begin{align}
	R_{\text{C}}(D,P)\leq R(D,P) \label{eq:restricted}
\end{align}
since the converse part in the proof of Theorem \ref{thm:RDP} is also applicable here.

We first establish the following technical lemma. For any $N'>0$, let $\mathbbm{B}^{(N')}=\{x\in\mathbbm{R}^L:\|x\|^2\leq N'\}$, $\overline{\mathbbm{B}^{(N')}}=\{x\in\mathbbm{R}^L:\|x\|^2> N'\}$ and define a deterministic mapping $\xi^{(N')}:\mathcal{X}\rightarrow\mathcal{X}^{(N')}$as follows.
\begin{enumerate}
	\item $\mathcal{X}\subseteq\mathbbm{B}(N')$: Let $\mathcal{X}^{(N')}=\mathcal{X}$ and $\xi^{(N')}(x)=x$ for all $x\in\mathcal{X}$.
	
	\item $\mathcal{X}\not\subseteq\mathbbm{B}(N')$: Let $\mathcal{X}^{(N')}=(\mathcal{X}\cap\mathbbm{B}^{(N')})\cup\{r^{(N')}\}$ and 
	\begin{align}
		\xi^{(N')}(x)=\begin{cases}
			x,& x\in\mathcal{X}\cap\mathbbm{B}^{(N')},\\
			r^{(N')},&x\in\mathcal{X}\cap\overline{\mathbbm{B}^{(N')}},
		\end{cases}
	\end{align}
	where $r^{(N')}$ is a fixed point in $\mathcal{X}\cap\overline{\mathbbm{B}^{(N')}}$ satisfying
	\begin{align}
		\|r^{(N')}\|^2\leq 2\inf\limits_{x\in\mathcal{X}\cap\overline{\mathbbm{B}^{(N')}}}\|x\|^2.
	\end{align}
\end{enumerate}
\begin{lemma}\label{lem:dominated}
	For any jointly distributed $X'$, $\hat{X}'$, and $U'$ with $\mathbbm{E}[\|X'\|^2]<\infty$ and $\mathbbm{E}[\|X'-\hat{X}'\|^2]<\infty$,
	\begin{align}
		&\lim\limits_{M\rightarrow\infty}\mathbbm{E}[\mathbbm{E}[\|\xi^{(N')}(X')-\xi^{(N')}(\hat{X}')\|^2|U']]\nonumber\\&=\mathbbm{E}[\mathbbm{E}[\|X'-\hat{X}'\|^2|U']].
	\end{align}
\end{lemma}
\begin{IEEEproof}
	Note that 
	\begin{align}
		&\mathbbm{E}[\mathbbm{E}[\|\xi^{(N')}(X')-\xi^{(N')}(\hat{X}')\|^2|U']]\nonumber\\&=\mathbbm{E}[\|\xi^{(N')}(X')-\xi^{(N')}(\hat{X}')\|^2]
		\end{align}
	and
	\begin{align}
		\mathbbm{E}[\mathbbm{E}[\|X'-\hat{X}'\|^2|U']]=\mathbbm{E}[\|X'-\hat{X}'\|^2].
	\end{align}
Since
\begin{align}
&\|\xi^{(N')}(X')-\xi^{(N')}(\hat{X}')\|^2\nonumber\\&\leq 2\|\xi^{(N')}(X')\|^2+2\|\xi^{(N')}(\hat{X}')\|^2\nonumber\\
&\leq 4\|X'\|^2+4\|\hat{X}'\|^2\nonumber\\
&\leq4\|X'\|^2+4\|X'-(X'-\hat{X}')\|^2\nonumber\\
&\leq 4\|X'\|^2+4(2\|X'\|^2+2\|X'-\hat{X}'\|^2)\nonumber\\
&=12\|X'\|^2+8\|X'-\hat{X}'\|^2
\end{align}
and
\begin{align}
	&\mathbbm{E}[12\|X'\|^2+8\|X'-\hat{X}'\|^2]\nonumber\\&=12\mathbbm{E}[\|X'\|^2]+8\mathbbm{E}[\|X'-\hat{X}'\|^2]\nonumber\\
	&<\infty,
\end{align}
invoking the dominated convergence theorem yields the desired result.
\end{IEEEproof}

Now we proceed to prove (\ref{eq:restricted}). It suffices to consider the case $P>0$ since the case $P=0$ is covered by the proof in Appendix \ref{app:continuous}. In light of Proposition \ref{prop:convexity}, $R(D,P)$ is continuous in $(D,P)$ for $D>0$ and $P>0$. So for any $\epsilon>0$, there exists $\rho>0$ satisfying $R(D-\rho,P-\rho)\leq R(D,P)+\epsilon$. By the definition of $R(D-\rho,P-\rho)$, we can find $p_{U\hat{X}|X}$ such that $I(X;U)\leq R(D-\rho,P-\rho)+\epsilon$, $X\leftrightarrow U\leftrightarrow\hat{X}$ form a Markov chain, $\mathbbm{E}[\|X-\hat{X}\|^2]\leq D-\rho$, and $\mathbbm{E}[W^2_2(p_{X|U}(\cdot|U),p_{\hat{X}|U}(\cdot|U))]\leq P-\rho$. 
Let $X^{(N')}=\xi^{(N')}(X)$ and $\hat{X}^{(N')}=\xi^{(N')}(\hat{X})$. 
It follows by Lemma \ref{lem:dominated} (with $U'$ being a constant) that
\begin{align}
	\lim\limits_{M\rightarrow\infty}\mathbbm{E}[\|X^{(N')}-\hat{X}^{(N')}\|^2]=\mathbbm{E}[\|X-\hat{X}\|^2].
\end{align}
Moreover, since every coupling of $p_{X|U}$ and $p_{\hat{X}|U}$ induces a coupling of $p_{X^{(N')}|U}$ and $p_{\hat{X}^{(N')}|U}$, it follows again by Lemma \ref{lem:dominated} that
\begin{align}
	&\limsup\limits_{M\rightarrow\infty}\mathbbm{E}[W^2_2(p_{X^{(N')}|U}(\cdot|U),p_{\hat{X}^{(N')}|U}(\cdot|U))]\nonumber\\&\leq\mathbbm{E}[W^2_2(p_{X|U}(\cdot|U),p_{\hat{X}|U}(\cdot|U))].
\end{align}
Therefore, we can choose a sufficiently large $M$ to ensure 
\begin{align}
	&\mathbbm{E}[\|X^{(N')}-\hat{X}^{(N')}\|^2]\leq D-\frac{\rho}{2},\\
	&\mathbbm{E}[W^2_2(p_{X^{(N')}|U}(\cdot|U),p_{\hat{X}^{(N')}|U}(\cdot|U))]\leq P-\frac{\rho}{2}.
\end{align}

Let $N$ be a postive integer such that $\mathcal{X}^{(N')}\subseteq\frac{1}{\sqrt{N}}[-N:N]^L$. For each cell $\prod_{j=1}^L[\frac{i_j}{\sqrt{N}},\frac{i_j+1}{\sqrt{N}}]$, pick some $x\in(\mathcal{X}^{(N')}\cap\prod_{j=1}^L[\frac{i_j}{\sqrt{N}},\frac{i_j+1}{\sqrt{N}}])\backslash\{r^{(M)}\}$ as its representative point\footnote{If $(\mathcal{X}^{(N')}\cap\prod_{j=1}^L[\frac{i_j}{\sqrt{N}},\frac{i_j+1}{\sqrt{N}}])\backslash\{r^{(M)}\}$ is empty, then this cell has no representative point.}, $(i_1,\cdots,i_L)\in[-N:N-1]^L$.  Let $\mathcal{Y}^{(N',N)}$ denote the set that consists of such representative points and $r^{(M)}$. Clearly, $\mathcal{Y}^{(N',N)}\subseteq\mathcal{X}^{(N')}\subseteq\mathcal{X}$. 
Construct $\xi^{(N',N)}:\mathcal{X}^{(N')}\rightarrow\mathcal{Y}^{(N',N)}$ that maps $r^{(N')}$ to itself and maps any other  $x\in\mathcal{X}^{(N')}$ to the representitive point of the cell that contains\footnote{If $x$ is contained in multiple cells, assign it to one of them. The assignment is done in a systematic manner to avoid measurability issues.} $x$. Let $Y=\xi^{(N',N)}(X^{(N')})$ and $\hat{Y}=\xi^{(N',N)}(\hat{X}^{(N')})$. Note that $Y\leftrightarrow X^{(N')}\leftrightarrow X\leftrightarrow U\leftrightarrow\hat{X}\leftrightarrow\hat{X}^{(N')}\leftrightarrow\hat{Y}$ form a Markov chain.  As a consequence,
\begin{align}
	I(Y;U)&\leq I(X^{(N')};U)\nonumber\\
	&\leq I(X;U)\nonumber\\
	&\leq R(D-\rho,P-\rho)+\epsilon\nonumber\\
	&\leq R(D,P)+2\epsilon.\label{eq:c1}
\end{align}
Moreover, we have
\begin{align}
	&\mathbbm{E}[\|Y-\hat{Y}\|^2]\nonumber\\&=\mathbbm{E}\left[\sum\limits_{\ell=1}^L|Y_{\ell}-\hat{Y}_{\ell}|^2\right]\nonumber\\ &\leq\mathbbm{E}\left[\sum\limits_{\ell=1}^L\left(|X^{(N')}_{\ell}-\hat{X}^{(N')}_{\ell}|+\frac{2}{\sqrt{N}}\right)^2\right]\nonumber\\
	&=\mathbbm{E}\left[\sum\limits_{\ell=1}^L\left(|X^{(N')}_{\ell}-\hat{X}^{(N')}_{\ell}|^2+\frac{4}{\sqrt{N}}|X^{(N')}_{\ell}-\hat{X}^{(N')}_{\ell}|\right)\right]\nonumber\\
	&\quad+\frac{4L}{N}\nonumber\\&\leq\mathbbm{E}[\|X^{(N')}-\hat{X}^{(N')}\|^2]+4\sqrt{\frac{L}{N}\mathbbm{E}[\|X^{(N')}-\hat{X}^{(N')}\|^2]}\nonumber\\&\quad+\frac{4L}{N}\label{eq:v}\\
	&\leq D-\frac{\rho}{2}+4\sqrt{\frac{L}{N}\left(D-\frac{\rho}{2}\right)}+\frac{4L}{N}\nonumber\\
	&=:\tilde{D}.\label{eq:c2}
\end{align} 
Since (\ref{eq:v}) is valid for any jointly distributed $(X^{(N')},\hat{X}^{(N')})$ and the induced $(Y,\hat{Y})$ (via mapping $\xi^{(N',N)}$), it follows that 
\begin{align}
	&W^2_2(p_{Y|U}(\cdot|u),p_{\hat{Y}|U}(\cdot|u))\nonumber\\&\leq W^2_2(p_{X^{(N')}|U}(\cdot|u),p_{\hat{X}^{(N')}|U}(\cdot|u))\nonumber\\&\quad+4\sqrt{\frac{L}{N}W^2_2(p_{X^{(N')}|U}(\cdot|u),p_{\hat{X}^{(N')}|U}(\cdot|u))}+\frac{4L}{N},\nonumber\\&\hspace{2.6in} u\in\mathcal{U}.
\end{align}
Therefore, we have
\begin{align}
	&\mathbbm{E}[W^2_2(p_{Y|U}(\cdot|U),p_{\hat{Y}|U}(\cdot|U))]\nonumber\\&\leq\mathbbm{E}[W^2_2(p_{X^{(N')}|U}(\cdot|U),p_{\hat{X}^{(N')}|U}(\cdot|U))]\nonumber\\&\quad+4\sqrt{\frac{L}{N}\mathbbm{E}[W^2_2(p_{X^{(N')}|U}(\cdot|U),p_{\hat{X}^{(N')}|U}(\cdot|U))]}+\frac{4L}{N} \nonumber\\
	&\leq P-\frac{\rho}{2}+4\sqrt{\frac{L}{N}\left(P-\frac{\rho}{2}\right)}+\frac{4L}{N}\nonumber\\
	&=:\tilde{P}.\label{eq:c3}
\end{align}
Combining (\ref{eq:c1}),  (\ref{eq:c2}), and (\ref{eq:c3}) yields
\begin{align}
R(\tilde{D},\tilde{P}|p_Y)\leq R(D,P)+2\epsilon,
\end{align}
where $R(\tilde{D},\tilde{P}|p_Y)$ denotes the counterpart of $R(\tilde{D},\tilde{P})$ for source $Y$.

According to Theorem \ref{thm:RDP}, there exists $p_{Y^nM\hat{Y}^n}=p^n_Yp_{M|Y^n}p_{\hat{Y}^n|M}$ such that $\frac{1}{n}\mathbbm{E}[\ell(M)]\leq R(D',P'|p_Y)+\epsilon$, $\frac{1}{n}\mathbbm{E}[\|Y^n-\hat{Y}^n\|^2]\leq D'$, and $\mathbbm{E}[W^2_2(p_{Y|U}(\cdot|U),p_{\hat{Y}|U}(\cdot|U))]\leq P'$. Construct $p_{X^nY^nM\hat{Y}^n}=p_{X^n|Y^n}p_{Y^M\hat{Y}^n}$ with $p_{X^n|Y^n}=p^n_{X|Y}$ and let $\hat{X}^n=\hat{Y}^n$. Note that
\begin{align}
	&\frac{1}{n}\mathbbm{E}[\|X^n-\hat{X}^n\|^2]\nonumber\\&=\frac{1}{n}\mathbbm{E}[\|X^n-\hat{Y}^n\|]\nonumber\\
	&=\frac{1}{n}\sum\limits_{t=1}^n\mathbbm{P}\{Y(t)\neq r^{(N')}\}\mathbbm{E}[\|X(t)-\hat{Y}(t)\|^2|Y(t)\neq r^{(N')}]\nonumber\\
	&\quad+\frac{1}{n}\sum\limits_{t=1}^n\mathbbm{P}\{Y(t)= r^{(N')}\}\mathbbm{E}[\|X(t)-\hat{Y}(t)\|^2|Y(t)= r^{(N')}].\label{eq:sub_0}
\end{align}
For $y\in\mathcal{Y}^{(N',N)}\backslash\{r^{(N')}\}$,
\begin{align}
	&\mathbbm{E}[\|X(t)-\hat{Y}(t)\|^2|Y(t)=y]\nonumber\\&=\mathbbm{E}[\|(X(t)-Y(t))+(Y(t)-\hat{Y}(t))\|^2|Y(t)=y]\nonumber\\
	&\leq\mathbbm{E}[\|Y(t)-\hat{Y}(t)\|^2|Y(t)=y]\nonumber\\&\quad+2\sqrt{\mathbbm{E}[\|X(t)-Y(t)\|^2\|Y(t)-\hat{Y}(t)\|^2|Y(t)=y]}\nonumber\\
	&\quad+\mathbbm{E}[\|X(t)-Y(t)\|^2|Y(t)=y]\nonumber\\
	&\leq\mathbbm{E}[\|Y(t)-\hat{Y}(t)\|^2|Y(t)=y]\nonumber\\&\quad+2\sqrt{\frac{L}{N}\mathbbm{E}[\|Y(t)-\hat{Y}(t)\|^2|Y(t)=y]}+\frac{L}{N}.
\end{align}
Therefore, 
\begin{align}
	&\mathbbm{P}\{Y(t)\neq r^{(N')}\}\mathbbm{E}[\|X(t)-\hat{Y}(t)\|^2|Y(t)\neq r^{(N')}]\nonumber\\
	&=\sum\limits_{y\in\mathcal{Y}^{(N',N)}\backslash\{r^{(N')}\}}p_{Y(t)}(y)	\mathbbm{E}[\|X(t)-\hat{Y}(t)\|^2|Y(t)=y]\nonumber\\
	&\leq\sum\limits_{y\in\mathcal{Y}^{(N',N)}}p_{Y(t)}(y)	\mathbbm{E}[\|X(t)-\hat{Y}(t)\|^2|Y(t)=y]\nonumber\\
	&\leq\mathbbm{E}[\|Y(t)-\hat{Y}(t)\|^2]+2\sqrt{\frac{L}{N}\mathbbm{E}[\|Y(t)-\hat{Y}(t)\|^2]}+\frac{L}{N}.\label{eq:sub_1}
\end{align}
Moreover, we have
\begin{align}
&\mathbbm{P}\{Y(t)= r^{(N')}\}\mathbbm{E}[\|X(t)-\hat{Y}(t)\|^2|Y(t)= r^{(N')}]\nonumber\\&\leq\mathbbm{P}\{Y(t)= r^{(N')}\}\mathbbm{E}[2\|X(t)\|^2+2\|\hat{Y}(t)\|^2|Y(t)= r^{(N')}]\nonumber\\
&\leq6\mathbbm{P}\{Y(t)= r^{(N')}\}\mathbbm{E}[\|X(t)\|^2|Y(t)= r^{(N')}]\nonumber\\
&=6\mathbbm{P}\{Y= r^{(N')}\}\mathbbm{E}[\|X\|^2|Y= r^{(N')}]\nonumber\\
&=6\tilde{\delta}_{N'},\label{eq:sub_2}
\end{align}
where
\begin{align}
	\tilde{\delta}_{N'}=\mathbbm{P}\{\|X\|^2>N'\}\mathbbm{E}[\|X\|^2|\|X\|^2>N'].
\end{align}
Substituting (\ref{eq:sub_1}) and (\ref{eq:sub_2}) into (\ref{eq:sub_0}) gives
\begin{align}
	&\frac{1}{n}\mathbbm{E}[\|X^n-\hat{X}^n\|^2]\nonumber\\&\leq\frac{1}{n}\sum\limits_{t=1}^n\Bigg(\mathbbm{E}[\|Y(t)-\hat{Y}(t)\|^2]+2\sqrt{\frac{L}{N}\mathbbm{E}[\|Y(t)-\hat{Y}(t)\|^2]}\nonumber\\&\hspace{0.63in}+\frac{L}{N}+6\tilde{\delta}_{N'}\Bigg)\nonumber\\
	&\leq\frac{1}{n}\mathbbm{E}[\|Y^n-\hat{Y}^n\|^2]+2\sqrt{\frac{L}{nN}\mathbbm{E}[\|Y^n-\hat{Y}^n\|^2]}+\frac{L}{N}+6\tilde{\delta}_{N'}\label{eq:restricted_valid}\\
	&\leq\tilde{D}+2\sqrt{\frac{L}{N}\tilde{D}}+\frac{L}{N}+6\tilde{\delta}_{N'}.
\end{align}
Since (\ref{eq:restricted_valid}) is valid for any jointly distributed $(Y^n,\hat{Y}^n)$ and the induced $(X^n,\hat{X}^n)$ (via Markov coupling $X^n\leftrightarrow Y^n\leftrightarrow\hat{Y}^n$ with $p_{X^n|Y^n}=p^n_{X|Y}$ and setting $\hat{X}^n=\hat{Y}^n$), it follows that 
\begin{align}
	&\frac{1}{n}W^2_2(p_{X^n|M}(\cdot|m),p_{\hat{X}^n|M}(\cdot|m))\nonumber\\
	&\leq \frac{1}{n}W^2_2(p_{Y^n|M}(\cdot|m),p_{\hat{Y}^n|M}(\cdot|m))\nonumber\\&\quad+2\sqrt{\frac{L}{nN}W^2_2(p_{Y^n|M}(\cdot|m),p_{\hat{Y}^n|M}(\cdot|m))}+\frac{L}{N}+6\tilde{\delta}_{N'},\nonumber\\ &\hspace{2.5in}m\in\mathcal{M}.
\end{align}
Therefore, we have
\begin{align}
	&\frac{1}{n}\mathbbm{E}[W^2_2(p_{X^n|M}(\cdot|M),p_{\hat{X}^n|M}(\cdot|M))]\nonumber\\
	&\leq \frac{1}{n}\mathbbm{E}[W^2_2(p_{Y^n|M}(\cdot|M),p_{\hat{Y}^n|M}(\cdot|M))]\nonumber\\&\quad+2\sqrt{\frac{L}{nN}\mathbbm{E}[W^2_2(p_{Y^n|M}(\cdot|M),p_{\hat{Y}^n|M}(\cdot|M))]}+\frac{L}{N}+6\tilde{\delta}_{N'}\nonumber\\
	&\leq \tilde{P}+2\sqrt{\frac{L}{N}\tilde{P}}+\frac{L}{N}+6\tilde{\delta}_{N'}.
\end{align}
Since $\mathbbm{E}[\|X\|^2]<\infty$, it follows by the dominated convergence theorem that $\tilde{\delta}_{N'}$ tends to zero as $N'\rightarrow\infty$. Moreover $\tilde{D}+2\sqrt{\frac{L}{N}\tilde{D}}+\frac{L}{N}$ ad $\tilde{P}+2\sqrt{\frac{L}{N}\tilde{P}}+\frac{L}{N}$ converge respectively to $D-\frac{\rho}{2}$ and $P-\frac{\rho}{2}$ as $N\rightarrow\infty$. So when $N'$ and $N$ are sufficiently large, we have
\begin{align}
	&\frac{1}{n}\mathbbm{E}[\|X^n-\hat{X}^n\|^2]\leq D,\label{eq:restricted_D}\\
	&\frac{1}{n}\mathbbm{E}[W^2_2(p_{X^n|M}(\cdot|M),p_{\hat{X}^n|M}(\cdot|M))]\leq P. \label{eq:restricted_P}
\end{align}
Combining (\ref{eq:restricted_D}) and (\ref{eq:restricted_P}) as well as the fact that
\begin{align}
	\frac{1}{n}\mathbbm{E}[\ell(M)]\leq R(\tilde{D},\tilde{P}|p_Y)+\epsilon\leq R(D,P)+3\epsilon
\end{align}
yields $R_{\text{C}}(D,P)\leq R(D,P)+3\epsilon$. Since $\epsilon>0$ is arbitrary, this proves (\ref{eq:restricted}) for $P>0$.

\section{Proof of Corollary \ref{cor:CM}}\label{app:CM}

For any $p_{\bar{U}|X}$ satisfying (\ref{eq:conventionalRD}), let $U=\mathbbm{E}[X|\bar{U}]$. It follows by the data processing inequality \cite[Theorem 2.8.1]{Cover1} that 
\begin{align}
	I(X;U)\leq I(X;\bar{U}).
\end{align}
Moreover, let $\hat{X}$ be jointly distributed with $(X,U)$ such that $X\leftrightarrow U\leftrightarrow\hat{X}$ form a Markov chain and $p_{\hat{X}|U}=p_{X|U}$. We have
\begin{align}
	\mathbbm{E}[\|X-\hat{X}\|^2]&=\mathbbm{E}[\|X-U\|^2]+\mathbbm{E}[\|\hat{X}-U\|^2]\nonumber\\&=2\mathbbm{E}[\|X-U\|^2]\nonumber\\
	&\leq2\mathbbm{E}[\|X-\bar{U}\|^2]\nonumber\\
	&\leq D.
\end{align}
This shows $R(D,0)\leq R(\frac{D}{2})$, which, in conjunction with (\ref{eq:continuousRDP}), implies
\begin{align}
	R_{\text{C}}(D,0)\leq R\left(\frac{D}{2}\right).\label{eq:show1}
\end{align}

It is known \cite[Section III.B]{Saldi}, \cite[Corollary 1]{wagner2022rate}, \cite[Equation (16)]{Jun-JSAIT} that (\ref{eq:RDP_M}) holds for the case $\mathbbm{E}[\|X\|^2]<\infty$ and $\Delta(x,\hat{x})=\|x-\hat{x}\|^2$. 
For any $p_{U\hat{X}|X}$ satisfying 
(\ref{eq:M1}), (\ref{eq:M2}), and (\ref{eq:M3}), we have
\begin{align}
	&I(X;U)\geq I(X;\mathbbm{E}[X|U]),\label{eq:invoke1}\\
	&I(\hat{X};U)\geq I(\hat{X};\mathbbm{E}[\hat{X}|U]),\label{eq:invoke2}\\
	&\mathbbm{E}[\|X-\hat{X}\|^2]\geq\mathbbm{E}[\|X-\mathbbm{E}[X|U]\|^2]+\mathbbm{E}[\|\hat{X}-\mathbbm{E}[\hat{X}|U]\|^2].
\end{align}
If $\mathbbm{E}[\|X-\mathbbm{E}[X|U]\|^2]\leq\mathbbm{E}[\|\hat{X}-\mathbbm{E}[\hat{X}|U]\|^2]$, then let $\bar{U}=\mathbbm{E}[X|U]$. Note that
\begin{align}
	&\mathbbm{E}[\|X-\bar{U}\|]\leq\frac{1}{2}\mathbbm{E}[\|X-\hat{X}\|^2]\leq\frac{D}{2},
\end{align}
which, together with (\ref{eq:invoke1}), implies 
\begin{align}
	R\left(\frac{D}{2}\right)\leq R_{\text{C}}(M,0).\label{eq:show2}
\end{align}
If $\mathbbm{E}[\|X-\mathbbm{E}[X|U]\|^2]>\mathbbm{E}[\|\hat{X}-\mathbbm{E}[\hat{X}|U]\|^2]$, then let $\bar{U}=\mathbbm{E}[\hat{X}|U]$. Note that
\begin{align}
	&\mathbbm{E}[\|\hat{X}-\bar{U}\|]\leq\frac{1}{2}\mathbbm{E}[\|X-\hat{X}\|^2]\leq\frac{D}{2},
\end{align}
which, together with (\ref{eq:invoke2}) and (\ref{eq:M3}), implies (\ref{eq:show2}) as well.

Combining (\ref{eq:show1}), (\ref{eq:show2}), and (\ref{eq:M<C}) proves Corollary \ref{cor:CM}.

\section{Proof of Theorem~\ref{thm:lowerbound}}\label{app:lowerbound}

We first establish the following technical lemma.

\begin{lemma}\label{lem:optimization}
	For $\sigma^2_{\ell}>0$, $\ell\in[1:L]$, $D>0$, and $P\geq 0$,
	\begin{align}
		\chi(D,P)=-\sum\limits_{\ell=1}^L\frac{1}{2}\log(2\pi e\omega_{\ell}),
	\end{align}
	where
	\begin{align}
		\chi(D,P)=&\min_{\{\gamma_{\ell},\hat{\gamma}_{\ell}\}_{\ell=1}^L}-\sum_{\ell=1}^L\frac{1}{2}\log (2\pi e\gamma_{\ell})\label{eq:optimization}\\
		&\hspace{0.4cm}\text{s.t.} \quad 0\leq  \gamma_{\ell}\leq \sigma^2_{\ell},\quad \ell\in[1:L],\label{eq:gamma1}\\%
		&\hspace{0.8cm}\quad 0\leq\hat{\gamma}_{\ell},\quad\ell\in[1:L],\label{eq:gamma2}\\%
		&\hspace{0.8cm}\quad\sum_{\ell=1}^L (\gamma_{\ell}+\hat{\gamma}_{\ell})\leq D,\label{eq:gamma3}\\%
		&\hspace{0.8cm}\quad\sum_{\ell=1}^L(\sqrt{\gamma_{\ell}}-\sqrt{\hat{\gamma}_{\ell}})^2\leq P.\label{eq:gamma4}
	\end{align}
	Moreover, the minimum value of the optimization problem in (\ref{eq:optimization}) is attained at 
	\begin{align}
		&\gamma_{\ell}=\gamma^*_{\ell}=\omega_{\ell},\quad\ell\in[1:L],\label{eq:optgamma}\\
		&\hat{\gamma}_{\ell}=\hat{\gamma}^*_{\ell}=\alpha\omega_{\ell},\quad\ell\in[1:L],\label{eq:optgammahat}
	\end{align}
	with
	\begin{align}
		\alpha=\begin{cases}
			\left(\frac{D-\sqrt{(2D-P)P}}{D-P}\right)^2,& D>P,\\
			0,&D\leq P.
		\end{cases}\label{eq:alpha}
	\end{align}
\end{lemma}
\begin{IEEEproof}
	It is easy to verify that the optimization problem in (\ref{eq:optimization}) is a convex program. The Lagrangian of this convex program is given by
	\begin{align}
		\zeta&=-\sum_{\ell=1}^L\frac{1}{2}\log(2\pi e\gamma_{\ell})+\nu_1 \left(\sum_{\ell=1}^L (\gamma_{\ell}+\hat{\gamma}_{\ell})-D\right)\nonumber\\&\quad+\nu_2\left(\sum_{\ell=1}^L(\sqrt{\gamma_{\ell}}-\sqrt{\hat{\gamma}_{\ell}})^2-P\right)+\sum_{\ell=1}^L\tau_{\ell}(\gamma_{\ell}-\sigma^2_{\ell})\nonumber\\&\quad-\sum_{\ell=1}^L \hat{\tau}_{\ell}\hat{\gamma}_{\ell}.
	\end{align} 
Note that
\begin{align}
	\frac{\partial \zeta}{\partial\gamma_{\ell}}=-\frac{1}{2\gamma_{\ell}}+\nu_1+\nu_2\left(1-\frac{\sqrt{\hat{\gamma}_{\ell}}}{\sqrt{\gamma_{\ell}}}\right)+\tau_{\ell},\quad\ell\in[1:L],
\end{align}
and
\begin{align}
\frac{\partial\zeta}{\partial\hat{\gamma}_{\ell}}=\nu_1+\nu_2\left(1-\frac{\sqrt{\gamma_{\ell}}}{\sqrt{\hat{\gamma}_{\ell}}}\right)-\hat{\tau}_{\ell},\quad\ell\in[1:L].
\end{align}
The minimum value of the optimization problem in (\ref{eq:optimization}) is attained at $\{\gamma_{\ell},\hat{\gamma}_{\ell}\}^L_{\ell=1}=\{\gamma^*_{\ell},\hat{\gamma}^*_{\ell}\}^L_{\ell=1}$
if and only if there exist nonnegative Lagrange multipliers $\nu_1$, $\nu_2$, $\tau_{\ell}$, $\hat{\tau}_{\ell}$, $\ell\in[1:L]$, such that
\begin{align}
	&0\leq  \gamma^*_{\ell}\leq \sigma^2_{\ell},\quad \ell\in[1:L],\label{eq:KKT1}\\%
	&0\leq\hat{\gamma}^*_{\ell},\quad\ell\in[1:L],\label{eq:KKT2}\\%
	&\sum_{\ell=1}^L (\gamma^*_{\ell}+\hat{\gamma}^*_{\ell})\leq D,\label{eq:KKT3}\\%
	&\sum_{\ell=1}^L(\sqrt{\gamma^*_{\ell}}-\sqrt{\hat{\gamma}^*_{\ell}})^2\leq P,\label{eq:KKT4}\\
	&-\frac{1}{2\gamma^*_{\ell}}+\nu_1+\nu_2\left(1-\frac{\sqrt{\hat{\gamma}^*_{\ell}}}{\sqrt{\gamma^*_{\ell}}}\right)+\tau_{\ell}=0,\quad\ell\in[1:L],\label{eq:KKT5}\\
	&\nu_1+\nu_2\left(1-\frac{\sqrt{\gamma^*_{\ell}}}{\sqrt{\hat{\gamma}^*_{\ell}}}\right)-\hat{\tau}_{\ell}=0,\quad\ell\in[1:L],\label{eq:KKT6}\\
	&\nu_1 \left(\sum_{\ell=1}^L (\gamma^*_{\ell}+\hat{\gamma}^*_{\ell})-D\right)=0,\label{eq:KKT7}\\
	&\nu_2\left(\sum_{\ell=1}^L(\sqrt{\gamma^*_{\ell}}-\sqrt{\hat{\gamma}^*_{\ell}})^2-P\right)=0,\label{eq:KKT8}\\
	&\tau_{\ell}(\gamma^*_{\ell}-\sigma^2_{\ell})=0,\quad\ell\in[1:L],\label{eq:KKT9}\\
	&\hat{\tau}_{\ell}\hat{\gamma}_{\ell}=0,\quad\ell\in[1:L].\label{eq:KKT10}
\end{align}
Therefore, it suffices to verify that the above 
Karush–Kuhn–Tucker conditions are satisfied by $\{\gamma^*_{\ell},\hat{\gamma}^*_{\ell}\}^L_{\ell=1}$ defined in (\ref{eq:optgamma}) and (\ref{eq:optgammahat}). We shall consider the following cases separately.
	
\begin{enumerate}
	\item  $P<D$ and $D+\sqrt{(2D-(D\wedge P))(D\wedge P)}<2\sum_{\ell}^L\sigma^2_{\ell}$ (i.e., $P<D$ and $D+\sqrt{(2D-P)P}<2\sum_{\ell}^L\sigma^2_{\ell}$):
	
	We have
	\begin{align}
		&\gamma^*_{\ell}=\omega\wedge\sigma^2_{\ell},\quad\ell\in[1:L],\\
		&\hat{\gamma}^*_{\ell}=	\left(\frac{D-\sqrt{(2D-P)P}}{D-P}\right)^2(\omega\wedge\sigma^2_{\ell}),\quad\ell\in[1:L],
	\end{align}
where $\omega$ is the unique solution to
\begin{align}
\sum\limits_{\ell=1}^L(\omega\wedge\sigma^2_{\ell})=\frac{D+\sqrt{(2D-P)P}}{2}.
\end{align}
Let
\begin{align}
	&\nu_1=\frac{P+\sqrt{(2D-P)P}}{4\omega\sqrt{(2D-P)P}},\\
	&\nu_2=\frac{D-P}{4\omega\sqrt{(2D-P)P}},\\
	&\tau_{\ell}=\frac{(\omega-\sigma^2_{\ell})\vee 0}{2\omega\sigma^2_{\ell}},\quad\ell\in[1:L],\\
	&\hat{\tau}_{\ell}=0,\quad\ell\in[1:L].
\end{align}
It is clear that the Lagrange multipliers defined above are nonnegative. Moreover, the Karush–Kuhn–Tucker conditions (\ref{eq:KKT1})--(\ref{eq:KKT10}) are satisfied.
In particular, 
\begin{align}
	&\sum\limits_{\ell=1}^L(\gamma^*_{\ell}+\hat{\gamma}^*_{\ell})\nonumber\\&=\left(1+\left(\frac{D-\sqrt{(2D-P)P}}{D-P}\right)^2\right)\sum\limits_{\ell=1}^L(\omega\wedge\sigma^2_{\ell})\nonumber\\
	&=\left(1+\left(\frac{D-\sqrt{(2D-P)P}}{D-P}\right)^2\right)\nonumber\\&\quad\times\frac{D+\sqrt{(2D-P)P}}{2}\nonumber\\
	&=D
\end{align}
and
\begin{align}
 &\sum\limits_{\ell=1}^L(\sqrt{\gamma^*_{\ell}}-\sqrt{\hat{\gamma}^2_{\ell}})^2\nonumber\\&=\left(1-\frac{D-\sqrt{(2D-P)P}}{D-P}\right)^2\sum\limits_{\ell=1}^L(\omega\wedge\sigma^2_{\ell})\nonumber\\
 &=\left(1-\frac{D-\sqrt{(2D-P)P}}{D-P}\right)^2\frac{D+\sqrt{(2D-P)P}}{2}\nonumber\\
 &=P.
\end{align}

	\item $P\geq D$ and $D+\sqrt{(2D-(D\wedge P))(D\wedge P)}<2\sum_{\ell}^L\sigma^2_{\ell}$ (i.e., $P\geq D$ and $D<\sum_{\ell}^L\sigma^2_{\ell}$):

	We have
	\begin{align}
		&\gamma^*_{\ell}=\omega\wedge\sigma^2_{\ell},\quad\ell\in[1:L],\\
		&\hat{\gamma}^*_{\ell}=0,\quad\ell\in[1:L],
	\end{align}
	where $\omega$ is the unique solution to
	\begin{align}
		\sum\limits_{\ell=1}^L(\omega\wedge\sigma^2_{\ell})=D.
	\end{align}
	Let
	\begin{align}
		&\nu_1=\frac{1}{2\omega},\\
		&\nu_2=0,\\
		&\tau_{\ell}=\frac{(\omega-\sigma^2_{\ell})\vee 0}{2\omega\sigma^2_{\ell}},\quad\ell\in[1:L],\\
		&\hat{\tau}_{\ell}=\frac{1}{2\omega},\quad\ell\in[1:L].
	\end{align}
It is clear that the Lagrange multipliers defined above are nonnegative. Moreover, the Karush–Kuhn–Tucker conditions (\ref{eq:KKT1})--(\ref{eq:KKT10}) are satisfied.
In particular, 
\begin{align}
	\sum\limits_{\ell=1}^L(\gamma^*_{\ell}+\hat{\gamma}^*_{\ell})=\sum\limits_{\ell=1}^L(\omega\wedge\sigma^2_{\ell})=D
\end{align}
and
\begin{align}
	\sum\limits_{\ell=1}^L(\sqrt{\gamma^*_{\ell}}-\sqrt{\hat{\gamma}^*_{\ell}})^2=\sum\limits_{\ell=1}^L(\omega\wedge\sigma^2_{\ell})=D\leq P.
\end{align}

	\item $P<D$ and $D+\sqrt{(2D-(D\wedge P))(D\wedge P)}\geq 2\sum_{\ell}^L\sigma^2_{\ell}$ (i.e., $P<D$ and $D+\sqrt{(2D-P)P}\geq 2\sum_{\ell}^L\sigma^2_{\ell}$):
	
	We have
	\begin{align}
		&\gamma^*_{\ell}=\sigma^2_{\ell},\quad\ell\in[1:L],\\
		&\hat{\gamma}^*_{\ell}=\left(\frac{D-\sqrt{(2D-P)P}}{D-P}\right)^2\sigma^2_{\ell},\quad\ell\in[1:L].
	\end{align}
	Let
	\begin{align}
		&\nu_1=0,\\
		&\nu_2=0,\\
		&\tau_{\ell}=\frac{1}{2\sigma^2_{\ell}},\quad\ell\in[1:L],\\
		&\hat{\tau}_{\ell}=0,\quad\ell\in[1:L].
	\end{align}
It is clear that the Lagrange multipliers defined above are nonnegative. Moreover, the Karush–Kuhn–Tucker conditions (\ref{eq:KKT1})--(\ref{eq:KKT10}) are satisfied.
In particular, 
\begin{align}
	&\sum\limits_{\ell=1}^L(\gamma^*_{\ell}+\hat{\gamma}^*_{\ell})\nonumber\\&=\left(1+\left(\frac{D-\sqrt{(2D-P)P}}{D-P}\right)^2\right)\sum\limits_{\ell=1}^L\sigma^2_{\ell}\nonumber\\
	&\leq\left(1+\left(\frac{D-\sqrt{(2D-P)P}}{D-P}\right)^2\right)\nonumber\\&\quad\times\frac{D+\sqrt{(2D-P)P}}{2}\nonumber\\
	&=D
\end{align}
and
\begin{align}
	&\sum\limits_{\ell=1}^L(\sqrt{\gamma^*_{\ell}}-\sqrt{\hat{\gamma}^*_{\ell}})^2\nonumber\\&=\left(1-\frac{D-\sqrt{(2D-P)P}}{D-P}\right)^2\sum\limits_{\ell=1}^L\sigma^2_{\ell}\nonumber\\
	&\leq \left(1-\frac{D-\sqrt{(2D-P)P}}{D-P}\right)^2\frac{D+\sqrt{(2D-P)P}}{2}\nonumber\\
	&=P.
\end{align}

\item $P\geq D$ and $D+\sqrt{(2D-(D\wedge P))(D\wedge P)}\geq 2\sum_{\ell}^L\sigma^2_{\ell}$ (i.e., $P\geq D$ and $D\geq\sum_{\ell}^L\sigma^2_{\ell}$):

We have
\begin{align}
	&\gamma^*_{\ell}=\sigma^2_{\ell},\quad\ell\in[1:L],\\
	&\hat{\gamma}^*_{\ell}=0,\quad\ell\in[1:L].
\end{align}
Let
\begin{align}
	&\nu_1=0,\\
	&\nu_2=0,\\
	&\tau_{\ell}=\frac{1}{2\sigma^2_{\ell}},\quad\ell\in[1:L],\\
	&\hat{\tau}_{\ell}=0,\quad\ell\in[1:L].
\end{align}
It is clear that the Lagrange multipliers defined above are nonnegative. Moreover, the Karush–Kuhn–Tucker conditions (\ref{eq:KKT1})--(\ref{eq:KKT10}) are satisfied.
In particular, 
\begin{align}
	\sum\limits_{\ell=1}^L(\gamma^*_{\ell}+\hat{\gamma}^*_{\ell})=\sum\limits_{\ell=1}^L\sigma^2_{\ell}\leq D
\end{align}
and
\begin{align}
	\sum\limits_{\ell=1}^L(\sqrt{\gamma^*_{\ell}}-\sqrt{\hat{\gamma}^*_{\ell}})^2=\sum\limits_{\ell=1}^L\sigma^2_{\ell}\leq D\leq P.
\end{align}
\end{enumerate}	
	
This completes the proof of Lemma \ref{lem:optimization}.
\end{IEEEproof}

Now we proceed to prove Theorem~\ref{thm:lowerbound}. 
For any $(U',\hat{X}')$ satisfying (\ref{eq:RDP'1})--(\ref{eq:RDP'4}), let $\gamma_{\ell}=\mathbbm{E}[V^2_{\ell}]$ and $\hat{\gamma}_{\ell}=\mathbbm{E}[\hat{V}^2_{\ell}]$, where $V_{\ell}$ and $\hat{V}_{\ell}$ denote the $\ell$-th component of $V$ and $\hat{V}$, respectively, $\ell\in[1:L]$. Clearly,  
\begin{align}
	&0\leq\gamma_{\ell}\leq\sigma^2_{\ell},\quad \ell\in[1:L],\label{eq:constraint1}\\
	&0\leq\hat{\gamma}_{\ell},\quad \ell\in[1:L].\label{eq:constraint2}
\end{align}
Note that	
\begin{align}
	I(X;U')&=h(X)-h(X|U')\nonumber\\
	&=h(X)-h(V|U')\nonumber\\
	&\geq h(X)-h(V)\nonumber\\
	&\geq h(X)-\sum\limits_{\ell=1}^Lh(V_{\ell})\nonumber\\
	&\stackrel{(a)}{\geq} h(X)-\sum\limits_{\ell=1}^L\frac{1}{2}\log(2\pi e\gamma_{\ell}),\label{eq:Gaussianmax}
\end{align}
where ($a$) is due to \cite[Theorem 9.6.5]{Cover1}. Moreover, we have	
\begin{align}
	D&\geq\mathbbm{E}[\|V\|^2]+\mathbbm{E}[\|\hat{V}\|^2]\nonumber\\
	&=\sum\limits_{\ell=1}^L(\gamma_{\ell}+\hat{\gamma}_{\ell})\label{eq:combdistortion}
\end{align}
and	
\begin{align}
	P&\geq\mathbbm{E}[W^2_2(p_{V|U'}(\cdot|U'),p_{\hat{V}|U'}(\cdot|U'))]\nonumber\\
	&\stackrel{(b)}{\geq} W^2_2(p_{V},p_{\hat{V}})\nonumber\\
	&\stackrel{(c)}{\geq}\sum\limits_{\ell}^LW^2_2(p_{V_{\ell}},p_{\hat{V}_{\ell}})\nonumber\\
	&\stackrel{(d)}{\geq}\sum\limits_{\ell=1}^{L}(\sqrt{\gamma_{\ell}}-\sqrt{\hat{\gamma}_{\ell}})^2,\label{eq:combperception}
\end{align}
where ($b$) is due to part 2) of Proposition \ref{prop:property}, ($c$) is due to part 1) of Proposition \ref{prop:property}, and ($d$) is due to \cite[Equation (6)]{DL82}. 	
Combining (\ref{eq:constraint1}), (\ref{eq:constraint2}), (\ref{eq:Gaussianmax}), (\ref{eq:combdistortion}), and (\ref{eq:combperception}) yields $R'(D,P)\geq h(X)+\chi(D,P)$. In light of Theorem \ref{thm:continuous} and Lemma \ref{lem:optimization},  the proof is complete.

\section{Proof of Corollary \ref{cor:Gaussianmixture}}\label{app:Gaussianmixture}
In light of Theorem \ref{thm:lowerbound}, 
\begin{align}
	R_{\text{C}}(D,P)\geq h(X)-\frac{L}{2}\log(2\pi e\omega)
\end{align}
if $\omega\leq\min\{\sigma^2_{\ell}\}_{\ell=1}^L$, where
\begin{align}
	\omega=\frac{D+\sqrt{(2D-(D\wedge P))(D\wedge P)}}{2L}.
\end{align}
Let $\Sigma$ denote the covariance matrix of $X$. Note that $\sigma^2_{\ell}$, $\ell\in[1:L]$, are the diagonal entries of $\Sigma$. 
We have
\begin{align}
	\min\{\sigma^2_{\ell}\}_{\ell=1}^L&\stackrel{(a)}{\geq}\lambda_{\min}(\Sigma)\nonumber\\
	&\stackrel{(b)}{\geq}\lambda_{\min}\left(\sum_{k=1}^K\beta_k\Sigma_k\right)\nonumber\\
	&\geq\min\{\lambda_{\min}(\Sigma_k)\}_{k=1}^K,
\end{align}
where ($a$) is due to \cite[Theorem 4.3.26]{HJ85}, and ($b$) is due to $\Sigma\succeq\sum_{k=1}^K\beta_k\Sigma_k$. 

Now it remains to show
\begin{align}
	R_{\text{C}}(D,P)\leq h(X)-\frac{L}{2}\log\left(2\pi e\omega\right)\label{eq:theother}
\end{align}
if $\omega\leq\min\{\lambda_{\min}(\Sigma_k)\}_{k=1}^K$. Note that $\omega\leq\min\{\lambda_{\min}(\Sigma_k)\}_{k=1}^K$ ensures $\Sigma_k-\omega I\succeq 0$, $k\in[1:K]$,  where $I$ denotes the $L\times L$ identity matrix.
We can write $X=U'+V$, where $U'\sim\sum_{k=1}^K\beta_k\mathcal{N}(\mu_k,\Sigma_k-\omega I)$ and $V\sim\mathcal{N}(0,\omega I)$ are independent. Moreover, let $\hat{X}'=U'+\hat{V}$ with  $\hat{V}\sim\mathcal{N}(0,\alpha\omega I)$ independent of $(U',V)$, where $\alpha$ is defined in (\ref{eq:alpha}). It is clear that (\ref{eq:RDP'1}) and (\ref{eq:RDP'2}) are satisfied. Moreover, (\ref{eq:RDP'3}) and (\ref{eq:RDP'4}) are also satisfied since
\begin{align}
	\mathbbm{E}[\|V\|^2]+\mathbbm{E}[\|\hat{V}\|^2]=L(1+\alpha)\omega=D
\end{align}
and 
\begin{align}
	&\mathbbm{E}[W^2_2(p_{V|U'}(\cdot|U'),p_{\hat{V}|U'}(\cdot|U'))]\nonumber\\&=W^2_2(p_{V},p_{\hat{V}})\nonumber\\
	&\stackrel{(c)}{=}L(\sqrt{\omega}-\sqrt{\alpha\omega})^2\nonumber\\
	&=\begin{cases}
		P,& D>P,\\
		D,& D\leq P,
	\end{cases}\nonumber\\
	&\leq P,
\end{align}
where ($c$) is due to \cite[Proposition 7]{GS84}.
Note that
\begin{align}
	I(X;U')&=h(X)-h(X|U')\nonumber\\
	&=h(X)-h(V)\nonumber\\
	&=h(X)-\frac{L}{2}\log(2\pi e\omega).
\end{align}
Therefore, when $\omega\leq\min\{\lambda_{\min}(\Sigma_k)\}_{k=1}^K$, we must have $R'(D,P)\leq h(X)-\frac{L}{2}\log(2\pi e\omega)$, which in light of Theorem \ref{thm:continuous} further implies
(\ref{eq:theother}).

\section{Proof of Corollary \ref{cor:Gaussian}}\label{app:Gaussian}

Let $Z=\Theta X$. We have $Z\sim\mathcal{N}(\Theta\mu,\Lambda)$ and consequently $h(Z)=\sum_{\ell=1}^L\frac{1}{2}\log(2\pi e\lambda_{\ell})$. Since unitary transformation is invertible and preserves the Euclidean norm, invoking  Theorem \ref{thm:lowerbound} with $X$ replaced by $Z$ proves
\begin{align}
	R_{\text{C}}(D,P)\geq\sum\limits_{\ell=1}^L\frac{1}{2}\log\frac{\lambda_{\ell}}{\omega_{\ell}}.
\end{align}



It remains to show
\begin{align}
	R_{\text{C}}(D,P)\leq\sum\limits_{\ell=1}^L\frac{1}{2}\log\frac{\lambda_{\ell}}{\omega_{\ell}}.\label{eq:Gaussianupper}
\end{align}
For any $\{\gamma_{\ell},\hat{\gamma}_{\ell}\}_{\ell=1}^L$ satisfying (\ref{eq:gamma1})--(\ref{eq:gamma4}) with $\sigma^2_{\ell}$ replaced by $\lambda_{\ell}$, $\ell\in[1:L]$, define two diagonal matrices $\Gamma$ and $\hat{\Gamma}$ with the $\ell$-th diagonal entry being $\gamma_{\ell}$ and $\hat{\gamma}_{\ell}$, respectively, $\ell\in[1:L]$. Note that (\ref{eq:gamma1}) and (\ref{eq:gamma2}) ensures $\Gamma\succeq 0$, $\hat{\Gamma}\succeq 0$, and $\Lambda-\Gamma\succeq 0$. We can write $X=U'+V$, where $U'\sim\mathcal{N}(0,\Theta^T(\Lambda-\Gamma)\Theta)$ and $V\sim\mathcal{N}(0,\Theta^T\Gamma\Theta)$ are independent. Moreover, let $\hat{X}'=U+\hat{V}'$ with $\hat{V}'\sim\mathcal{N}(0,\Theta^T\hat{\Gamma}\Theta)$ independent of $(U',V)$. It is clear that (\ref{eq:RDP'1}) and (\ref{eq:RDP'2}) are satisfied. Moreover, (\ref{eq:RDP'3}) and (\ref{eq:RDP'4}) are also satisfied since
\begin{align}
	\mathbbm{E}[\|V\|]+\mathbbm{E}[\|\hat{V}\|^2]=\sum\limits_{\ell=1}^L(\gamma_{\ell}+\hat{\gamma}_{\ell})\leq D
\end{align}
and
\begin{align}
	&\mathbbm{E}[W^2_2(p_{V|U'}(\cdot|U'),p_{\hat{V}|U'}(\cdot|U))]\nonumber\\&=W^2_2(p_{V},p_{\hat{V}})\nonumber\\
	&\stackrel{(a)}{=}\sum\limits_{\ell=1}^{L}(\sqrt{\gamma_{\ell}}-\sqrt{\hat{\gamma}_{\ell}})^2\nonumber\\
	&\leq P,
\end{align}
where ($a$) is due to \cite[Proposition 7]{GS84}.
Note that
\begin{align}
	I(X;U')&=h(X)-h(X|U')\nonumber\\
	&=h(X)-h(V)\nonumber\\
	&=h(X)-\sum\limits_{\ell=1}^L\frac{1}{2}\log(2\pi e\gamma_{\ell}).
\end{align}
Therefore, we must have $R'(P,D)\leq h(X)+\chi(D,P)$ with $\sigma^2_{\ell}$ replaced by $\lambda_{\ell}$, $\ell\in[1:L]$, in the definition of $\chi(D,P)$, which in light of Lemma \ref{lem:optimization} and Theorem \ref{thm:continuous} as well as the fact $h(X)=h(Z)=\sum_{\ell=1}^L\frac{1}{2}\log(2\pi e\lambda_{\ell})$ further implies (\ref{eq:Gaussianupper}).

\bibliographystyle{IEEEtran}
\bibliography{IT_references}

\begin{thebibliography}{10}
\providecommand{\url}[1]{#1}
\csname url@samestyle\endcsname
\providecommand{\newblock}{\relax}
\providecommand{\bibinfo}[2]{#2}
\providecommand{\BIBentrySTDinterwordspacing}{\spaceskip=0pt\relax}
\providecommand{\BIBentryALTinterwordstretchfactor}{4}
\providecommand{\BIBentryALTinterwordspacing}{\spaceskip=\fontdimen2\font plus
\BIBentryALTinterwordstretchfactor\fontdimen3\font minus
  \fontdimen4\font\relax}
\providecommand{\BIBforeignlanguage}[2]{{%
\expandafter\ifx\csname l@#1\endcsname\relax
\typeout{** WARNING: IEEEtran.bst: No hyphenation pattern has been}%
\typeout{** loaded for the language `#1'. Using the pattern for}%
\typeout{** the default language instead.}%
\else
\language=\csname l@#1\endcsname
\fi
#2}}
\providecommand{\BIBdecl}{\relax}
\BIBdecl

\bibitem{blau2019rethinking}
Y.~Blau and T.~Michaeli, ``Rethinking lossy compression: The
  rate-distortion-perception tradeoff,'' in \emph{Proc. Int. Conf. on Mach.
  Learn. (ICML)}.\hskip 1em plus 0.5em minus 0.4em\relax PMLR, 2019, pp.
  675--685.

\bibitem{Cover1}
T.~M. Cover and J.~A. Thomas, \emph{Elements of Information Theory}.\hskip 1em
  plus 0.5em minus 0.4em\relax Wiley, 1991.

\bibitem{blau2018perception}
Y.~Blau and T.~Michaeli, ``The perception-distortion tradeoff,'' in \emph{Proc.
  IEEE Conf. Comput. Vis. Pattern Recognit. (CVPR)}, 2018, pp. 6228--6237.

\bibitem{Theis-Wagner}
L.~Theis and A.~Wagner, ``A coding theorem for the rate-distortion-perception
  function,'' in \emph{Neural Compression Workshop of Int. Conf. Learn.
  Represent. (ICLR)}, 2021.

\bibitem{Jun-Ashish2021}
G.~Zhang, J.~Qian, J.~Chen, and A.~Khisti, ``Universal
  rate-distortion-perception representations for lossy compression,'' in
  \emph{Proc. Adv. Neural Inf. Process. Syst. (NeurIPS)}, 2021, pp.
  11\,517--11\,529.

\bibitem{Matsumoto18}
R.~Matsumoto, ``Introducing the perception-distortion tradeoff into the
  rate-distortion theory of general information sources,'' \emph{IEICE Comm.
  Express}, vol.~7, no.~11, pp. 427--431, 2018.

\bibitem{Matsumoto19}
------, ``Rate-distortion-perception tradeoff of variable-length source coding
  for general information sources,'' \emph{IEICE Comm. Express}, vol.~8, no.~2,
  pp. 38--42, 2019.

\bibitem{wagner2022rate}
A.~B. Wagner, ``The rate-distortion-perception tradeoff: The role of common
  randomness,'' \emph{arXiv:2202.04147}, 2022.

\bibitem{Jun-JSAIT}
J.~Chen, L.~Yu, J.~Wang, W.~Shi, Y.~Ge, and W.~Tong, ``On the
  rate-distortion-perception function,'' \emph{IEEE J. Sel. Areas Inf. Theory},
  vol.~3, no.~4, pp. 664--673, 2022.

\bibitem{Saldi}
N.~Saldi, T.~Linder, and S.~Y\"{u}ksel, ``Output constrained lossy source
  coding with limited common randomness,'' \emph{IEEE Trans.~Inf.~Theory},
  vol.~61, no.~9, pp. 4984--4998, 2015.

\bibitem{gunduz1}
X.~Niu, D.~G{\"u}nd{\"u}z, B.~Bai, and W.~Han, ``Conditional
  rate-distortion-perception trade-off,'' \emph{arXiv:2305.09318}, 2023.

\bibitem{gunduz2}
Y.~Hamdi and D.~G{\"u}nd{\"u}z, ``The rate-distortion-perception trade-off with
  side information,'' \emph{arXiv:2305.13116}, 2023.

\bibitem{video1}
F.~Mentzer, E.~Agustsson, J.~Ball\'e, D.~Minnen, N.~Johnston, and G.~Toderici,
  ``Neural video compression using gans for detail synthesis and propagation,''
  in \emph{Proc. Europ. Conf. Comput Vis. (ECCV)}, 2022.

\bibitem{image-comp1}
E.~Agustsson, M.~Tschannen, F.~Mentzer, R.~Timofte, and L.~Van~Gool,
  ``Generative adversarial networks for extreme learned image compression,'' in
  \emph{Proc. IEEE Conf. Comput. Vis. Pattern Recognit. (CVPR)}, 2019, pp.
  221--231.

\bibitem{image-comp2}
J.~Ball\'e, V.~Laparra, and E.~P. Simoncelli, ``End-to-end optimized image
  compression,'' in \emph{Proc. Int. Conf. on Learn. Represent. (ICLR)}, 2017.

\bibitem{image-comp3}
L.~Theis, W.~Shi, A.~Cunningham, and F.~Husz\'ar, ``Lossy image compression
  with compressive autoencoders,'' in \emph{Proc. Int. Conf. Learn. Represent.
  (ICLR)}, 2017.

\bibitem{image-comp4}
F.~Mentzer, E.~Agustsson, M.~Tschannen, R.~Timofte, and L.~V. Gool,
  ``Conditional probability models for deep image compression,'' in \emph{Proc.
  IEEE Conf. Comput. Vis. Pattern Recognit. (CVPR)}, 2018.

\bibitem{GAN}
F.~Mentzer, G.~Toderici, M.~Tschannen, and E.~Agustsson, ``High-fidelity
  generative image compression,'' in \emph{Proc. Adv. Neural Inf. Process.
  Syst. (NeurIPS)}, 2020.

\bibitem{SSIM}
A.~Golinski, R.~Pourreza, Y.~Yang, G.~Sautiere, and T.~S. Cohen, ``Feedback
  recurrent autoencoder for video compression,'' in \emph{Proc. Asian Conf.
  Comput. Vis.}, 2020.

\bibitem{Huan-Liu}
H.~Liu, G.~Zhang, J.~Chen, and A.~Khisti, ``Lossy compression with distribution
  shift as entropy constrained optimal transport,'' in \emph{Proc. Int. Conf.
  Learn. Represent. (ICLR)}, 2022.

\bibitem{xu2023conditional}
T.~Xu, Q.~Zhang, Y.~Li, D.~He, Z.~Wang, Y.~Wang, H.~Qin, Y.~Wang, J.~Liu, and
  Y.~Q. Zhang, ``Conditional perceptual quality preserving image compression,''
  \emph{arXiv: 2308.08154}, 2023.

\bibitem{ASP23}
T.~A. Atif, M.~A. Sohail, and S.~S. Pradhan, ``Lossy quantum source coding with
  a global error criterion based on a posterior reference map,'' \emph{arXiv:
  2302.00625}, 2023.

\bibitem{YWYML21}
Z.~Yan, F.~Wen, R.~Ying, C.~Ma, and P.~Liu, ``On perceptual lossy compression:
  The cost of perceptual reconstruction and an optimal training framework,'' in
  \emph{Proc. Int. Conf. Mach. Learn. (ICML)}, 2021.

\bibitem{TA21}
L.~Theis and E.~Agustsson, ``On the advantages of stochastic encoders,'' in
  \emph{Neural Compression Workshop of Int. Conf. Learn. Represent. (ICLR)},
  2021.

\bibitem{ElGamal}
A.~{El Gamal} and Y.~H. Kim, \emph{Network Information Theory}.\hskip 1em plus
  0.5em minus 0.4em\relax Cambridge University Press, 2011.

\bibitem{freirich2021theory}
D.~Freirich, T.~Michaeli, and R.~Meir, ``A theory of the distortion-perception
  tradeoff in wasserstein space,'' \emph{Proc. Adv. in Neural Inf. Process.
  Syst. (NeurIPS)}, vol.~34, pp. 25\,661--25\,672, 2021.

\bibitem{Yan22}
Z.~Yan, F.~Wen, and P.~Liu, ``Optimally controllable perceptual lossy
  compression,'' in \emph{Proc. Int. Conf. Mach. Learn. (ICML)}, 2022.

\bibitem{vectorGaussianRDP}
J.~Qian, S.~S. Salehkalaibar, J.~Chen, A.~Khisti, W.~Yu, W.~Shi, Y.~Ge, and
  W.~Tong, ``Rate-distortion-perception tradeoff for gaussian vector sources,''
  \emph{arXiv:2406.18008}, 2024.

\bibitem{DL82}
D.~C. Dowson and B.~V. Landau, ``The fr\'echet distance between multivariate
  normal distributions,'' \emph{J. Multivariate Anal.}, vol.~12, no.~3, pp.
  450--455, 1982.

\bibitem{HJ85}
R.~A. Horn and C.~R. Johnson, \emph{Matrix Analysis}.\hskip 1em plus 0.5em
  minus 0.4em\relax Cambridge University Press, 1985.

\bibitem{GS84}
C.~R. Givens and R.~M. Shortt, ``A class of wasserstein metrics for
  probability,'' \emph{Michigan Math. J.}, vol.~31, no.~2, pp. 231--240, 1984.

\end{thebibliography}

	
	


\appendices






\begin{IEEEbiographynophoto}{Sadaf Salehkalaibar} (Senior Member, IEEE) received her B.Sc., M.Sc., and Ph.D. degrees in Electrical Engineering from Sharif University of Technology, Tehran, Iran, in 2008, 2010, and 2014, respectively. She was a postdoctoral fellow at Telecom ParisTech, Paris, France, in 2015 and 2017. From 2016 to 2022, she served as an assistant professor in the Electrical and Computer Engineering Department at the University of Tehran. During this period, she held visiting positions at Telecom ParisTech and the National University of Singapore in 2018 and 2019. Dr. Salehkalaibar also worked as a research fellow at McMaster University in 2022 and later as a research associate at the University of Toronto until July 2024.

She is currently an assistant professor in the Department of Computer Science at the University of Manitoba. Her research interests include machine learning methods for communication systems, signal processing, network information theory, and the development of efficient algorithms for explainable artificial intelligence.
\end{IEEEbiographynophoto}

\begin{IEEEbiographynophoto}{Jun Chen} (Senior Member, IEEE) received the B.E. degree in communication engineering from Shanghai Jiao Tong University, Shanghai, China, in 2001, and the M.S. and Ph.D. degrees in electrical and computer engineering from Cornell University, Ithaca, NY, USA, in 2004 and 2006, respectively. 
	
	From September 2005 to July 2006, he was a Post-Doctoral Research Associate with the Coordinated Science Laboratory, University of Illinois at Urbana-Champaign, Urbana, IL, USA, and a Post-Doctoral Fellow with the IBM Thomas J. Watson Research Center, Yorktown Heights, NY, USA, from July 2006 to August 2007. Since September 2007, he has been with the Department of Electrical and Computer Engineering, McMaster University, Hamilton, ON, Canada, where he is currently a Professor. His research interests include information theory, machine learning, wireless communications, and signal processing. 
	
	Dr. Chen was a recipient of the Josef Raviv Memorial Postdoctoral Fellowship in 2006, the Early Researcher Award from the Province of Ontario in 2010, the IBM Faculty Award in 2010, the ICC Best Paper Award in 2020, and the JSPS Invitational Fellowship in 2021. He held the title of the Barber-Gennum Chair of information technology from 2008 to 2013 and the title of the Joseph Ip Distinguished Engineering Fellow from 2016 to 2018. He was an Associate Editor of the IEEE Transactions on Information Theory (2014 - 2016) and an Editor of the IEEE Transactions on Green Communications and Networking (2020 - 2021). He is currently serving as an Associate Editor of the IEEE Transactions on Information Theory and an Associate Editor of the IEEE Transactions on Communications.
\end{IEEEbiographynophoto}

\begin{IEEEbiographynophoto}{Ashish Khisti}  (Member, IEEE) received the B.A.Sc.
degree from the Engineering Science Program, University of Toronto, in 2002, and the master’s and
Ph.D. degrees from the Department of Electrical Engineering and Computer Science, Massachusetts
Institute of Technology (MIT), Cambridge, MA, USA, in 2004 and 2008, respectively. Since 2009,
he has been on the faculty with the Electrical and Computer Engineering (ECE) Department, University of Toronto, where he was an Assistant Professor from 2009 to 2015, an Associate Professor from 2015 to 2019, and is currently a Full Professor. He also holds the Canada Research Chair of information theory with the ECE Department. His current research interests include theory and applications of machine learning and communication networks. He is also interested in interdisciplinary research involving engineering and healthcare.
\end{IEEEbiographynophoto}

\begin{IEEEbiographynophoto}{Wei Yu} (Fellow, IEEE) received the B.A.Sc. degree in computer engineering and mathematics from the University of Waterloo, Waterloo, ON, Canada, and the M.S. and Ph.D. degrees in electrical engineering from Stanford University, Stanford, CA, USA. He is currently a Professor in the Electrical and Computer Engineering Department at the University of Toronto, Toronto, ON, Canada, where he holds a Canada Research Chair (Tier 1) in Information Theory and Wireless Communications. He is a Fellow of the Canadian Academy of Engineering and a member of the College of New Scholars, Artists, and Scientists of the Royal Society of Canada. Prof. Wei Yu was the President of the IEEE Information Theory Society in 2021 and served on its Board of Governors in 2015-2023. He served as the Chair of the Signal Processing for Communications and Networking Technical Committee of the IEEE Signal Processing Society in 2017-2018. He was an IEEE Communications Society Distinguished Lecturer in 2015-2016. He served as an Area Editor of the IEEE Transactions on Wireless Communications, as an Associate Editor for IEEE Transactions on Information Theory, and as an Editor for the IEEE Transactions on Communications and IEEE Transactions on Wireless Communications. Prof. Wei Yu received the IEEE Communications Society and Information Theory Society Joint Paper Award in 2024, the IEEE Signal Processing Society Best Paper Award in 2021, 2017, and 2008, the IEEE Marconi Prize Paper Award in Wireless Communications in 2019, the IEEE Communications Society Award for Advances in Communication in 2019, the Journal of Communications and Networks Best Paper Award in 2017, the IEEE Communications Society Best Tutorial Paper Award in 2015, and the Steacie Memorial Fellowship in 2015. He is a Clarivate Highly Cited Researcher.
\end{IEEEbiographynophoto}

\end{document}